\pdfoutput=1
\documentclass[twocolumn,pra,superscriptaddress,longbibliography]{revtex4-2}
\usepackage{graphicx,amsmath,amssymb}
\usepackage[colorlinks=true, citecolor=blue, urlcolor=blue, linkcolor=red]{hyperref}

\begin{document}
\title{Engineering exotic second-order topological semimetals by periodic driving}
\author{Bao-Qin Wang}\thanks{These two authors contributed equally.}
\affiliation{Lanzhou Center for Theoretical Physics, Key Laboratory of Theoretical Physics of Gansu Province, Lanzhou University, Lanzhou 730000, China}
\author{Hong Wu}\thanks{These two authors contributed equally.}
\affiliation{Lanzhou Center for Theoretical Physics, Key Laboratory of Theoretical Physics of Gansu Province, Lanzhou University, Lanzhou 730000, China}
\author{Jun-Hong An}
\email{anjhong@lzu.edu.cn}
\affiliation{Lanzhou Center for Theoretical Physics, Key Laboratory of Theoretical Physics of Gansu Province, Lanzhou University, Lanzhou 730000, China}
\begin{abstract}
Second-order topological semimetals (SOTSMs) are featured with hinge Fermi arcs. How to generate them in different systems has attracted much attention. We propose a scheme to create exotic SOTSMs by periodic driving. Novel Dirac SOTSMs, with a widely tunable number of nodes and hinge Fermi arcs, the adjacent nodes having the same chirality, and the coexisting nodal points and loops, are generated at ease by the periodic driving. When the time-reversal symmetry is broken, our scheme permits us to realize hybrid-order Weyl semimetals with the coexisting hinge and surface Fermi arcs. Our Weyl semimetals possess a rich hybrid of 2D sliced zero- and $\pi/T$-mode topological phases, which may be any combination of the normal insulator, Chern insulator, and second-order topological insulator. Enriching the family of topological semimetals, our scheme supplies a convenient way to artificially synthesize exotic topological phases by periodic driving.
\end{abstract}
\maketitle
\section{Introduction}
Topological quantum matters \cite{RevModPhys.82.3045,RevModPhys.83.1057,RevModPhys.88.035005,RevModPhys.90.015001,RevModPhys.93.025002} including topological insulator, superconductor, and semimetal enrich the paradigm of condensed matter physics. The finding of higher-order topological phases opens up a new frontier of topological physics  \cite{Benalcazar61,PhysRevLett.119.246401,PhysRevLett.119.246402,Schindlereaat0346,PhysRevLett.121.096803,PhysRevLett.121.116801,PhysRevLett.122.086804,PhysRevLett.123.167001,PhysRevLett.123.216803,PhysRevLett.123.256402,PhysRevLett.124.046801,PhysRevLett.124.166804,PhysRevB.101.241104,PhysRevB.92.085126}. Featured with hinge and corner states for three (3D)- and two-dimensional (2D) systems, second-order topological insulators (SOTIs) with some fantastic applications \cite{Zhang2020} have been realized in various systems \cite{SerraGarcia2018,Peterson2018,Bismuth,Imhof2018,PhysRevB.102.104109,PhysRevLett.122.204301,PhysRevLett.122.233902,PhysRevLett.122.233903,PhysRevLett.122.204301,PhysRevLett.125.255502,PhysRevLett.126.156801}.
On the other hand, topological Dirac \cite{Liu864,Neupane2014,PhysRevLett.113.027603,Xu294,PhysRevLett.108.140405,PhysRevB.85.195320,PhysRevB.88.125427,PhysRevLett.115.176404,Liu2016} and Weyl \cite{PhysRevB.83.205101,PhysRevLett.107.127205,PhysRevX.5.011029,Huang2015,xu2015discovery,PhysRevX.5.031013,Yang2015,PhysRevX.5.031023,Xu2015,Shekhar2015,Lu2013,Lu622,Soluyanov2015,PhysRevLett.115.265304,PhysRevLett.124.236401} semimetals also have been widely studied due to their chiral anomaly and close connection with diverse
topological phases \cite{PhysRevB.84.235126,PhysRevB.96.041103,PhysRevB.96.201305,PhysRevLett.126.076603,PhysRevLett.125.146402,PhysRevLett.124.236401}. Second-order topological semimetals (SOTSMs) in both Dirac \cite{PhysRevB.98.241103,Wieder2020,PhysRevB.101.205134,PhysRevB.101.220506,PhysRevLett.125.126403} and Weyl types \cite{PhysRevLett.125.146401,PhysRevLett.125.266804} were recently proposed. Different from surface Fermi arc in first-order semimetals, SOTSMs manifest in hinge Fermi arc \cite{PhysRevLett.125.146401,PhysRevLett.125.266804,PhysRevLett.125.126403}. Although they have been simulated in classical acoustic metamaterials \cite{Luo2021,Wei2021}, their observation of SOTSMs in electronic materials is hard. One of the difficulties is that the control ways to various of interactions in natural materials are limited because their features would not be switched once they are fabricated.

Coherent control via periodic driving dubbed Floquet engineering has become a versatile tool in creating topological phases in systems of ultracold atoms \cite{RevModPhys.89.011004,PhysRevLett.116.205301}, photonics \cite{Rechtsman2013,PhysRevLett.122.173901}, superconductor qubits \cite{Roushan2017}, and graphene \cite{McIver2020}. It permits us to artificially synthesize a variety of exotic topological phases \cite{PhysRevB.93.184306,PhysRevA.100.023622,PhysRevLett.121.036401,PhysRevB.102.041119,PhysRevLett.113.236803}. Greatly increasing the controllability and reducing the realistic difficulty in generating topological phases by adjusting intrinsic parameters of static systems, Floquet engineering supplies an extra dimension in exploring topological matters. Following the theme of condensed-matter physics in discovering novel quantum matter, one generally desires to know what exotic features can be created in the recently proposed SOTSMs by this novel dimension. Further, one also expects Floquet engineering to enrich the control way to complicated interactions in natural materials such that the difficulty in observing SOTSMs there could be reduced. Although some studies on Floquet engineering to SOTIs have been performed \cite{PhysRevB.103.L041115,PhysRevLett.123.016806,PhysRevB.99.045441,PhysRevB.100.115403,PhysRevLett.124.057001,PhysRevLett.124.216601,PhysRevB.103.045424,PhysRevLett.124.216601,PhysRevB.103.115308,PhysRevB.100.085138,PhysRevB.103.184510}, those for SOTSMs are still lacking. A challenge is how to establish the bulk-corner correspondence (BCC) under the fact that the periodic-driving induced symmetry breaking and the presence of novel $\pi/T$-mode corner states invalidates the BCC of the original static system.

\begin{figure}[tbp]
\centering
\includegraphics[width=1 \columnwidth]{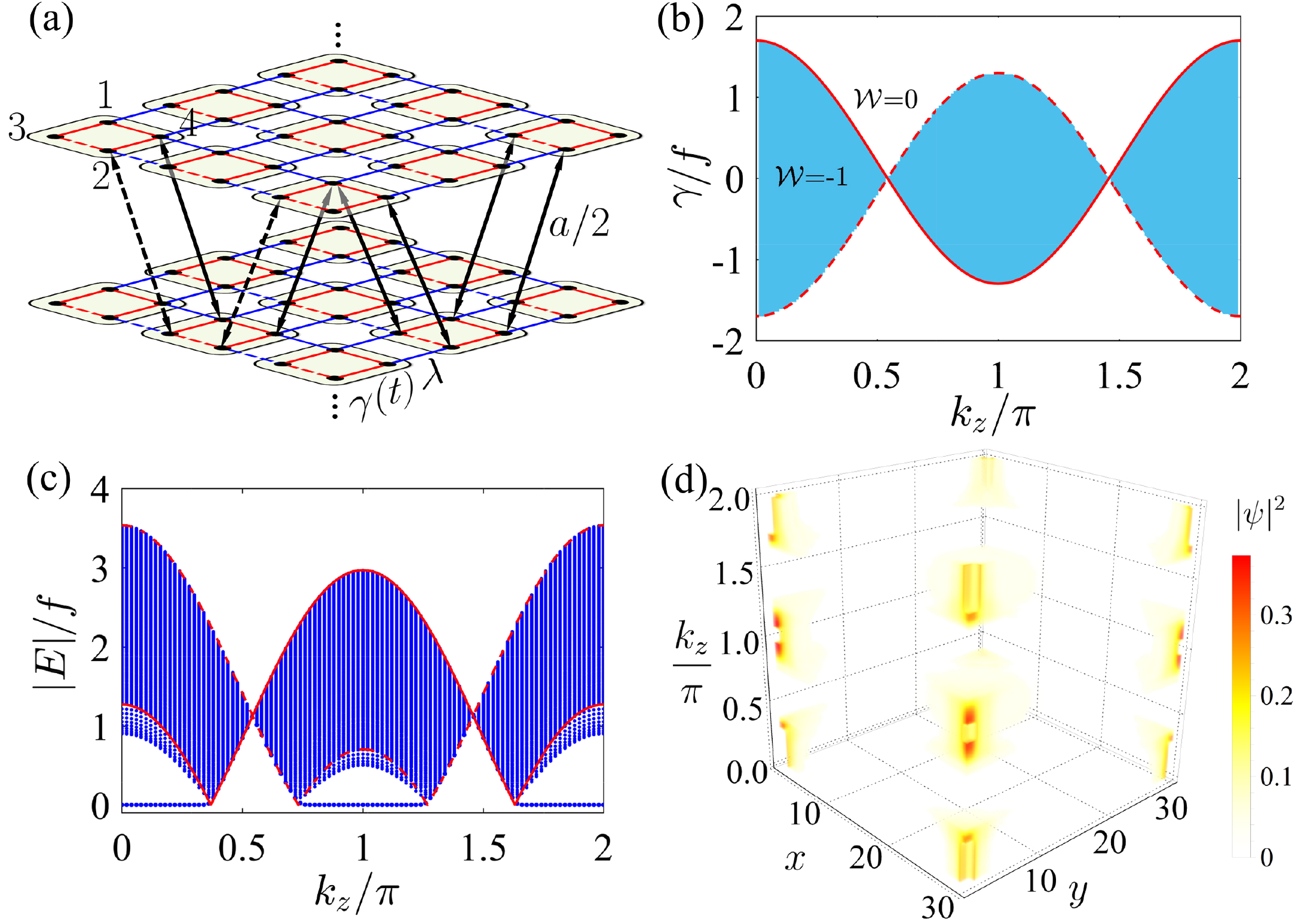}
 \caption{(a) Schematics of SOTSM on a cubic lattice with an intracell hopping rate $\gamma$, and intercell ones $\lambda$ in the single layer and $a/2$ between two neighboring layers, respectively. The dashed lines denote the hopping rates with a $\pi$-phase difference from their solid counterparts. (b) Static phase diagram described by the winding number $\mathcal{W}$. (c) Energy spectrum and (d) hinge Fermi arcs under the $x,y$-direction open boundary condition when $\gamma=0.8f$. The red solid (dashed) lines in (b) and (c) are the dispersion relations of $\theta=\pi$ ($0$). We use $\lambda=0.2f$, $a=1.5f$, and the lattice numbers $L_x=L_y=30$.}
 \label{statr}
\end{figure}

We propose a scheme to artificially create exotic SOTSMs by Floquet engineering. A complete BCC to the SOTSMs induced by periodic driving is established. Taking a system of spinless fermions moving on a cubic lattice as an example, we find diverse Dirac SOTSMs with a widely tunable number of nodes and hinge Fermi arcs, the adjacent nodes appearing in pair of same chirality, and the coexisting second-order nodal points and lines, which are absent in the static system. By adding a perturbation to break the time-reversal symmetry, hybrid-order Weyl SOTSMs manifesting in the coexisting hinge and surface Fermi arcs are created by the driving. Our work highlights Floquet engineering as a convenient way to control and explore novel SOTSMs.

\section{Static system}
We investigate a system of spinless fermions moving on a 3D lattice [see Fig. \ref{statr}(a)]. Its Hamiltonian \cite{SMP} reads $\hat{H}=\sum_{\bf k}\hat{\bf C}_{\bf k}^\dag\mathcal{H}({\bf k})\hat{\bf C}_{\bf k}$, with $\hat{\bf C}^\dag_{\bf k}=( \begin{array}{cccc} \hat{c}_{{\bf k},1}^\dag & \hat{c}_{{\bf k},2}^\dag  & \hat{c}_{{\bf k},3}^\dag  & \hat{c}_{{\bf k},4}^\dag \end{array})$ and \cite{Benalcazar61}
\begin{eqnarray}
\mathcal{H}(\mathbf{k})&=&[\gamma+\chi(k_{z})\cos k_{x}]\Gamma_5-\chi(k_{z})\sin k_{x}\Gamma_3\nonumber\\
&&-[\gamma+\chi(k_{z})\cos k_{y}]\Gamma_2-\chi(k_{z})\sin k_{y}\Gamma_1,\label{bloch}
\end{eqnarray}
where $\lambda$, $\gamma$, and $a$ are, respectively, the intercell, intracell, and interlayer hopping rates, $\chi(k_z)=\lambda+a\cos k_z$, $\Gamma_{1/2/3}=\tau_{y}\sigma_{x/y/z}$, $\Gamma_4=\tau_z\sigma_0$, and $\Gamma_5=\tau_{x}\sigma_{0}$, with $\tau_i$ and $\sigma_i$ being Pauli matrices, $\tau_0$ and $\sigma_0$ being identity matrices. It is a 3D generalization of 2D SOTIs in the Benalcazar-Bernevig-Hughes model \cite{PhysRevB.96.245115} by considering the interlayer hopping. It is different from the one of Ref. \cite{PhysRevLett.125.266804} only in the interlayer-hopping way.

The SOTSM is sliced into a family of 2D $k_z$-dependent SOTIs and normal insulators. As a prerequisite for Dirac semimetal, the time-reversal $\mathcal{T}=K$, with $K$ being the complex conjugation, and inversion $\mathcal{P}=\tau_0\sigma_y$ symmetries are respected. The system also has the mirror-rotation $\mathcal{M}_{xy}=[(\tau_0-\tau_z)\sigma_x-(\tau_0+\tau_z)\sigma_z]/2$ and chiral $\mathcal{S}=\tau_z\sigma_0$ symmetries. Thus, the 2D SOTIs are well described by $\mathcal{H}(\theta,\theta,k_z)$ along the high-symmetry line $k_x=k_y\equiv \theta$, which is diagonalized into $\text{diag}[\mathcal{H}^+(\theta,k_z),\mathcal{H}^-(\theta,k_z)]$ with $\mathcal{H}^\pm(\theta,k_z)={\bf h}^\pm\cdot{\pmb\sigma}$ and ${\bf h}^\pm=\sqrt{2}[\gamma+\chi(k_{z})\cos \theta,\pm\chi(k_{z})\sin \theta,0]$. Its topology is characterized by the mirror-graded winding numbers $\mathcal{W}(k_z)=(\mathcal{W}_{+}-\mathcal{W}_{-})/2$, where $\mathcal{W}_\pm$ are the winding numbers of $\mathcal{H}^\pm(\theta,k_z)$ \cite{Imhof2018,SMP}. The phase diagram in Fig. \ref{statr}(b) reveals a phase transition at $|\gamma|=|\chi(k_{z})|$, where $\mathcal{W}(k_z) = -1$ signifies the formation of a SOTI. The energy spectrum in Fig. \ref{statr}(c) confirms the presence of a $4|\mathcal{W}(k_z)|$-fold degenerate zero-mode state distributing at the corners. The corner state contributes to the hinge Fermi arcs [see Fig. \ref{statr}(d)]. The family of 2D SOTIs forms a 3D SOTSM which hosts the Dirac nodes at the critical points $k_z=\arccos[-(\lambda\pm\gamma)/a]$ of 2D SOTI transition. Each Dirac node carries a chirality $\mathcal{Q}$ \cite{PhysRevB.84.235126}. The chirality of the Dirac node $k_{z,0}$ equals to the difference of the winding numbers of its separated phases, i.e., $\mathcal{Q}=\mathcal{W}(k_{z,0}+\delta)-\mathcal{W}(k_{z,0}-\delta)$, with $\delta>0$ being an infinitesimal \cite{SMP}. Figure \ref{statr}(b) shows the adjacent Dirac nodes have the opposite $\mathcal{Q}$, which explains that only one four-fold degenerate corner state at most is formed.

\section{Periodic-driving induced exotic SOTSMs}
\subsection{Dirac SOTSMs via Floquet engineering}
We consider that the intracell hopping $\gamma$ periodically changes its strength in a step-like manner within the respective time duration $T_1$ and $T_2$ as
\begin{equation}
\gamma(t) =\begin{cases}\gamma_1=q_1f , ~t\in\lbrack mT, mT+T_1)\\\gamma_2=q_2f,~ t\in\lbrack mT+T_1, (m+1)T), \end{cases}
\label{procotol}
\end{equation}
where $m\in \mathbb{Z}$, $T=T_{1}+T_{2}$ is the driving period, and $f$ is an energy scale to make the amplitudes $q_j$ dimensionless. Determined by the overlap of spatial wave-function of the fermion on the relevant sites, $\gamma(t)$ could be realized by applying electric field via gate voltage. A periodic system $\hat{H}(t) $ does not have well-defined energies. According to Floquet theorem, the one-period evolution operator $\hat{U}_T=\mathbb{T}e^{-i\int_{0}^{T}\hat{H}(t)dt}$ defines an effective Hamiltonian $\hat{H}_\text{eff}\equiv {i\over T}\ln \hat{U}_T$ whose eigenvalues are called the quasienergies \cite{PhysRevA.7.2203,PhysRevA.91.052122}. The SOTSMs of our periodic system are defined in such quasienergy spectrum. Applying Floquet theorem on a general four-band $\mathcal{H}_j({\bf k})=\mathbf{n}_j\cdot\mathbf{\Gamma}$, we have $\mathcal{H}_\text{eff}({\bf k})={i\over T}\ln [e^{-i\mathcal{H}_2({\bf k})T_2}e^{-i\mathcal{H}_1({\bf k})T_1}]$ \cite{SMP}. First, we obtain from $\mathcal{H}_\text{eff}({\bf k})$ that the bands close for $\mathbf{k}$ and driving parameters satisfying either
\begin{eqnarray}
&&T_{j}E_{j}=z_{j}\pi ,\label{ejj1} \\
\text{or}~ &&
\begin{cases}
\underline{\mathbf{n}}_{1}\cdot \underline{\mathbf{n}}_{2}=\pm 1, \\
T_{1}{E}_{1}\pm T_{2}{E}_{2}=z\pi ,%
\end{cases}\label{ejj2}
\end{eqnarray}
with $\underline{\mathbf{n}}_{j}\equiv\mathbf{n}_{j}/|\mathbf{n}_{j}|$, at the quasienergy zero (or $\pi/T$) when $z_j$ are integers with same (or different) parities and $z$ is even (or odd) number. Giving the positions of Dirac nodes, Eqs. \eqref{ejj1} and \eqref{ejj2} provide a guideline to control the driving parameters for engineering various Dirac nodes at will. Deriving ${\bf n}_j$ from Eq. \eqref{bloch} with $\gamma$ driven as Eq. \eqref{procotol}, we obtain the conditions for the Dirac nodes as follows.
\\ \textbf{Case I:} Equation \eqref{ejj1} results in that the Dirac nodes present at ${\bf k}$ satisfying
\begin{equation}
\sqrt{2}[\gamma^2_j+\chi^2(k_z)+\gamma_j\chi(k_z)(\cos k_x+\cos k_y)]^{1\over 2}T_j=z_j\pi.\label{nlp}
\end{equation}Satisfied by three independent parameters $(k_x,k_y,k_z)$, the two constraints in Eqs. \eqref{nlp} result in the band-touching points to form a loop instead of discrete points. Thus, it generally gives a nodal-line semimetal. Defined in the full Brillouin zone (BZ), Eqs. \eqref{nlp} describe the physics beyond the high-symmetry lines.
\\ \textbf{Case II:} $\underline{\mathbf{n}}_1\cdot\underline{\mathbf{n}}_2=\pm1$ needs the high-symmetry line $\theta=0$ or $\pi$. According to Eq. \eqref{ejj2}, the Dirac nodes present when
\begin{equation}
\sqrt{2}\big[|\gamma_1+\chi(k_z)e^{i\theta}|T_1\pm|\gamma_2+\chi(k_z)e^{i\theta}|T_2\big]=z_{\theta,\pm}\pi, \label{npcd}
\end{equation}for sgn[$\prod_{j=1}^2(\gamma_j+\chi(k_z)e^{i\theta})$]$=\pm1$. Satisfied by discrete $\theta$ and $k_z$, it gives a nodal-point semimetal.

It is interesting to see that we not only can control the number and the position of the Dirac nodes but also can create nodal-line semimetal from the static nodal-point one by virtue of the periodic driving.

\begin{figure}[tbp]
\centering
\includegraphics[width=0.98\columnwidth]{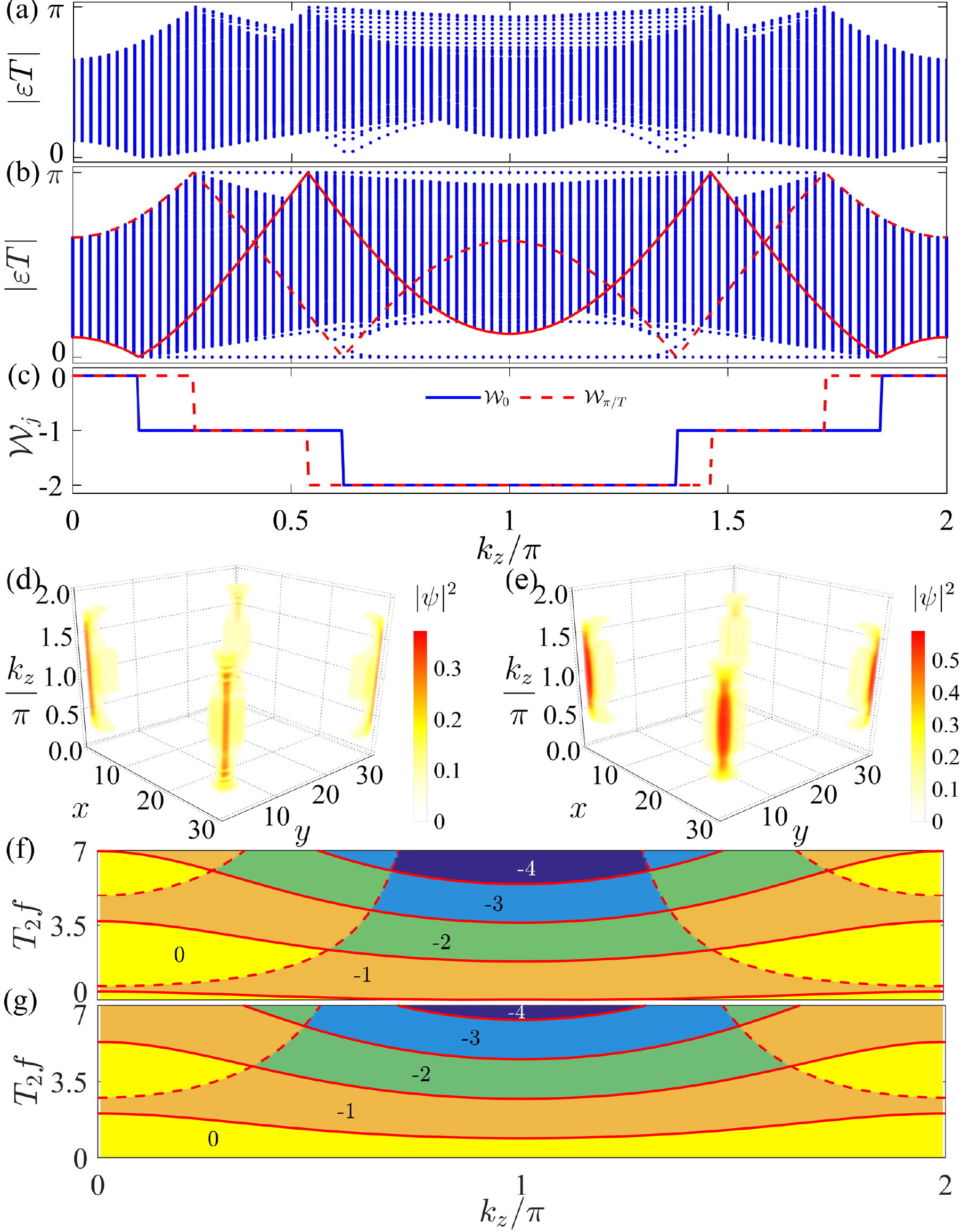}
\caption{Quasienergy spectra as the change of $k_{z}$ under $x$- (a) and $x,y$-direction (b) open boundary conditions. The red solid (dashed) line is the dispersion relation along the high-symmetry line $\theta=\pi$ ($0$). (c) Mirror-graded winding numbers for the zero- and $\pi/T$-mode corner states as the change of $k_{z}$. Hinge Fermi arcs contributed by the zero-(d) and $\pi/T$-mode (e) corner states. Phase diagram characterized by $\mathcal{W}_0$ (f) and $\mathcal{W}_{\pi/T}$ (g). Phase boundaries obtained from Eq. \eqref{npcd} with $z_{\pi,-}=0$, $-2$, $-4$, and $-6$ in (f) and $-1$, $-3$, $-5$, and $-7$ in (g) by solid lines and with $z_{0,-}=0$ and $-2$ in (f) and $-1$ in (g) by dashed lines. We use $T_1=0.5f^{-1}$, $T_2=3.5f^{-1}$, $q_{2}=-q_{1}=1.2$, $\lambda=-0.7f$, and $a=0.55f$. } \label{traj}
\end{figure}
Second, we establish the BCC from $\mathcal{H}_\text{eff}({\bf k})$ for our system. Although $\mathcal{H}_\text{eff}({\bf k})$ inherits the inversion and mirror-rotation symmetries, it does not have the time-reversal and chiral symmetries \cite{SMP}. To recover the symmetries, we make two unitary transformations $G_j({\bf k})=e^{i(-1)^j\mathcal{H}_j({\bf k})T_j/2}$ and obtain $\tilde{\mathcal{H}}_{\text{eff},j}({\bf k})=G_j({\bf k})\mathcal{H}_{\text{eff}}({\bf k})G_j^{-1}({\bf k})$ \cite{PhysRevA.100.023622}. Respecting the time-reversal and chiral symmetries, $\tilde{\mathcal{H}}_{\text{eff},j}({\bf k})$ have well-defined mirror-graded winding numbers $\mathcal{W}_j$. Since the transformations do not change the quasienergies, the SOTIs in $\mathcal{H}_\text{eff}({\bf k})$ at the quasienergies zero and $\pi/T$ are described by $\mathcal{W}_j$ as
\begin{equation}
\mathcal{W}_{0}=(\mathcal{W}_{1}+\mathcal{W}_{2})/2,~~\mathcal{W}_{\pi/T}=(\mathcal{W}_{1}-\mathcal{W}_{2})/2.\label{Arise}
\end{equation}Reflecting the firm BCC, the numbers of the zero- and $\pi/T$-mode corner states equal to $4|\mathcal{W}_0|$ and $4|\mathcal{W}_{\pi/T}|$. The $k_z$-dependent SOTIs form a 3D SOTSMs, which hosts second-order Dirac nodes at the critical points of a 2D topological phase transition and hinge Fermi arcs from the distribution of the corner states. Note that it is equivalent to deem that $\mathcal{H}_{\text{eff}}({\bf k})$ has hidden time-reversal and chiral symmetries under the redefined operations $G^{-1}_j(-{\bf k})\mathcal{T}G_j({\bf k})$ and $G^{-1}_j({\bf k})\mathcal{S}G_j({\bf k})$ \cite{SMP}.

We demonstrate the constructive role of the periodic driving in generating novel Dirac nodal-point SOTSMs in Fig. \ref{traj}. The quasienergy spectrum under the $x$-direction open boundary in Fig. \ref{traj}(a) shows a topologically trivial phase, while the one under the $x,y$-direction open boundary in Fig. \ref{traj}(b) shows rich topological phases in both the quasienergies zero and $\pi/T$. It signifies the diverse topological phases trivial in the first order but nontrivial in the second order. The Dirac nodal points in Fig. \ref{traj}(b) at $k_z=0.15\pi$, $0.54\pi$, $0.28\pi$, and $0.62\pi$ are well explained by Eq. \eqref{npcd} with $z_{\pi,-}=-2$, $-3$, $z_{0,-}=-1$ and 0, respectively. Compared with the static case, the number of the nodal points is dramatically enhanced. It verifies the periodic driving as a useful way to manipulate the nodal points. The 2D SOTIs are completely characterized by the winding number $\mathcal{W}_j$ defined in $\tilde{\mathcal{H}}_{\text{eff},j}$. The numbers $4|\mathcal{W}_0|$ and $4|\mathcal{W}_{\pi/T}|$ correctly count the zero- and $\pi/T$-mode corner states [see Fig. \ref{traj}(c)]. Another interesting result is that the chiralities of the adjacent Dirac nodal points possess the same sign instead of the opposite sign in the static case. This explains why more corner states than the static case are created by the periodic driving. It also endows the Dirac nodal points in our periodic system with robustness to the perturbation-induced annihilation, which is only sensitive to the nodal points with opposite chiralities \cite{PhysRevB.84.235126}. Both of the zero- and $\pi/T$-mode corner states contribute the hinge Fermi arcs [see Figs. \ref{traj}(d) and \ref{traj}(e)] of the SOTSMs.

\begin{figure}[tbp]
\centering
\includegraphics[width=0.98\columnwidth]{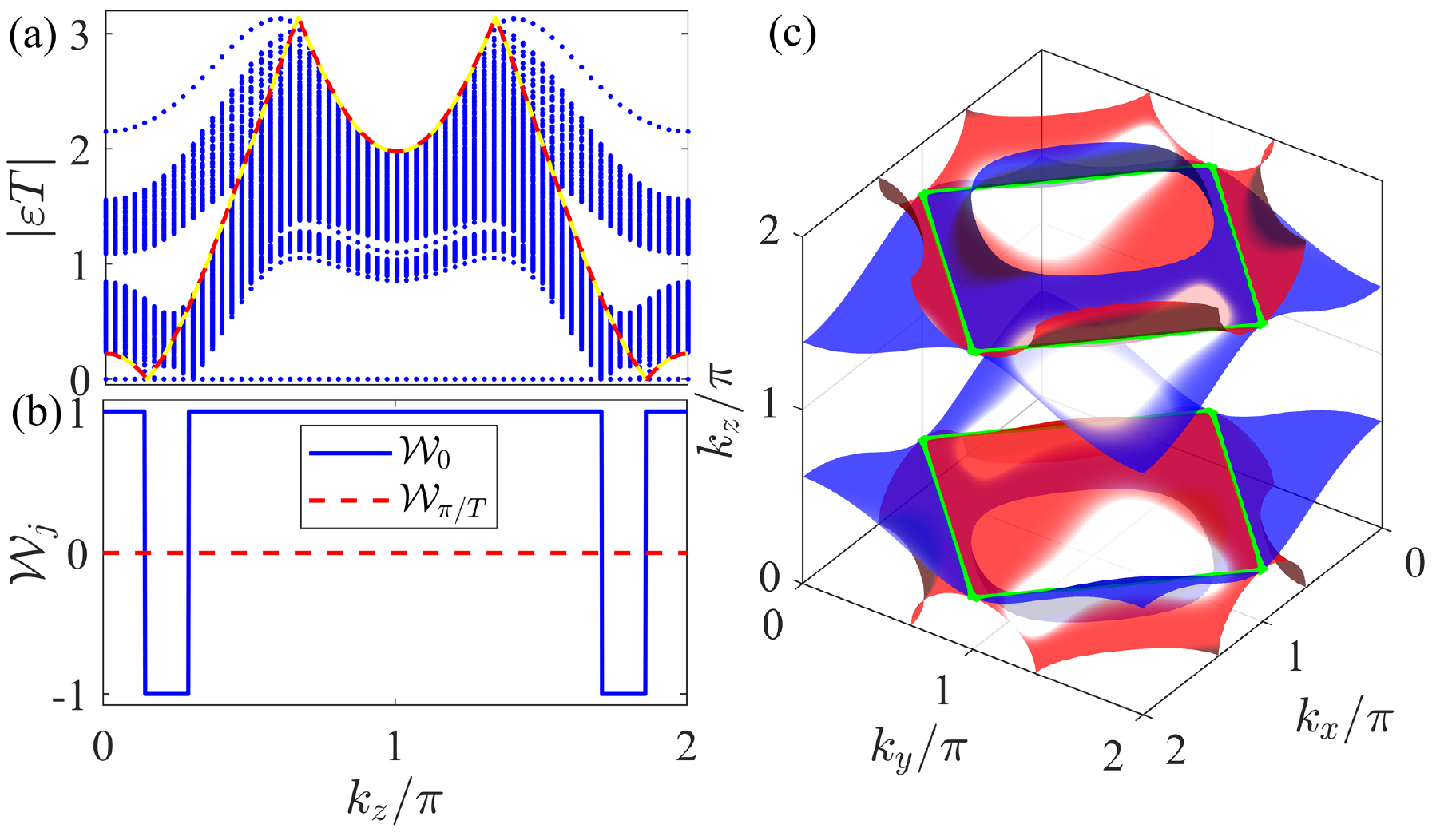}
\caption{(a) Quasienergy spectrum under the $x,y$-direction open boundary condition. The dispersion relations of $\theta=\pi$ (yellow solid line) and $0$ (red dashed line) determine the Dirac nodal points. (b) Mirror-graded winding numbers for the zero- and $\pi/T$-mode corner states. (c) Surfaces in BZ satisfying Eqs. \eqref{nlp} with $z_j=1$ for $j=1$ (red) and $2$ (blue). The green solid line is the intersecting line of the two surfaces. We use $T_1=T_2=f^{-1}$, $q_2=-q_1=1$, $\lambda=1.5f$, and $a=0.8f$.} \label{5}
\end{figure}

%Equations \eqref{nlp} and \eqref{npcd} reveal that we can manipulate the Dirac nodes by tuning the driving parameters. It supplies a convenient way to explore the second-order topology without resorting to the change of the intrinsic system-parameters.
To give a global picture of the Dirac nodal-point SOTSMs in our periodic system, we plot in Figs. \ref{traj}(f) and \ref{traj}(g) the phase diagram characterized by $\mathcal{W}_0$ and $\mathcal{W}_{\pi/T}$ in the $k_{z}$-$T_{2}$ plane. Much richer 2D-sliced SOTIs with a widely tunable number of zero- and $\pi/T$-mode corner states than the static case in Fig. \ref{statr}(b) are created by the periodic driving. The phase boundaries well described by Eq. \eqref{npcd} correspond to the Dirac nodal points of the SOTSMs. Different from the static case, where the Dirac nodal points separate the trivial and SOTIs, the ones in our periodic system also separate the SOTIs with a different number of corner states. A clear tendency with the increase of the period $T_2$ is that the number of the Dirac nodal points increases, which can be analytically understood from Eq. \eqref{npcd}.

Next, we create the Dirac nodal-loop SOTSMs from the static nodal-point ones via engineering the periodic driving to satisfy Eqs. \eqref{nlp}. The quasienergy spectrum in Fig. \ref{5}(a) reveals that, besides the zero-mode Dirac nodal points at $k_z=0.14\pi$ and $1.86\pi$, recoverable by Eq. \eqref{npcd} with $z_{\pi/0,+}=2$, and the $\pi/T$-mode ones at $k_z=0.66\pi$ and $1.34\pi$, recoverable by Eq. \eqref{npcd} with $z_{\pi/0,+}=1$, there are two extra band-touching points at $k_z=0.29\pi$ and $1.71\pi$. Plotting the two surfaces governed by Eqs. \eqref{nlp} in the BZ in Fig. \ref{5}(c), we really see two closed intersecting lines at $k_z=0.29\pi$ and $1.71\pi$. It confirms the presence of two parallel nodal loops. The associated mirror-graded winding numbers in Fig. \ref{5}(b) show that both of the nodal points and loops cause the second-order topological phase transition, which endows them with the second-order feature. All these results verify the formation of a novel SOTSM with coexisting nodal points and loops via periodically driving a static Dirac nodal-point one. Such phase has not been found in static systems. Although a similar semimetal with coexisting nodal points and loops was reported in the static system \cite{LU20202080}, it is in the first-order Weyl type. However, ours is in the second-order Dirac type and protected by both $\mathcal{P}$ and $\mathcal{T}$ symmetries. The result proves the distinguished role of the periodic driving in creating exotic matters absent in static systems.

\subsection{Weyl SOTSMs via Floquet engineering}
\begin{figure}[tbp]
\centering
\includegraphics[width=0.98\columnwidth]{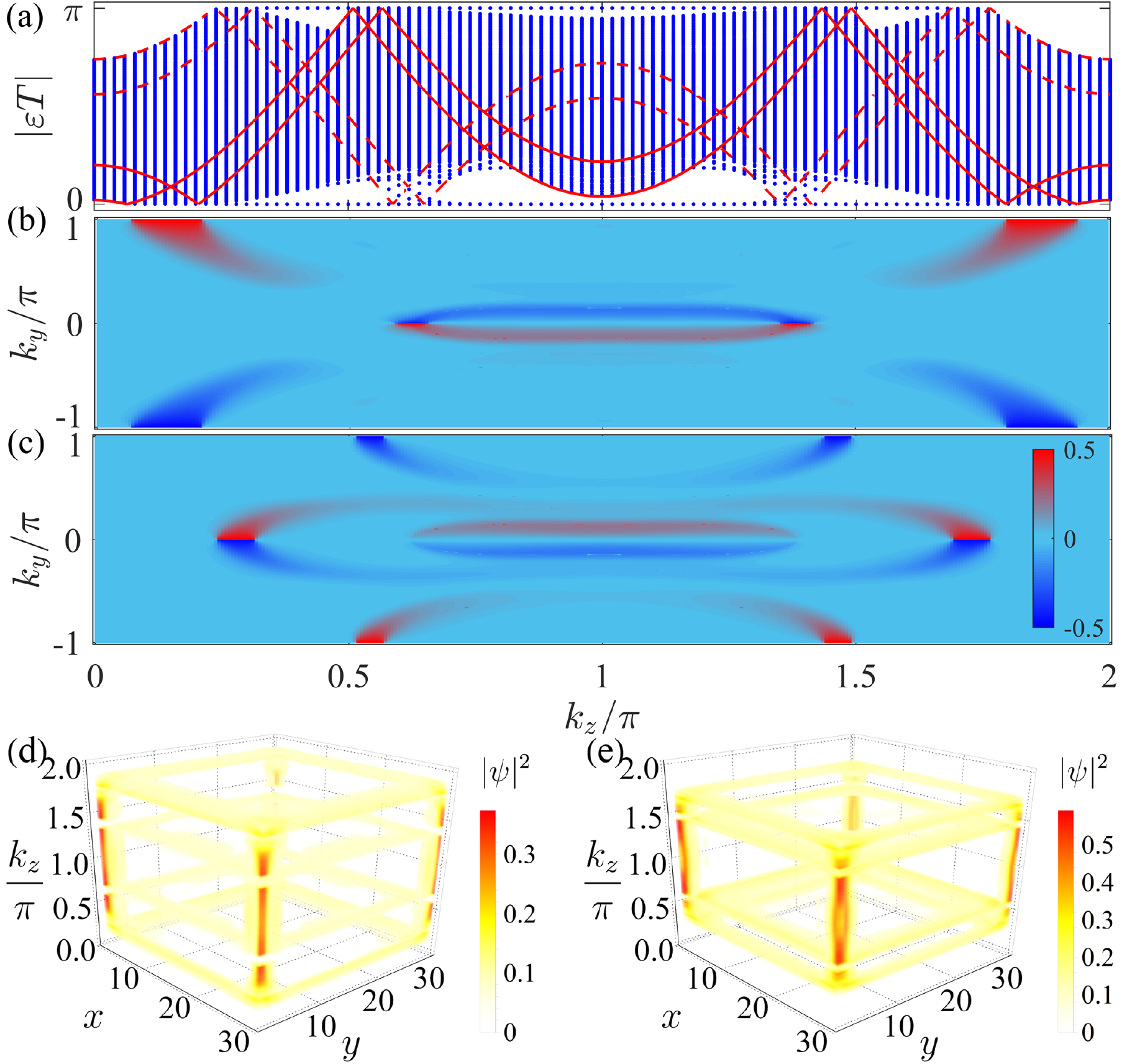}
\caption{(a) Quasienergy spectra under the $x,y$-direction open boundary condition. Wannier centers for the zero (b) and $\pi/T$ (c) quasienergy gaps. Coexisting surface and hinge Fermi arcs contributed by the zero-(d) and $\pi/T$-mode (e) first-order boundary and second-order corner states. We use $p=0.07f$ and the others being same as Fig. \ref{traj}. } \label{4}
\end{figure}
Our periodic driving scheme \eqref{procotol} can be used to create novel Weyl SOTSMs by introducing a perturbation $\Delta \mathcal{H}=ip\Gamma_{1}\Gamma_{3}$ to break the $\mathcal{T}$ symmetry. The quasienergy spectrum in Fig. \ref{4}(a) reveals that each Dirac point in Fig. \ref{traj}(a) splits into two Weyl points with a Chern insulator formed between them. Each Weyl point can be analytically explained by our band-touching condition \cite{SMP}. The Chern insulator is signified by the gapless chiral boundary states, which can be topologically witnessed by the Wannier center \cite{PhysRevLett.125.146401}. The Wannier center of the zero- and $\pi/T$-mode gaps jumping from $-0.5$ to $0.5$ when $k_y$ runs from $-\pi$ to $\pi$ verifies the formation of a Chern band [see Figs. \ref{4}(b) and \ref{4}(c)], which contributes the surface Fermi arcs. Therefore, we have realized a hybrid-order Weyl semimetal, which is featured with the coexisting first- and second-order Weyl points as well as the surface and hinge Fermi arcs [see Figs. \ref{4}(d) and \ref{4}(e)]. This result dramatically enriches the family of the recently proposed purely second-order Weyl semimetal in static systems \cite{PhysRevLett.125.146401,PhysRevLett.125.266804}. Another novel character of our periodic system is that it possesses a rich hybrid of topological phases in the zero and $\pi/T$ modes for a given $k_z$ in the 2D-sliced subsystem, which can be any combination of normal insulator, Chern insulator, and SOTI in the zero and $\pi/T$ modes. This is also absent in static systems. All of these results confirm again the superiority of the periodic driving in freely tuning and synthesizing exotic topological matters. Note that the hybrid-order nature of the Weyl semimetal in our periodic system does not depend on the form of the $\mathcal{T}$-symmetry breaking perturbation. To other forms, the explicit locations on forming the Weyl points may be different, but the conclusion that a Chern band is present between each pair of Weyl points does not change.
\section{Discussion and conclusion}
Although we only show the generation of the same order of semimetals as the static case, the periodic driving also has the ability to create the SOTSMs from the static first-order semimetal or even normal insulator \cite{SMP}. The step like driving protocol is considered just for analytical solvability. Our scheme is generalizable to other driving forms. The SOTSMs have been predicted in Cd$_3$As$_2$, KMgBi, and PtO$_2$ \cite{PhysRevLett.124.156601,Wieder2020} and realized in classical acoustic metamaterials \cite{Luo2021,Wei2021}. Periodic driving has exhibited its power in engineering exotic phases in electronic material \cite{Mahmood2016,McIver2020}, ultracold atoms \cite{Wintersperger2020}, superconductor qubits \cite{Roushan2017}, and photonics \cite{Rechtsman2013,Mukherjee2017,Maczewsky2017,PhysRevLett.122.173901}. The progress indicates that our result is realizable in the recent experimental state of the art.

In summary, we have investigated the exotic SOTSMs induced by periodic driving. It is revealed that the periodic driving provides a sufficient freedom in creating novel SOTSMs absent in static systems. The discovered widely tunable number of nodes and hinge Fermi arcs, the adjacent nodes with same chirality, and the coexisting nodal points and nodal loops in Dirac SOTSMs and the hybrid-order Weyl semimetals with the coexisting hinge and surface Fermi arcs dramatically enrich the family of topological semimetals in natural materials. Our result indicates that the periodic driving supplies a feasible and convenient way to explore the exotic semimetal physics by adding the time periodicity as a novel control dimension. This significantly reduces the difficulties in fabricating specific material structure in static systems.

\section*{Acknowledgments}
The work is supported by the National Natural Science Foundation (Grants No. 11875150, No. 11834005, and No. 12047501).
\bibliography{references}

\begin{thebibliography}{99}%
\makeatletter
\providecommand \@ifxundefined [1]{%
 \@ifx{#1\undefined}
}%
\providecommand \@ifnum [1]{%
 \ifnum #1\expandafter \@firstoftwo
 \else \expandafter \@secondoftwo
 \fi
}%
\providecommand \@ifx [1]{%
 \ifx #1\expandafter \@firstoftwo
 \else \expandafter \@secondoftwo
 \fi
}%
\providecommand \natexlab [1]{#1}%
\providecommand \enquote  [1]{``#1''}%
\providecommand \bibnamefont  [1]{#1}%
\providecommand \bibfnamefont [1]{#1}%
\providecommand \citenamefont [1]{#1}%
\providecommand \href@noop [0]{\@secondoftwo}%
\providecommand \href [0]{\begingroup \@sanitize@url \@href}%
\providecommand \@href[1]{\@@startlink{#1}\@@href}%
\providecommand \@@href[1]{\endgroup#1\@@endlink}%
\providecommand \@sanitize@url [0]{\catcode `\\12\catcode `\$12\catcode
  `\&12\catcode `\#12\catcode `\^12\catcode `\_12\catcode `\%12\relax}%
\providecommand \@@startlink[1]{}%
\providecommand \@@endlink[0]{}%
\providecommand \url  [0]{\begingroup\@sanitize@url \@url }%
\providecommand \@url [1]{\endgroup\@href {#1}{\urlprefix }}%
\providecommand \urlprefix  [0]{URL }%
\providecommand \Eprint [0]{\href }%
\providecommand \doibase [0]{https://doi.org/}%
\providecommand \selectlanguage [0]{\@gobble}%
\providecommand \bibinfo  [0]{\@secondoftwo}%
\providecommand \bibfield  [0]{\@secondoftwo}%
\providecommand \translation [1]{[#1]}%
\providecommand \BibitemOpen [0]{}%
\providecommand \bibitemStop [0]{}%
\providecommand \bibitemNoStop [0]{.\EOS\space}%
\providecommand \EOS [0]{\spacefactor3000\relax}%
\providecommand \BibitemShut  [1]{\csname bibitem#1\endcsname}%
\let\auto@bib@innerbib\@empty
%</preamble>
\bibitem [{\citenamefont {Hasan}\ and\ \citenamefont
  {Kane}(2010)}]{RevModPhys.82.3045}%
  \BibitemOpen
  \bibfield  {author} {\bibinfo {author} {\bibfnamefont {M.~Z.}\ \bibnamefont
  {Hasan}}\ and\ \bibinfo {author} {\bibfnamefont {C.~L.}\ \bibnamefont
  {Kane}},\ }\bibfield  {title} {\bibinfo {title} {Colloquium: Topological
  insulators},\ }\href {https://doi.org/10.1103/RevModPhys.82.3045} {\bibfield
  {journal} {\bibinfo  {journal} {Rev. Mod. Phys.}\ }\textbf {\bibinfo {volume}
  {82}},\ \bibinfo {pages} {3045} (\bibinfo {year} {2010})}\BibitemShut
  {NoStop}%
\bibitem [{\citenamefont {Qi}\ and\ \citenamefont
  {Zhang}(2011)}]{RevModPhys.83.1057}%
  \BibitemOpen
  \bibfield  {author} {\bibinfo {author} {\bibfnamefont {X.-L.}\ \bibnamefont
  {Qi}}\ and\ \bibinfo {author} {\bibfnamefont {S.-C.}\ \bibnamefont {Zhang}},\
  }\bibfield  {title} {\bibinfo {title} {Topological insulators and
  superconductors},\ }\href {https://doi.org/10.1103/RevModPhys.83.1057}
  {\bibfield  {journal} {\bibinfo  {journal} {Rev. Mod. Phys.}\ }\textbf
  {\bibinfo {volume} {83}},\ \bibinfo {pages} {1057} (\bibinfo {year}
  {2011})}\BibitemShut {NoStop}%
\bibitem [{\citenamefont {Chiu}\ \emph {et~al.}(2016)\citenamefont {Chiu},
  \citenamefont {Teo}, \citenamefont {Schnyder},\ and\ \citenamefont
  {Ryu}}]{RevModPhys.88.035005}%
  \BibitemOpen
  \bibfield  {author} {\bibinfo {author} {\bibfnamefont {C.-K.}\ \bibnamefont
  {Chiu}}, \bibinfo {author} {\bibfnamefont {J.~C.~Y.}\ \bibnamefont {Teo}},
  \bibinfo {author} {\bibfnamefont {A.~P.}\ \bibnamefont {Schnyder}},\ and\
  \bibinfo {author} {\bibfnamefont {S.}~\bibnamefont {Ryu}},\ }\bibfield
  {title} {\bibinfo {title} {Classification of topological quantum matter with
  symmetries},\ }\href {https://doi.org/10.1103/RevModPhys.88.035005}
  {\bibfield  {journal} {\bibinfo  {journal} {Rev. Mod. Phys.}\ }\textbf
  {\bibinfo {volume} {88}},\ \bibinfo {pages} {035005} (\bibinfo {year}
  {2016})}\BibitemShut {NoStop}%
\bibitem [{\citenamefont {Armitage}\ \emph {et~al.}(2018)\citenamefont
  {Armitage}, \citenamefont {Mele},\ and\ \citenamefont
  {Vishwanath}}]{RevModPhys.90.015001}%
  \BibitemOpen
  \bibfield  {author} {\bibinfo {author} {\bibfnamefont {N.~P.}\ \bibnamefont
  {Armitage}}, \bibinfo {author} {\bibfnamefont {E.~J.}\ \bibnamefont {Mele}},\
  and\ \bibinfo {author} {\bibfnamefont {A.}~\bibnamefont {Vishwanath}},\
  }\bibfield  {title} {\bibinfo {title} {{W}eyl and {D}irac semimetals in
  three-dimensional solids},\ }\href
  {https://doi.org/10.1103/RevModPhys.90.015001} {\bibfield  {journal}
  {\bibinfo  {journal} {Rev. Mod. Phys.}\ }\textbf {\bibinfo {volume} {90}},\
  \bibinfo {pages} {015001} (\bibinfo {year} {2018})}\BibitemShut {NoStop}%
\bibitem [{\citenamefont {Lv}\ \emph {et~al.}(2021)\citenamefont {Lv},
  \citenamefont {Qian},\ and\ \citenamefont {Ding}}]{RevModPhys.93.025002}%
  \BibitemOpen
  \bibfield  {author} {\bibinfo {author} {\bibfnamefont {B.~Q.}\ \bibnamefont
  {Lv}}, \bibinfo {author} {\bibfnamefont {T.}~\bibnamefont {Qian}},\ and\
  \bibinfo {author} {\bibfnamefont {H.}~\bibnamefont {Ding}},\ }\bibfield
  {title} {\bibinfo {title} {Experimental perspective on three-dimensional
  topological semimetals},\ }\href
  {https://doi.org/10.1103/RevModPhys.93.025002} {\bibfield  {journal}
  {\bibinfo  {journal} {Rev. Mod. Phys.}\ }\textbf {\bibinfo {volume} {93}},\
  \bibinfo {pages} {025002} (\bibinfo {year} {2021})}\BibitemShut {NoStop}%
\bibitem [{\citenamefont {Benalcazar}\ \emph
  {et~al.}(2017{\natexlab{a}})\citenamefont {Benalcazar}, \citenamefont
  {Bernevig},\ and\ \citenamefont {Hughes}}]{Benalcazar61}%
  \BibitemOpen
  \bibfield  {author} {\bibinfo {author} {\bibfnamefont {W.~A.}\ \bibnamefont
  {Benalcazar}}, \bibinfo {author} {\bibfnamefont {B.~A.}\ \bibnamefont
  {Bernevig}},\ and\ \bibinfo {author} {\bibfnamefont {T.~L.}\ \bibnamefont
  {Hughes}},\ }\bibfield  {title} {\bibinfo {title} {Quantized electric
  multipole insulators},\ }\href {https://doi.org/10.1126/science.aah6442}
  {\bibfield  {journal} {\bibinfo  {journal} {Science}\ }\textbf {\bibinfo
  {volume} {357}},\ \bibinfo {pages} {61} (\bibinfo {year}
  {2017}{\natexlab{a}})}\BibitemShut {NoStop}%
\bibitem [{\citenamefont {Langbehn}\ \emph {et~al.}(2017)\citenamefont
  {Langbehn}, \citenamefont {Peng}, \citenamefont {Trifunovic}, \citenamefont
  {von Oppen},\ and\ \citenamefont {Brouwer}}]{PhysRevLett.119.246401}%
  \BibitemOpen
  \bibfield  {author} {\bibinfo {author} {\bibfnamefont {J.}~\bibnamefont
  {Langbehn}}, \bibinfo {author} {\bibfnamefont {Y.}~\bibnamefont {Peng}},
  \bibinfo {author} {\bibfnamefont {L.}~\bibnamefont {Trifunovic}}, \bibinfo
  {author} {\bibfnamefont {F.}~\bibnamefont {von Oppen}},\ and\ \bibinfo
  {author} {\bibfnamefont {P.~W.}\ \bibnamefont {Brouwer}},\ }\bibfield
  {title} {\bibinfo {title} {Reflection-symmetric second-order topological
  insulators and superconductors},\ }\href
  {https://doi.org/10.1103/PhysRevLett.119.246401} {\bibfield  {journal}
  {\bibinfo  {journal} {Phys. Rev. Lett.}\ }\textbf {\bibinfo {volume} {119}},\
  \bibinfo {pages} {246401} (\bibinfo {year} {2017})}\BibitemShut {NoStop}%
\bibitem [{\citenamefont {Song}\ \emph {et~al.}(2017)\citenamefont {Song},
  \citenamefont {Fang},\ and\ \citenamefont {Fang}}]{PhysRevLett.119.246402}%
  \BibitemOpen
  \bibfield  {author} {\bibinfo {author} {\bibfnamefont {Z.}~\bibnamefont
  {Song}}, \bibinfo {author} {\bibfnamefont {Z.}~\bibnamefont {Fang}},\ and\
  \bibinfo {author} {\bibfnamefont {C.}~\bibnamefont {Fang}},\ }\bibfield
  {title} {\bibinfo {title} {$(d\ensuremath{-}2)$-dimensional edge states of
  rotation symmetry protected topological states},\ }\href
  {https://doi.org/10.1103/PhysRevLett.119.246402} {\bibfield  {journal}
  {\bibinfo  {journal} {Phys. Rev. Lett.}\ }\textbf {\bibinfo {volume} {119}},\
  \bibinfo {pages} {246402} (\bibinfo {year} {2017})}\BibitemShut {NoStop}%
\bibitem [{\citenamefont {Schindler}\ \emph
  {et~al.}(2018{\natexlab{a}})\citenamefont {Schindler}, \citenamefont {Cook},
  \citenamefont {Vergniory}, \citenamefont {Wang}, \citenamefont {Parkin},
  \citenamefont {Bernevig},\ and\ \citenamefont {Neupert}}]{Schindlereaat0346}%
  \BibitemOpen
  \bibfield  {author} {\bibinfo {author} {\bibfnamefont {F.}~\bibnamefont
  {Schindler}}, \bibinfo {author} {\bibfnamefont {A.~M.}\ \bibnamefont {Cook}},
  \bibinfo {author} {\bibfnamefont {M.~G.}\ \bibnamefont {Vergniory}}, \bibinfo
  {author} {\bibfnamefont {Z.}~\bibnamefont {Wang}}, \bibinfo {author}
  {\bibfnamefont {S.~S.~P.}\ \bibnamefont {Parkin}}, \bibinfo {author}
  {\bibfnamefont {B.~A.}\ \bibnamefont {Bernevig}},\ and\ \bibinfo {author}
  {\bibfnamefont {T.}~\bibnamefont {Neupert}},\ }\bibfield  {title} {\bibinfo
  {title} {Higher-order topological insulators},\ }\href
  {https://doi.org/10.1126/sciadv.aat0346} {\bibfield  {journal} {\bibinfo
  {journal} {Science Advances}\ }\textbf {\bibinfo {volume} {4}},\ \bibinfo
  {pages} {eaat0346} (\bibinfo {year} {2018}{\natexlab{a}})}\BibitemShut
  {NoStop}%
\bibitem [{\citenamefont {Yan}\ \emph {et~al.}(2018)\citenamefont {Yan},
  \citenamefont {Song},\ and\ \citenamefont {Wang}}]{PhysRevLett.121.096803}%
  \BibitemOpen
  \bibfield  {author} {\bibinfo {author} {\bibfnamefont {Z.}~\bibnamefont
  {Yan}}, \bibinfo {author} {\bibfnamefont {F.}~\bibnamefont {Song}},\ and\
  \bibinfo {author} {\bibfnamefont {Z.}~\bibnamefont {Wang}},\ }\bibfield
  {title} {\bibinfo {title} {Majorana corner modes in a high-temperature
  platform},\ }\href {https://doi.org/10.1103/PhysRevLett.121.096803}
  {\bibfield  {journal} {\bibinfo  {journal} {Phys. Rev. Lett.}\ }\textbf
  {\bibinfo {volume} {121}},\ \bibinfo {pages} {096803} (\bibinfo {year}
  {2018})}\BibitemShut {NoStop}%
\bibitem [{\citenamefont {Ezawa}(2018)}]{PhysRevLett.121.116801}%
  \BibitemOpen
  \bibfield  {author} {\bibinfo {author} {\bibfnamefont {M.}~\bibnamefont
  {Ezawa}},\ }\bibfield  {title} {\bibinfo {title} {Topological switch between
  second-order topological insulators and topological crystalline insulators},\
  }\href {https://doi.org/10.1103/PhysRevLett.121.116801} {\bibfield  {journal}
  {\bibinfo  {journal} {Phys. Rev. Lett.}\ }\textbf {\bibinfo {volume} {121}},\
  \bibinfo {pages} {116801} (\bibinfo {year} {2018})}\BibitemShut {NoStop}%
\bibitem [{\citenamefont {Liu}\ \emph {et~al.}(2019{\natexlab{a}})\citenamefont
  {Liu}, \citenamefont {Deng},\ and\ \citenamefont
  {Wakabayashi}}]{PhysRevLett.122.086804}%
  \BibitemOpen
  \bibfield  {author} {\bibinfo {author} {\bibfnamefont {F.}~\bibnamefont
  {Liu}}, \bibinfo {author} {\bibfnamefont {H.-Y.}\ \bibnamefont {Deng}},\ and\
  \bibinfo {author} {\bibfnamefont {K.}~\bibnamefont {Wakabayashi}},\
  }\bibfield  {title} {\bibinfo {title} {Helical topological edge states in a
  quadrupole phase},\ }\href {https://doi.org/10.1103/PhysRevLett.122.086804}
  {\bibfield  {journal} {\bibinfo  {journal} {Phys. Rev. Lett.}\ }\textbf
  {\bibinfo {volume} {122}},\ \bibinfo {pages} {086804} (\bibinfo {year}
  {2019}{\natexlab{a}})}\BibitemShut {NoStop}%
\bibitem [{\citenamefont {Zhang}\ \emph {et~al.}(2019)\citenamefont {Zhang},
  \citenamefont {Cole}, \citenamefont {Wu},\ and\ \citenamefont
  {Das~Sarma}}]{PhysRevLett.123.167001}%
  \BibitemOpen
  \bibfield  {author} {\bibinfo {author} {\bibfnamefont {R.-X.}\ \bibnamefont
  {Zhang}}, \bibinfo {author} {\bibfnamefont {W.~S.}\ \bibnamefont {Cole}},
  \bibinfo {author} {\bibfnamefont {X.}~\bibnamefont {Wu}},\ and\ \bibinfo
  {author} {\bibfnamefont {S.}~\bibnamefont {Das~Sarma}},\ }\bibfield  {title}
  {\bibinfo {title} {Higher-order topology and nodal topological
  superconductivity in {F}e({S}e,{T}e) heterostructures},\ }\href
  {https://doi.org/10.1103/PhysRevLett.123.167001} {\bibfield  {journal}
  {\bibinfo  {journal} {Phys. Rev. Lett.}\ }\textbf {\bibinfo {volume} {123}},\
  \bibinfo {pages} {167001} (\bibinfo {year} {2019})}\BibitemShut {NoStop}%
\bibitem [{\citenamefont {Park}\ \emph {et~al.}(2019)\citenamefont {Park},
  \citenamefont {Kim}, \citenamefont {Cho},\ and\ \citenamefont
  {Lee}}]{PhysRevLett.123.216803}%
  \BibitemOpen
  \bibfield  {author} {\bibinfo {author} {\bibfnamefont {M.~J.}\ \bibnamefont
  {Park}}, \bibinfo {author} {\bibfnamefont {Y.}~\bibnamefont {Kim}}, \bibinfo
  {author} {\bibfnamefont {G.~Y.}\ \bibnamefont {Cho}},\ and\ \bibinfo {author}
  {\bibfnamefont {S.}~\bibnamefont {Lee}},\ }\bibfield  {title} {\bibinfo
  {title} {Higher-order topological insulator in twisted bilayer graphene},\
  }\href {https://doi.org/10.1103/PhysRevLett.123.216803} {\bibfield  {journal}
  {\bibinfo  {journal} {Phys. Rev. Lett.}\ }\textbf {\bibinfo {volume} {123}},\
  \bibinfo {pages} {216803} (\bibinfo {year} {2019})}\BibitemShut {NoStop}%
\bibitem [{\citenamefont {Sheng}\ \emph {et~al.}(2019)\citenamefont {Sheng},
  \citenamefont {Chen}, \citenamefont {Liu}, \citenamefont {Chen},
  \citenamefont {Yu}, \citenamefont {Zhao},\ and\ \citenamefont
  {Yang}}]{PhysRevLett.123.256402}%
  \BibitemOpen
  \bibfield  {author} {\bibinfo {author} {\bibfnamefont {X.-L.}\ \bibnamefont
  {Sheng}}, \bibinfo {author} {\bibfnamefont {C.}~\bibnamefont {Chen}},
  \bibinfo {author} {\bibfnamefont {H.}~\bibnamefont {Liu}}, \bibinfo {author}
  {\bibfnamefont {Z.}~\bibnamefont {Chen}}, \bibinfo {author} {\bibfnamefont
  {Z.-M.}\ \bibnamefont {Yu}}, \bibinfo {author} {\bibfnamefont {Y.~X.}\
  \bibnamefont {Zhao}},\ and\ \bibinfo {author} {\bibfnamefont {S.~A.}\
  \bibnamefont {Yang}},\ }\bibfield  {title} {\bibinfo {title} {Two-dimensional
  second-order topological insulator in graphdiyne},\ }\href
  {https://doi.org/10.1103/PhysRevLett.123.256402} {\bibfield  {journal}
  {\bibinfo  {journal} {Phys. Rev. Lett.}\ }\textbf {\bibinfo {volume} {123}},\
  \bibinfo {pages} {256402} (\bibinfo {year} {2019})}\BibitemShut {NoStop}%
\bibitem [{\citenamefont {Tiwari}\ \emph {et~al.}(2020)\citenamefont {Tiwari},
  \citenamefont {Li}, \citenamefont {Bernevig}, \citenamefont {Neupert},\ and\
  \citenamefont {Parameswaran}}]{PhysRevLett.124.046801}%
  \BibitemOpen
  \bibfield  {author} {\bibinfo {author} {\bibfnamefont {A.}~\bibnamefont
  {Tiwari}}, \bibinfo {author} {\bibfnamefont {M.-H.}\ \bibnamefont {Li}},
  \bibinfo {author} {\bibfnamefont {B.~A.}\ \bibnamefont {Bernevig}}, \bibinfo
  {author} {\bibfnamefont {T.}~\bibnamefont {Neupert}},\ and\ \bibinfo {author}
  {\bibfnamefont {S.~A.}\ \bibnamefont {Parameswaran}},\ }\bibfield  {title}
  {\bibinfo {title} {Unhinging the surfaces of higher-order topological
  insulators and superconductors},\ }\href
  {https://doi.org/10.1103/PhysRevLett.124.046801} {\bibfield  {journal}
  {\bibinfo  {journal} {Phys. Rev. Lett.}\ }\textbf {\bibinfo {volume} {124}},\
  \bibinfo {pages} {046801} (\bibinfo {year} {2020})}\BibitemShut {NoStop}%
\bibitem [{\citenamefont {Ren}\ \emph {et~al.}(2020)\citenamefont {Ren},
  \citenamefont {Qiao},\ and\ \citenamefont {Niu}}]{PhysRevLett.124.166804}%
  \BibitemOpen
  \bibfield  {author} {\bibinfo {author} {\bibfnamefont {Y.}~\bibnamefont
  {Ren}}, \bibinfo {author} {\bibfnamefont {Z.}~\bibnamefont {Qiao}},\ and\
  \bibinfo {author} {\bibfnamefont {Q.}~\bibnamefont {Niu}},\ }\bibfield
  {title} {\bibinfo {title} {Engineering corner states from two-dimensional
  topological insulators},\ }\href
  {https://doi.org/10.1103/PhysRevLett.124.166804} {\bibfield  {journal}
  {\bibinfo  {journal} {Phys. Rev. Lett.}\ }\textbf {\bibinfo {volume} {124}},\
  \bibinfo {pages} {166804} (\bibinfo {year} {2020})}\BibitemShut {NoStop}%
\bibitem [{\citenamefont {Zeng}\ \emph {et~al.}(2020)\citenamefont {Zeng},
  \citenamefont {Yang},\ and\ \citenamefont {Xu}}]{PhysRevB.101.241104}%
  \BibitemOpen
  \bibfield  {author} {\bibinfo {author} {\bibfnamefont {Q.-B.}\ \bibnamefont
  {Zeng}}, \bibinfo {author} {\bibfnamefont {Y.-B.}\ \bibnamefont {Yang}},\
  and\ \bibinfo {author} {\bibfnamefont {Y.}~\bibnamefont {Xu}},\ }\bibfield
  {title} {\bibinfo {title} {Higher-order topological insulators and semimetals
  in generalized {A}ubry-{A}ndr\'e-{H}arper models},\ }\href
  {https://doi.org/10.1103/PhysRevB.101.241104} {\bibfield  {journal} {\bibinfo
   {journal} {Phys. Rev. B}\ }\textbf {\bibinfo {volume} {101}},\ \bibinfo
  {pages} {241104(R)} (\bibinfo {year} {2020})}\BibitemShut {NoStop}%
\bibitem [{\citenamefont {Slager}\ \emph {et~al.}(2015)\citenamefont {Slager},
  \citenamefont {Rademaker}, \citenamefont {Zaanen},\ and\ \citenamefont
  {Balents}}]{PhysRevB.92.085126}%
  \BibitemOpen
  \bibfield  {author} {\bibinfo {author} {\bibfnamefont {R.-J.}\ \bibnamefont
  {Slager}}, \bibinfo {author} {\bibfnamefont {L.}~\bibnamefont {Rademaker}},
  \bibinfo {author} {\bibfnamefont {J.}~\bibnamefont {Zaanen}},\ and\ \bibinfo
  {author} {\bibfnamefont {L.}~\bibnamefont {Balents}},\ }\bibfield  {title}
  {\bibinfo {title} {Impurity-bound states and {G}reen's function zeros as
  local signatures of topology},\ }\href
  {https://doi.org/10.1103/PhysRevB.92.085126} {\bibfield  {journal} {\bibinfo
  {journal} {Phys. Rev. B}\ }\textbf {\bibinfo {volume} {92}},\ \bibinfo
  {pages} {085126} (\bibinfo {year} {2015})}\BibitemShut {NoStop}%
\bibitem [{\citenamefont {Zhang}\ \emph {et~al.}(2020)\citenamefont {Zhang},
  \citenamefont {Xie}, \citenamefont {Hao}, \citenamefont {Dang}, \citenamefont
  {Xiao}, \citenamefont {Shi}, \citenamefont {Ni}, \citenamefont {Niu},
  \citenamefont {Wang}, \citenamefont {Jin}, \citenamefont {Zhang},\ and\
  \citenamefont {Xu}}]{Zhang2020}%
  \BibitemOpen
  \bibfield  {author} {\bibinfo {author} {\bibfnamefont {W.}~\bibnamefont
  {Zhang}}, \bibinfo {author} {\bibfnamefont {X.}~\bibnamefont {Xie}}, \bibinfo
  {author} {\bibfnamefont {H.}~\bibnamefont {Hao}}, \bibinfo {author}
  {\bibfnamefont {J.}~\bibnamefont {Dang}}, \bibinfo {author} {\bibfnamefont
  {S.}~\bibnamefont {Xiao}}, \bibinfo {author} {\bibfnamefont {S.}~\bibnamefont
  {Shi}}, \bibinfo {author} {\bibfnamefont {H.}~\bibnamefont {Ni}}, \bibinfo
  {author} {\bibfnamefont {Z.}~\bibnamefont {Niu}}, \bibinfo {author}
  {\bibfnamefont {C.}~\bibnamefont {Wang}}, \bibinfo {author} {\bibfnamefont
  {K.}~\bibnamefont {Jin}}, \bibinfo {author} {\bibfnamefont {X.}~\bibnamefont
  {Zhang}},\ and\ \bibinfo {author} {\bibfnamefont {X.}~\bibnamefont {Xu}},\
  }\bibfield  {title} {\bibinfo {title} {Low-threshold topological nanolasers
  based on the second-order corner state},\ }\href
  {https://doi.org/10.1038/s41377-020-00352-1} {\bibfield  {journal} {\bibinfo
  {journal} {Light. Sci. Appl.}\ }\textbf {\bibinfo {volume} {9}},\ \bibinfo
  {pages} {109} (\bibinfo {year} {2020})}\BibitemShut {NoStop}%
\bibitem [{\citenamefont {Serra-Garcia}\ \emph {et~al.}(2018)\citenamefont
  {Serra-Garcia}, \citenamefont {Peri}, \citenamefont {Süsstrunk},
  \citenamefont {Bilal}, \citenamefont {Larsen}, \citenamefont {Villanueva},\
  and\ \citenamefont {Huber}}]{SerraGarcia2018}%
  \BibitemOpen
  \bibfield  {author} {\bibinfo {author} {\bibfnamefont {M.}~\bibnamefont
  {Serra-Garcia}}, \bibinfo {author} {\bibfnamefont {V.}~\bibnamefont {Peri}},
  \bibinfo {author} {\bibfnamefont {R.}~\bibnamefont {Süsstrunk}}, \bibinfo
  {author} {\bibfnamefont {O.~R.}\ \bibnamefont {Bilal}}, \bibinfo {author}
  {\bibfnamefont {T.}~\bibnamefont {Larsen}}, \bibinfo {author} {\bibfnamefont
  {L.~G.}\ \bibnamefont {Villanueva}},\ and\ \bibinfo {author} {\bibfnamefont
  {S.~D.}\ \bibnamefont {Huber}},\ }\bibfield  {title} {\bibinfo {title}
  {Observation of a phononic quadrupole topological insulator},\ }\href
  {https://doi.org/10.1038/nature25156} {\bibfield  {journal} {\bibinfo
  {journal} {Nature}\ }\textbf {\bibinfo {volume} {555}},\ \bibinfo {pages}
  {342} (\bibinfo {year} {2018})}\BibitemShut {NoStop}%
\bibitem [{\citenamefont {Peterson}\ \emph {et~al.}(2018)\citenamefont
  {Peterson}, \citenamefont {Benalcazar}, \citenamefont {Hughes},\ and\
  \citenamefont {Bahl}}]{Peterson2018}%
  \BibitemOpen
  \bibfield  {author} {\bibinfo {author} {\bibfnamefont {C.~W.}\ \bibnamefont
  {Peterson}}, \bibinfo {author} {\bibfnamefont {W.~A.}\ \bibnamefont
  {Benalcazar}}, \bibinfo {author} {\bibfnamefont {T.~L.}\ \bibnamefont
  {Hughes}},\ and\ \bibinfo {author} {\bibfnamefont {G.}~\bibnamefont {Bahl}},\
  }\bibfield  {title} {\bibinfo {title} {A quantized microwave quadrupole
  insulator with topologically protected corner states},\ }\href
  {https://doi.org/10.1038/nature25777} {\bibfield  {journal} {\bibinfo
  {journal} {Nature}\ }\textbf {\bibinfo {volume} {555}},\ \bibinfo {pages}
  {346} (\bibinfo {year} {2018})}\BibitemShut {NoStop}%
\bibitem [{\citenamefont {Schindler}\ \emph
  {et~al.}(2018{\natexlab{b}})\citenamefont {Schindler}, \citenamefont {Wang},
  \citenamefont {Vergniory}, \citenamefont {Cook}, \citenamefont {Murani},
  \citenamefont {Sengupta}, \citenamefont {Kasumov}, \citenamefont {Deblock},
  \citenamefont {Jeon}, \citenamefont {Drozdov}, \citenamefont {Bouchiat},
  \citenamefont {Guéron}, \citenamefont {Yazdani}, \citenamefont {Bernevig},\
  and\ \citenamefont {Neupert}}]{Bismuth}%
  \BibitemOpen
  \bibfield  {author} {\bibinfo {author} {\bibfnamefont {F.}~\bibnamefont
  {Schindler}}, \bibinfo {author} {\bibfnamefont {Z.}~\bibnamefont {Wang}},
  \bibinfo {author} {\bibfnamefont {M.~G.}\ \bibnamefont {Vergniory}}, \bibinfo
  {author} {\bibfnamefont {A.~M.}\ \bibnamefont {Cook}}, \bibinfo {author}
  {\bibfnamefont {A.}~\bibnamefont {Murani}}, \bibinfo {author} {\bibfnamefont
  {S.}~\bibnamefont {Sengupta}}, \bibinfo {author} {\bibfnamefont {A.~Y.}\
  \bibnamefont {Kasumov}}, \bibinfo {author} {\bibfnamefont {R.}~\bibnamefont
  {Deblock}}, \bibinfo {author} {\bibfnamefont {S.}~\bibnamefont {Jeon}},
  \bibinfo {author} {\bibfnamefont {I.}~\bibnamefont {Drozdov}}, \bibinfo
  {author} {\bibfnamefont {H.}~\bibnamefont {Bouchiat}}, \bibinfo {author}
  {\bibfnamefont {S.}~\bibnamefont {Guéron}}, \bibinfo {author} {\bibfnamefont
  {A.}~\bibnamefont {Yazdani}}, \bibinfo {author} {\bibfnamefont {B.~A.}\
  \bibnamefont {Bernevig}},\ and\ \bibinfo {author} {\bibfnamefont
  {T.}~\bibnamefont {Neupert}},\ }\bibfield  {title} {\bibinfo {title}
  {Higher-order topology in bismuth},\ }\href
  {https://doi.org/10.1038/s41567-018-0224-7} {\bibfield  {journal} {\bibinfo
  {journal} {Nature Physics}\ }\textbf {\bibinfo {volume} {14}},\ \bibinfo
  {pages} {918} (\bibinfo {year} {2018}{\natexlab{b}})}\BibitemShut {NoStop}%
\bibitem [{\citenamefont {Imhof}\ \emph {et~al.}(2018)\citenamefont {Imhof},
  \citenamefont {Berger}, \citenamefont {Bayer}, \citenamefont {Brehm},
  \citenamefont {Molenkamp}, \citenamefont {Kiessling}, \citenamefont
  {Schindler}, \citenamefont {Lee}, \citenamefont {Greiter}, \citenamefont
  {Neupert},\ and\ \citenamefont {Thomale}}]{Imhof2018}%
  \BibitemOpen
  \bibfield  {author} {\bibinfo {author} {\bibfnamefont {S.}~\bibnamefont
  {Imhof}}, \bibinfo {author} {\bibfnamefont {C.}~\bibnamefont {Berger}},
  \bibinfo {author} {\bibfnamefont {F.}~\bibnamefont {Bayer}}, \bibinfo
  {author} {\bibfnamefont {J.}~\bibnamefont {Brehm}}, \bibinfo {author}
  {\bibfnamefont {L.~W.}\ \bibnamefont {Molenkamp}}, \bibinfo {author}
  {\bibfnamefont {T.}~\bibnamefont {Kiessling}}, \bibinfo {author}
  {\bibfnamefont {F.}~\bibnamefont {Schindler}}, \bibinfo {author}
  {\bibfnamefont {C.~H.}\ \bibnamefont {Lee}}, \bibinfo {author} {\bibfnamefont
  {M.}~\bibnamefont {Greiter}}, \bibinfo {author} {\bibfnamefont
  {T.}~\bibnamefont {Neupert}},\ and\ \bibinfo {author} {\bibfnamefont
  {R.}~\bibnamefont {Thomale}},\ }\bibfield  {title} {\bibinfo {title}
  {Topolectrical-circuit realization of topological corner modes},\ }\href
  {https://doi.org/10.1038/s41567-018-0246-1} {\bibfield  {journal} {\bibinfo
  {journal} {Nature Physics}\ }\textbf {\bibinfo {volume} {14}},\ \bibinfo
  {pages} {925} (\bibinfo {year} {2018})}\BibitemShut {NoStop}%
\bibitem [{\citenamefont {Wu}\ \emph {et~al.}(2020{\natexlab{a}})\citenamefont
  {Wu}, \citenamefont {Huang}, \citenamefont {Lu}, \citenamefont {Wu},
  \citenamefont {Deng}, \citenamefont {Li},\ and\ \citenamefont
  {Liu}}]{PhysRevB.102.104109}%
  \BibitemOpen
  \bibfield  {author} {\bibinfo {author} {\bibfnamefont {J.}~\bibnamefont
  {Wu}}, \bibinfo {author} {\bibfnamefont {X.}~\bibnamefont {Huang}}, \bibinfo
  {author} {\bibfnamefont {J.}~\bibnamefont {Lu}}, \bibinfo {author}
  {\bibfnamefont {Y.}~\bibnamefont {Wu}}, \bibinfo {author} {\bibfnamefont
  {W.}~\bibnamefont {Deng}}, \bibinfo {author} {\bibfnamefont {F.}~\bibnamefont
  {Li}},\ and\ \bibinfo {author} {\bibfnamefont {Z.}~\bibnamefont {Liu}},\
  }\bibfield  {title} {\bibinfo {title} {Observation of corner states in
  second-order topological electric circuits},\ }\href
  {https://doi.org/10.1103/PhysRevB.102.104109} {\bibfield  {journal} {\bibinfo
   {journal} {Phys. Rev. B}\ }\textbf {\bibinfo {volume} {102}},\ \bibinfo
  {pages} {104109} (\bibinfo {year} {2020}{\natexlab{a}})}\BibitemShut
  {NoStop}%
\bibitem [{\citenamefont {Fan}\ \emph {et~al.}(2019)\citenamefont {Fan},
  \citenamefont {Xia}, \citenamefont {Tong}, \citenamefont {Zheng},\ and\
  \citenamefont {Yu}}]{PhysRevLett.122.204301}%
  \BibitemOpen
  \bibfield  {author} {\bibinfo {author} {\bibfnamefont {H.}~\bibnamefont
  {Fan}}, \bibinfo {author} {\bibfnamefont {B.}~\bibnamefont {Xia}}, \bibinfo
  {author} {\bibfnamefont {L.}~\bibnamefont {Tong}}, \bibinfo {author}
  {\bibfnamefont {S.}~\bibnamefont {Zheng}},\ and\ \bibinfo {author}
  {\bibfnamefont {D.}~\bibnamefont {Yu}},\ }\bibfield  {title} {\bibinfo
  {title} {Elastic higher-order topological insulator with topologically
  protected corner states},\ }\href
  {https://doi.org/10.1103/PhysRevLett.122.204301} {\bibfield  {journal}
  {\bibinfo  {journal} {Phys. Rev. Lett.}\ }\textbf {\bibinfo {volume} {122}},\
  \bibinfo {pages} {204301} (\bibinfo {year} {2019})}\BibitemShut {NoStop}%
\bibitem [{\citenamefont {Chen}\ \emph {et~al.}(2019)\citenamefont {Chen},
  \citenamefont {Deng}, \citenamefont {Shi}, \citenamefont {Zhao},
  \citenamefont {Chen},\ and\ \citenamefont {Dong}}]{PhysRevLett.122.233902}%
  \BibitemOpen
  \bibfield  {author} {\bibinfo {author} {\bibfnamefont {X.-D.}\ \bibnamefont
  {Chen}}, \bibinfo {author} {\bibfnamefont {W.-M.}\ \bibnamefont {Deng}},
  \bibinfo {author} {\bibfnamefont {F.-L.}\ \bibnamefont {Shi}}, \bibinfo
  {author} {\bibfnamefont {F.-L.}\ \bibnamefont {Zhao}}, \bibinfo {author}
  {\bibfnamefont {M.}~\bibnamefont {Chen}},\ and\ \bibinfo {author}
  {\bibfnamefont {J.-W.}\ \bibnamefont {Dong}},\ }\bibfield  {title} {\bibinfo
  {title} {Direct observation of corner states in second-order topological
  photonic crystal slabs},\ }\href
  {https://doi.org/10.1103/PhysRevLett.122.233902} {\bibfield  {journal}
  {\bibinfo  {journal} {Phys. Rev. Lett.}\ }\textbf {\bibinfo {volume} {122}},\
  \bibinfo {pages} {233902} (\bibinfo {year} {2019})}\BibitemShut {NoStop}%
\bibitem [{\citenamefont {Xie}\ \emph {et~al.}(2019)\citenamefont {Xie},
  \citenamefont {Su}, \citenamefont {Wang}, \citenamefont {Su}, \citenamefont
  {Shen}, \citenamefont {Zhan}, \citenamefont {Lu}, \citenamefont {Wang},\ and\
  \citenamefont {Chen}}]{PhysRevLett.122.233903}%
  \BibitemOpen
  \bibfield  {author} {\bibinfo {author} {\bibfnamefont {B.-Y.}\ \bibnamefont
  {Xie}}, \bibinfo {author} {\bibfnamefont {G.-X.}\ \bibnamefont {Su}},
  \bibinfo {author} {\bibfnamefont {H.-F.}\ \bibnamefont {Wang}}, \bibinfo
  {author} {\bibfnamefont {H.}~\bibnamefont {Su}}, \bibinfo {author}
  {\bibfnamefont {X.-P.}\ \bibnamefont {Shen}}, \bibinfo {author}
  {\bibfnamefont {P.}~\bibnamefont {Zhan}}, \bibinfo {author} {\bibfnamefont
  {M.-H.}\ \bibnamefont {Lu}}, \bibinfo {author} {\bibfnamefont {Z.-L.}\
  \bibnamefont {Wang}},\ and\ \bibinfo {author} {\bibfnamefont {Y.-F.}\
  \bibnamefont {Chen}},\ }\bibfield  {title} {\bibinfo {title} {Visualization
  of higher-order topological insulating phases in two-dimensional dielectric
  photonic crystals},\ }\href {https://doi.org/10.1103/PhysRevLett.122.233903}
  {\bibfield  {journal} {\bibinfo  {journal} {Phys. Rev. Lett.}\ }\textbf
  {\bibinfo {volume} {122}},\ \bibinfo {pages} {233903} (\bibinfo {year}
  {2019})}\BibitemShut {NoStop}%
\bibitem [{\citenamefont {Yang}\ \emph {et~al.}(2020)\citenamefont {Yang},
  \citenamefont {Li}, \citenamefont {Peng}, \citenamefont {Zou},\ and\
  \citenamefont {Cheng}}]{PhysRevLett.125.255502}%
  \BibitemOpen
  \bibfield  {author} {\bibinfo {author} {\bibfnamefont {Z.-Z.}\ \bibnamefont
  {Yang}}, \bibinfo {author} {\bibfnamefont {X.}~\bibnamefont {Li}}, \bibinfo
  {author} {\bibfnamefont {Y.-Y.}\ \bibnamefont {Peng}}, \bibinfo {author}
  {\bibfnamefont {X.-Y.}\ \bibnamefont {Zou}},\ and\ \bibinfo {author}
  {\bibfnamefont {J.-C.}\ \bibnamefont {Cheng}},\ }\bibfield  {title} {\bibinfo
  {title} {Helical higher-order topological states in an acoustic crystalline
  insulator},\ }\href {https://doi.org/10.1103/PhysRevLett.125.255502}
  {\bibfield  {journal} {\bibinfo  {journal} {Phys. Rev. Lett.}\ }\textbf
  {\bibinfo {volume} {125}},\ \bibinfo {pages} {255502} (\bibinfo {year}
  {2020})}\BibitemShut {NoStop}%
\bibitem [{\citenamefont {Yang}\ \emph {et~al.}(2021)\citenamefont {Yang},
  \citenamefont {Lu}, \citenamefont {Yan}, \citenamefont {Huang}, \citenamefont
  {Deng},\ and\ \citenamefont {Liu}}]{PhysRevLett.126.156801}%
  \BibitemOpen
  \bibfield  {author} {\bibinfo {author} {\bibfnamefont {Y.}~\bibnamefont
  {Yang}}, \bibinfo {author} {\bibfnamefont {J.}~\bibnamefont {Lu}}, \bibinfo
  {author} {\bibfnamefont {M.}~\bibnamefont {Yan}}, \bibinfo {author}
  {\bibfnamefont {X.}~\bibnamefont {Huang}}, \bibinfo {author} {\bibfnamefont
  {W.}~\bibnamefont {Deng}},\ and\ \bibinfo {author} {\bibfnamefont
  {Z.}~\bibnamefont {Liu}},\ }\bibfield  {title} {\bibinfo {title}
  {Hybrid-order topological insulators in a phononic crystal},\ }\href
  {https://doi.org/10.1103/PhysRevLett.126.156801} {\bibfield  {journal}
  {\bibinfo  {journal} {Phys. Rev. Lett.}\ }\textbf {\bibinfo {volume} {126}},\
  \bibinfo {pages} {156801} (\bibinfo {year} {2021})}\BibitemShut {NoStop}%
\bibitem [{\citenamefont {Liu}\ \emph {et~al.}(2014)\citenamefont {Liu},
  \citenamefont {Zhou}, \citenamefont {Zhang}, \citenamefont {Wang},
  \citenamefont {Weng}, \citenamefont {Prabhakaran}, \citenamefont {Mo},
  \citenamefont {Shen}, \citenamefont {Fang}, \citenamefont {Dai},
  \citenamefont {Hussain},\ and\ \citenamefont {Chen}}]{Liu864}%
  \BibitemOpen
  \bibfield  {author} {\bibinfo {author} {\bibfnamefont {Z.~K.}\ \bibnamefont
  {Liu}}, \bibinfo {author} {\bibfnamefont {B.}~\bibnamefont {Zhou}}, \bibinfo
  {author} {\bibfnamefont {Y.}~\bibnamefont {Zhang}}, \bibinfo {author}
  {\bibfnamefont {Z.~J.}\ \bibnamefont {Wang}}, \bibinfo {author}
  {\bibfnamefont {H.~M.}\ \bibnamefont {Weng}}, \bibinfo {author}
  {\bibfnamefont {D.}~\bibnamefont {Prabhakaran}}, \bibinfo {author}
  {\bibfnamefont {S.-K.}\ \bibnamefont {Mo}}, \bibinfo {author} {\bibfnamefont
  {Z.~X.}\ \bibnamefont {Shen}}, \bibinfo {author} {\bibfnamefont
  {Z.}~\bibnamefont {Fang}}, \bibinfo {author} {\bibfnamefont {X.}~\bibnamefont
  {Dai}}, \bibinfo {author} {\bibfnamefont {Z.}~\bibnamefont {Hussain}},\ and\
  \bibinfo {author} {\bibfnamefont {Y.~L.}\ \bibnamefont {Chen}},\ }\bibfield
  {title} {\bibinfo {title} {Discovery of a three-dimensional topological
  {D}irac semimetal, {N}a$_3${B}i},\ }\href
  {https://doi.org/10.1126/science.1245085} {\bibfield  {journal} {\bibinfo
  {journal} {Science}\ }\textbf {\bibinfo {volume} {343}},\ \bibinfo {pages}
  {864} (\bibinfo {year} {2014})}\BibitemShut {NoStop}%
\bibitem [{\citenamefont {Neupane}\ \emph {et~al.}(2014)\citenamefont
  {Neupane}, \citenamefont {Xu}, \citenamefont {Sankar}, \citenamefont
  {Alidoust}, \citenamefont {Bian}, \citenamefont {Liu}, \citenamefont
  {Belopolski}, \citenamefont {Chang}, \citenamefont {Jeng}, \citenamefont
  {Lin}, \citenamefont {Bansil}, \citenamefont {Chou},\ and\ \citenamefont
  {Hasan}}]{Neupane2014}%
  \BibitemOpen
  \bibfield  {author} {\bibinfo {author} {\bibfnamefont {M.}~\bibnamefont
  {Neupane}}, \bibinfo {author} {\bibfnamefont {S.-Y.}\ \bibnamefont {Xu}},
  \bibinfo {author} {\bibfnamefont {R.}~\bibnamefont {Sankar}}, \bibinfo
  {author} {\bibfnamefont {N.}~\bibnamefont {Alidoust}}, \bibinfo {author}
  {\bibfnamefont {G.}~\bibnamefont {Bian}}, \bibinfo {author} {\bibfnamefont
  {C.}~\bibnamefont {Liu}}, \bibinfo {author} {\bibfnamefont {I.}~\bibnamefont
  {Belopolski}}, \bibinfo {author} {\bibfnamefont {T.-R.}\ \bibnamefont
  {Chang}}, \bibinfo {author} {\bibfnamefont {H.-T.}\ \bibnamefont {Jeng}},
  \bibinfo {author} {\bibfnamefont {H.}~\bibnamefont {Lin}}, \bibinfo {author}
  {\bibfnamefont {A.}~\bibnamefont {Bansil}}, \bibinfo {author} {\bibfnamefont
  {F.}~\bibnamefont {Chou}},\ and\ \bibinfo {author} {\bibfnamefont {M.~Z.}\
  \bibnamefont {Hasan}},\ }\bibfield  {title} {\bibinfo {title} {Observation of
  a three-dimensional topological {D}irac semimetal phase in high-mobility
  {C}d$_3${A}s$_2$},\ }\href {https://doi.org/10.1038/ncomms4786} {\bibfield
  {journal} {\bibinfo  {journal} {Nature Communications}\ }\textbf {\bibinfo
  {volume} {5}},\ \bibinfo {pages} {3786} (\bibinfo {year} {2014})}\BibitemShut
  {NoStop}%
\bibitem [{\citenamefont {Borisenko}\ \emph {et~al.}(2014)\citenamefont
  {Borisenko}, \citenamefont {Gibson}, \citenamefont {Evtushinsky},
  \citenamefont {Zabolotnyy}, \citenamefont {B\"uchner},\ and\ \citenamefont
  {Cava}}]{PhysRevLett.113.027603}%
  \BibitemOpen
  \bibfield  {author} {\bibinfo {author} {\bibfnamefont {S.}~\bibnamefont
  {Borisenko}}, \bibinfo {author} {\bibfnamefont {Q.}~\bibnamefont {Gibson}},
  \bibinfo {author} {\bibfnamefont {D.}~\bibnamefont {Evtushinsky}}, \bibinfo
  {author} {\bibfnamefont {V.}~\bibnamefont {Zabolotnyy}}, \bibinfo {author}
  {\bibfnamefont {B.}~\bibnamefont {B\"uchner}},\ and\ \bibinfo {author}
  {\bibfnamefont {R.~J.}\ \bibnamefont {Cava}},\ }\bibfield  {title} {\bibinfo
  {title} {Experimental realization of a three-dimensional {D}irac semimetal},\
  }\href {https://doi.org/10.1103/PhysRevLett.113.027603} {\bibfield  {journal}
  {\bibinfo  {journal} {Phys. Rev. Lett.}\ }\textbf {\bibinfo {volume} {113}},\
  \bibinfo {pages} {027603} (\bibinfo {year} {2014})}\BibitemShut {NoStop}%
\bibitem [{\citenamefont {Xu}\ \emph {et~al.}(2015{\natexlab{a}})\citenamefont
  {Xu}, \citenamefont {Liu}, \citenamefont {Kushwaha}, \citenamefont {Sankar},
  \citenamefont {Krizan}, \citenamefont {Belopolski}, \citenamefont {Neupane},
  \citenamefont {Bian}, \citenamefont {Alidoust}, \citenamefont {Chang},
  \citenamefont {Jeng}, \citenamefont {Huang}, \citenamefont {Tsai},
  \citenamefont {Lin}, \citenamefont {Shibayev}, \citenamefont {Chou},
  \citenamefont {Cava},\ and\ \citenamefont {Hasan}}]{Xu294}%
  \BibitemOpen
  \bibfield  {author} {\bibinfo {author} {\bibfnamefont {S.-Y.}\ \bibnamefont
  {Xu}}, \bibinfo {author} {\bibfnamefont {C.}~\bibnamefont {Liu}}, \bibinfo
  {author} {\bibfnamefont {S.~K.}\ \bibnamefont {Kushwaha}}, \bibinfo {author}
  {\bibfnamefont {R.}~\bibnamefont {Sankar}}, \bibinfo {author} {\bibfnamefont
  {J.~W.}\ \bibnamefont {Krizan}}, \bibinfo {author} {\bibfnamefont
  {I.}~\bibnamefont {Belopolski}}, \bibinfo {author} {\bibfnamefont
  {M.}~\bibnamefont {Neupane}}, \bibinfo {author} {\bibfnamefont
  {G.}~\bibnamefont {Bian}}, \bibinfo {author} {\bibfnamefont {N.}~\bibnamefont
  {Alidoust}}, \bibinfo {author} {\bibfnamefont {T.-R.}\ \bibnamefont {Chang}},
  \bibinfo {author} {\bibfnamefont {H.-T.}\ \bibnamefont {Jeng}}, \bibinfo
  {author} {\bibfnamefont {C.-Y.}\ \bibnamefont {Huang}}, \bibinfo {author}
  {\bibfnamefont {W.-F.}\ \bibnamefont {Tsai}}, \bibinfo {author}
  {\bibfnamefont {H.}~\bibnamefont {Lin}}, \bibinfo {author} {\bibfnamefont
  {P.~P.}\ \bibnamefont {Shibayev}}, \bibinfo {author} {\bibfnamefont {F.-C.}\
  \bibnamefont {Chou}}, \bibinfo {author} {\bibfnamefont {R.~J.}\ \bibnamefont
  {Cava}},\ and\ \bibinfo {author} {\bibfnamefont {M.~Z.}\ \bibnamefont
  {Hasan}},\ }\bibfield  {title} {\bibinfo {title} {Observation of {F}ermi arc
  surface states in a topological metal},\ }\href
  {https://doi.org/10.1126/science.1256742} {\bibfield  {journal} {\bibinfo
  {journal} {Science}\ }\textbf {\bibinfo {volume} {347}},\ \bibinfo {pages}
  {294} (\bibinfo {year} {2015}{\natexlab{a}})}\BibitemShut {NoStop}%
\bibitem [{\citenamefont {Young}\ \emph {et~al.}(2012)\citenamefont {Young},
  \citenamefont {Zaheer}, \citenamefont {Teo}, \citenamefont {Kane},
  \citenamefont {Mele},\ and\ \citenamefont {Rappe}}]{PhysRevLett.108.140405}%
  \BibitemOpen
  \bibfield  {author} {\bibinfo {author} {\bibfnamefont {S.~M.}\ \bibnamefont
  {Young}}, \bibinfo {author} {\bibfnamefont {S.}~\bibnamefont {Zaheer}},
  \bibinfo {author} {\bibfnamefont {J.~C.~Y.}\ \bibnamefont {Teo}}, \bibinfo
  {author} {\bibfnamefont {C.~L.}\ \bibnamefont {Kane}}, \bibinfo {author}
  {\bibfnamefont {E.~J.}\ \bibnamefont {Mele}},\ and\ \bibinfo {author}
  {\bibfnamefont {A.~M.}\ \bibnamefont {Rappe}},\ }\bibfield  {title} {\bibinfo
  {title} {{D}irac semimetal in three dimensions},\ }\href
  {https://doi.org/10.1103/PhysRevLett.108.140405} {\bibfield  {journal}
  {\bibinfo  {journal} {Phys. Rev. Lett.}\ }\textbf {\bibinfo {volume} {108}},\
  \bibinfo {pages} {140405} (\bibinfo {year} {2012})}\BibitemShut {NoStop}%
\bibitem [{\citenamefont {Wang}\ \emph {et~al.}(2012)\citenamefont {Wang},
  \citenamefont {Sun}, \citenamefont {Chen}, \citenamefont {Franchini},
  \citenamefont {Xu}, \citenamefont {Weng}, \citenamefont {Dai},\ and\
  \citenamefont {Fang}}]{PhysRevB.85.195320}%
  \BibitemOpen
  \bibfield  {author} {\bibinfo {author} {\bibfnamefont {Z.}~\bibnamefont
  {Wang}}, \bibinfo {author} {\bibfnamefont {Y.}~\bibnamefont {Sun}}, \bibinfo
  {author} {\bibfnamefont {X.-Q.}\ \bibnamefont {Chen}}, \bibinfo {author}
  {\bibfnamefont {C.}~\bibnamefont {Franchini}}, \bibinfo {author}
  {\bibfnamefont {G.}~\bibnamefont {Xu}}, \bibinfo {author} {\bibfnamefont
  {H.}~\bibnamefont {Weng}}, \bibinfo {author} {\bibfnamefont {X.}~\bibnamefont
  {Dai}},\ and\ \bibinfo {author} {\bibfnamefont {Z.}~\bibnamefont {Fang}},\
  }\bibfield  {title} {\bibinfo {title} {{D}irac semimetal and topological
  phase transitions in ${A}_{3}${B}i (${A}=\text{Na}$, {K}, {R}b)},\ }\href
  {https://doi.org/10.1103/PhysRevB.85.195320} {\bibfield  {journal} {\bibinfo
  {journal} {Phys. Rev. B}\ }\textbf {\bibinfo {volume} {85}},\ \bibinfo
  {pages} {195320} (\bibinfo {year} {2012})}\BibitemShut {NoStop}%
\bibitem [{\citenamefont {Wang}\ \emph {et~al.}(2013)\citenamefont {Wang},
  \citenamefont {Weng}, \citenamefont {Wu}, \citenamefont {Dai},\ and\
  \citenamefont {Fang}}]{PhysRevB.88.125427}%
  \BibitemOpen
  \bibfield  {author} {\bibinfo {author} {\bibfnamefont {Z.}~\bibnamefont
  {Wang}}, \bibinfo {author} {\bibfnamefont {H.}~\bibnamefont {Weng}}, \bibinfo
  {author} {\bibfnamefont {Q.}~\bibnamefont {Wu}}, \bibinfo {author}
  {\bibfnamefont {X.}~\bibnamefont {Dai}},\ and\ \bibinfo {author}
  {\bibfnamefont {Z.}~\bibnamefont {Fang}},\ }\bibfield  {title} {\bibinfo
  {title} {Three-dimensional {D}irac semimetal and quantum transport in
  {C}d${}_{3}${A}s${}_{2}$},\ }\href
  {https://doi.org/10.1103/PhysRevB.88.125427} {\bibfield  {journal} {\bibinfo
  {journal} {Phys. Rev. B}\ }\textbf {\bibinfo {volume} {88}},\ \bibinfo
  {pages} {125427} (\bibinfo {year} {2013})}\BibitemShut {NoStop}%
\bibitem [{\citenamefont {Chen}\ \emph
  {et~al.}(2015{\natexlab{a}})\citenamefont {Chen}, \citenamefont {Chen},
  \citenamefont {Song}, \citenamefont {Schneeloch}, \citenamefont {Gu},
  \citenamefont {Wang},\ and\ \citenamefont {Wang}}]{PhysRevLett.115.176404}%
  \BibitemOpen
  \bibfield  {author} {\bibinfo {author} {\bibfnamefont {R.~Y.}\ \bibnamefont
  {Chen}}, \bibinfo {author} {\bibfnamefont {Z.~G.}\ \bibnamefont {Chen}},
  \bibinfo {author} {\bibfnamefont {X.-Y.}\ \bibnamefont {Song}}, \bibinfo
  {author} {\bibfnamefont {J.~A.}\ \bibnamefont {Schneeloch}}, \bibinfo
  {author} {\bibfnamefont {G.~D.}\ \bibnamefont {Gu}}, \bibinfo {author}
  {\bibfnamefont {F.}~\bibnamefont {Wang}},\ and\ \bibinfo {author}
  {\bibfnamefont {N.~L.}\ \bibnamefont {Wang}},\ }\bibfield  {title} {\bibinfo
  {title} {Magnetoinfrared spectroscopy of landau levels and zeeman splitting
  of three-dimensional massless {D}irac {F}ermions in {Z}r{T}e$_5$},\ }\href
  {https://doi.org/10.1103/PhysRevLett.115.176404} {\bibfield  {journal}
  {\bibinfo  {journal} {Phys. Rev. Lett.}\ }\textbf {\bibinfo {volume} {115}},\
  \bibinfo {pages} {176404} (\bibinfo {year} {2015}{\natexlab{a}})}\BibitemShut
  {NoStop}%
\bibitem [{\citenamefont {Liu}\ \emph {et~al.}(2016)\citenamefont {Liu},
  \citenamefont {Yuan}, \citenamefont {Zhang}, \citenamefont {Jin},
  \citenamefont {Narayan}, \citenamefont {Luo}, \citenamefont {Chen},
  \citenamefont {Yang}, \citenamefont {Zou}, \citenamefont {Wu}, \citenamefont
  {Sanvito}, \citenamefont {Xia}, \citenamefont {Li}, \citenamefont {Wang},\
  and\ \citenamefont {Xiu}}]{Liu2016}%
  \BibitemOpen
  \bibfield  {author} {\bibinfo {author} {\bibfnamefont {Y.}~\bibnamefont
  {Liu}}, \bibinfo {author} {\bibfnamefont {X.}~\bibnamefont {Yuan}}, \bibinfo
  {author} {\bibfnamefont {C.}~\bibnamefont {Zhang}}, \bibinfo {author}
  {\bibfnamefont {Z.}~\bibnamefont {Jin}}, \bibinfo {author} {\bibfnamefont
  {A.}~\bibnamefont {Narayan}}, \bibinfo {author} {\bibfnamefont
  {C.}~\bibnamefont {Luo}}, \bibinfo {author} {\bibfnamefont {Z.}~\bibnamefont
  {Chen}}, \bibinfo {author} {\bibfnamefont {L.}~\bibnamefont {Yang}}, \bibinfo
  {author} {\bibfnamefont {J.}~\bibnamefont {Zou}}, \bibinfo {author}
  {\bibfnamefont {X.}~\bibnamefont {Wu}}, \bibinfo {author} {\bibfnamefont
  {S.}~\bibnamefont {Sanvito}}, \bibinfo {author} {\bibfnamefont
  {Z.}~\bibnamefont {Xia}}, \bibinfo {author} {\bibfnamefont {L.}~\bibnamefont
  {Li}}, \bibinfo {author} {\bibfnamefont {Z.}~\bibnamefont {Wang}},\ and\
  \bibinfo {author} {\bibfnamefont {F.}~\bibnamefont {Xiu}},\ }\bibfield
  {title} {\bibinfo {title} {Zeeman splitting and dynamical mass generation in
  {D}irac semimetal {Z}r{T}e$_5$},\ }\href
  {https://doi.org/10.1038/ncomms12516} {\bibfield  {journal} {\bibinfo
  {journal} {Nature Communications}\ }\textbf {\bibinfo {volume} {7}},\
  \bibinfo {pages} {12516} (\bibinfo {year} {2016})}\BibitemShut {NoStop}%
\bibitem [{\citenamefont {Wan}\ \emph {et~al.}(2011)\citenamefont {Wan},
  \citenamefont {Turner}, \citenamefont {Vishwanath},\ and\ \citenamefont
  {Savrasov}}]{PhysRevB.83.205101}%
  \BibitemOpen
  \bibfield  {author} {\bibinfo {author} {\bibfnamefont {X.}~\bibnamefont
  {Wan}}, \bibinfo {author} {\bibfnamefont {A.~M.}\ \bibnamefont {Turner}},
  \bibinfo {author} {\bibfnamefont {A.}~\bibnamefont {Vishwanath}},\ and\
  \bibinfo {author} {\bibfnamefont {S.~Y.}\ \bibnamefont {Savrasov}},\
  }\bibfield  {title} {\bibinfo {title} {Topological semimetal and {F}ermi-arc
  surface states in the electronic structure of pyrochlore iridates},\ }\href
  {https://doi.org/10.1103/PhysRevB.83.205101} {\bibfield  {journal} {\bibinfo
  {journal} {Phys. Rev. B}\ }\textbf {\bibinfo {volume} {83}},\ \bibinfo
  {pages} {205101} (\bibinfo {year} {2011})}\BibitemShut {NoStop}%
\bibitem [{\citenamefont {Burkov}\ and\ \citenamefont
  {Balents}(2011)}]{PhysRevLett.107.127205}%
  \BibitemOpen
  \bibfield  {author} {\bibinfo {author} {\bibfnamefont {A.~A.}\ \bibnamefont
  {Burkov}}\ and\ \bibinfo {author} {\bibfnamefont {L.}~\bibnamefont
  {Balents}},\ }\bibfield  {title} {\bibinfo {title} {{W}eyl semimetal in a
  topological insulator multilayer},\ }\href
  {https://doi.org/10.1103/PhysRevLett.107.127205} {\bibfield  {journal}
  {\bibinfo  {journal} {Phys. Rev. Lett.}\ }\textbf {\bibinfo {volume} {107}},\
  \bibinfo {pages} {127205} (\bibinfo {year} {2011})}\BibitemShut {NoStop}%
\bibitem [{\citenamefont {Weng}\ \emph {et~al.}(2015)\citenamefont {Weng},
  \citenamefont {Fang}, \citenamefont {Fang}, \citenamefont {Bernevig},\ and\
  \citenamefont {Dai}}]{PhysRevX.5.011029}%
  \BibitemOpen
  \bibfield  {author} {\bibinfo {author} {\bibfnamefont {H.}~\bibnamefont
  {Weng}}, \bibinfo {author} {\bibfnamefont {C.}~\bibnamefont {Fang}}, \bibinfo
  {author} {\bibfnamefont {Z.}~\bibnamefont {Fang}}, \bibinfo {author}
  {\bibfnamefont {B.~A.}\ \bibnamefont {Bernevig}},\ and\ \bibinfo {author}
  {\bibfnamefont {X.}~\bibnamefont {Dai}},\ }\bibfield  {title} {\bibinfo
  {title} {{W}eyl semimetal phase in noncentrosymmetric transition-metal
  monophosphides},\ }\href {https://doi.org/10.1103/PhysRevX.5.011029}
  {\bibfield  {journal} {\bibinfo  {journal} {Phys. Rev. X}\ }\textbf {\bibinfo
  {volume} {5}},\ \bibinfo {pages} {011029} (\bibinfo {year}
  {2015})}\BibitemShut {NoStop}%
\bibitem [{\citenamefont {Huang}\ \emph
  {et~al.}(2015{\natexlab{a}})\citenamefont {Huang}, \citenamefont {Xu},
  \citenamefont {Belopolski}, \citenamefont {Lee}, \citenamefont {Chang},
  \citenamefont {Wang}, \citenamefont {Alidoust}, \citenamefont {Bian},
  \citenamefont {Neupane}, \citenamefont {Zhang}, \citenamefont {Jia},
  \citenamefont {Bansil}, \citenamefont {Lin},\ and\ \citenamefont
  {Hasan}}]{Huang2015}%
  \BibitemOpen
  \bibfield  {author} {\bibinfo {author} {\bibfnamefont {S.-M.}\ \bibnamefont
  {Huang}}, \bibinfo {author} {\bibfnamefont {S.-Y.}\ \bibnamefont {Xu}},
  \bibinfo {author} {\bibfnamefont {I.}~\bibnamefont {Belopolski}}, \bibinfo
  {author} {\bibfnamefont {C.-C.}\ \bibnamefont {Lee}}, \bibinfo {author}
  {\bibfnamefont {G.}~\bibnamefont {Chang}}, \bibinfo {author} {\bibfnamefont
  {B.}~\bibnamefont {Wang}}, \bibinfo {author} {\bibfnamefont {N.}~\bibnamefont
  {Alidoust}}, \bibinfo {author} {\bibfnamefont {G.}~\bibnamefont {Bian}},
  \bibinfo {author} {\bibfnamefont {M.}~\bibnamefont {Neupane}}, \bibinfo
  {author} {\bibfnamefont {C.}~\bibnamefont {Zhang}}, \bibinfo {author}
  {\bibfnamefont {S.}~\bibnamefont {Jia}}, \bibinfo {author} {\bibfnamefont
  {A.}~\bibnamefont {Bansil}}, \bibinfo {author} {\bibfnamefont
  {H.}~\bibnamefont {Lin}},\ and\ \bibinfo {author} {\bibfnamefont {M.~Z.}\
  \bibnamefont {Hasan}},\ }\bibfield  {title} {\bibinfo {title} {A {W}eyl
  {F}ermion semimetal with surface {F}ermi arcs in the transition metal
  monopnictide {T}a{A}s class},\ }\href {https://doi.org/10.1038/ncomms8373}
  {\bibfield  {journal} {\bibinfo  {journal} {Nature Communications}\ }\textbf
  {\bibinfo {volume} {6}},\ \bibinfo {pages} {7373} (\bibinfo {year}
  {2015}{\natexlab{a}})}\BibitemShut {NoStop}%
\bibitem [{\citenamefont {Xu}\ \emph {et~al.}(2015{\natexlab{b}})\citenamefont
  {Xu}, \citenamefont {Belopolski}, \citenamefont {Alidoust}, \citenamefont
  {Neupane}, \citenamefont {Bian}, \citenamefont {Zhang}, \citenamefont
  {Sankar}, \citenamefont {Chang}, \citenamefont {Yuan}, \citenamefont {Lee},
  \citenamefont {Huang}, \citenamefont {Zheng}, \citenamefont {Ma},
  \citenamefont {Sanchez}, \citenamefont {Wang}, \citenamefont {Bansil},
  \citenamefont {Chou}, \citenamefont {Shibayev}, \citenamefont {Lin},
  \citenamefont {Jia},\ and\ \citenamefont {Hasan}}]{xu2015discovery}%
  \BibitemOpen
  \bibfield  {author} {\bibinfo {author} {\bibfnamefont {S.-Y.}\ \bibnamefont
  {Xu}}, \bibinfo {author} {\bibfnamefont {I.}~\bibnamefont {Belopolski}},
  \bibinfo {author} {\bibfnamefont {N.}~\bibnamefont {Alidoust}}, \bibinfo
  {author} {\bibfnamefont {M.}~\bibnamefont {Neupane}}, \bibinfo {author}
  {\bibfnamefont {G.}~\bibnamefont {Bian}}, \bibinfo {author} {\bibfnamefont
  {C.}~\bibnamefont {Zhang}}, \bibinfo {author} {\bibfnamefont
  {R.}~\bibnamefont {Sankar}}, \bibinfo {author} {\bibfnamefont
  {G.}~\bibnamefont {Chang}}, \bibinfo {author} {\bibfnamefont
  {Z.}~\bibnamefont {Yuan}}, \bibinfo {author} {\bibfnamefont {C.-C.}\
  \bibnamefont {Lee}}, \bibinfo {author} {\bibfnamefont {S.-M.}\ \bibnamefont
  {Huang}}, \bibinfo {author} {\bibfnamefont {H.}~\bibnamefont {Zheng}},
  \bibinfo {author} {\bibfnamefont {J.}~\bibnamefont {Ma}}, \bibinfo {author}
  {\bibfnamefont {D.~S.}\ \bibnamefont {Sanchez}}, \bibinfo {author}
  {\bibfnamefont {B.}~\bibnamefont {Wang}}, \bibinfo {author} {\bibfnamefont
  {A.}~\bibnamefont {Bansil}}, \bibinfo {author} {\bibfnamefont
  {F.}~\bibnamefont {Chou}}, \bibinfo {author} {\bibfnamefont {P.~P.}\
  \bibnamefont {Shibayev}}, \bibinfo {author} {\bibfnamefont {H.}~\bibnamefont
  {Lin}}, \bibinfo {author} {\bibfnamefont {S.}~\bibnamefont {Jia}},\ and\
  \bibinfo {author} {\bibfnamefont {M.~Z.}\ \bibnamefont {Hasan}},\ }\bibfield
  {title} {\bibinfo {title} {Discovery of a {W}eyl {F}ermion semimetal and
  topological {F}ermi arcs},\ }\href {https://doi.org/10.1126/science.aaa9297}
  {\bibfield  {journal} {\bibinfo  {journal} {Science}\ }\textbf {\bibinfo
  {volume} {349}},\ \bibinfo {pages} {613} (\bibinfo {year}
  {2015}{\natexlab{b}})}\BibitemShut {NoStop}%
\bibitem [{\citenamefont {Lv}\ \emph {et~al.}(2015)\citenamefont {Lv},
  \citenamefont {Weng}, \citenamefont {Fu}, \citenamefont {Wang}, \citenamefont
  {Miao}, \citenamefont {Ma}, \citenamefont {Richard}, \citenamefont {Huang},
  \citenamefont {Zhao}, \citenamefont {Chen}, \citenamefont {Fang},
  \citenamefont {Dai}, \citenamefont {Qian},\ and\ \citenamefont
  {Ding}}]{PhysRevX.5.031013}%
  \BibitemOpen
  \bibfield  {author} {\bibinfo {author} {\bibfnamefont {B.~Q.}\ \bibnamefont
  {Lv}}, \bibinfo {author} {\bibfnamefont {H.~M.}\ \bibnamefont {Weng}},
  \bibinfo {author} {\bibfnamefont {B.~B.}\ \bibnamefont {Fu}}, \bibinfo
  {author} {\bibfnamefont {X.~P.}\ \bibnamefont {Wang}}, \bibinfo {author}
  {\bibfnamefont {H.}~\bibnamefont {Miao}}, \bibinfo {author} {\bibfnamefont
  {J.}~\bibnamefont {Ma}}, \bibinfo {author} {\bibfnamefont {P.}~\bibnamefont
  {Richard}}, \bibinfo {author} {\bibfnamefont {X.~C.}\ \bibnamefont {Huang}},
  \bibinfo {author} {\bibfnamefont {L.~X.}\ \bibnamefont {Zhao}}, \bibinfo
  {author} {\bibfnamefont {G.~F.}\ \bibnamefont {Chen}}, \bibinfo {author}
  {\bibfnamefont {Z.}~\bibnamefont {Fang}}, \bibinfo {author} {\bibfnamefont
  {X.}~\bibnamefont {Dai}}, \bibinfo {author} {\bibfnamefont {T.}~\bibnamefont
  {Qian}},\ and\ \bibinfo {author} {\bibfnamefont {H.}~\bibnamefont {Ding}},\
  }\bibfield  {title} {\bibinfo {title} {Experimental discovery of {W}eyl
  semimetal {T}a{A}s},\ }\href {https://doi.org/10.1103/PhysRevX.5.031013}
  {\bibfield  {journal} {\bibinfo  {journal} {Phys. Rev. X}\ }\textbf {\bibinfo
  {volume} {5}},\ \bibinfo {pages} {031013} (\bibinfo {year}
  {2015})}\BibitemShut {NoStop}%
\bibitem [{\citenamefont {Yang}\ \emph {et~al.}(2015)\citenamefont {Yang},
  \citenamefont {Liu}, \citenamefont {Sun}, \citenamefont {Peng}, \citenamefont
  {Yang}, \citenamefont {Zhang}, \citenamefont {Zhou}, \citenamefont {Zhang},
  \citenamefont {Guo}, \citenamefont {Rahn}, \citenamefont {Prabhakaran},
  \citenamefont {Hussain}, \citenamefont {Mo}, \citenamefont {Felser},
  \citenamefont {Yan},\ and\ \citenamefont {Chen}}]{Yang2015}%
  \BibitemOpen
  \bibfield  {author} {\bibinfo {author} {\bibfnamefont {L.~X.}\ \bibnamefont
  {Yang}}, \bibinfo {author} {\bibfnamefont {Z.~K.}\ \bibnamefont {Liu}},
  \bibinfo {author} {\bibfnamefont {Y.}~\bibnamefont {Sun}}, \bibinfo {author}
  {\bibfnamefont {H.}~\bibnamefont {Peng}}, \bibinfo {author} {\bibfnamefont
  {H.~F.}\ \bibnamefont {Yang}}, \bibinfo {author} {\bibfnamefont
  {T.}~\bibnamefont {Zhang}}, \bibinfo {author} {\bibfnamefont
  {B.}~\bibnamefont {Zhou}}, \bibinfo {author} {\bibfnamefont {Y.}~\bibnamefont
  {Zhang}}, \bibinfo {author} {\bibfnamefont {Y.~F.}\ \bibnamefont {Guo}},
  \bibinfo {author} {\bibfnamefont {M.}~\bibnamefont {Rahn}}, \bibinfo {author}
  {\bibfnamefont {D.}~\bibnamefont {Prabhakaran}}, \bibinfo {author}
  {\bibfnamefont {Z.}~\bibnamefont {Hussain}}, \bibinfo {author} {\bibfnamefont
  {S.-K.}\ \bibnamefont {Mo}}, \bibinfo {author} {\bibfnamefont
  {C.}~\bibnamefont {Felser}}, \bibinfo {author} {\bibfnamefont
  {B.}~\bibnamefont {Yan}},\ and\ \bibinfo {author} {\bibfnamefont {Y.~L.}\
  \bibnamefont {Chen}},\ }\bibfield  {title} {\bibinfo {title} {{W}eyl
  semimetal phase in the non-centrosymmetric compound taas},\ }\href
  {https://doi.org/10.1038/nphys3425} {\bibfield  {journal} {\bibinfo
  {journal} {Nature Physics}\ }\textbf {\bibinfo {volume} {11}},\ \bibinfo
  {pages} {728} (\bibinfo {year} {2015})}\BibitemShut {NoStop}%
\bibitem [{\citenamefont {Huang}\ \emph
  {et~al.}(2015{\natexlab{b}})\citenamefont {Huang}, \citenamefont {Zhao},
  \citenamefont {Long}, \citenamefont {Wang}, \citenamefont {Chen},
  \citenamefont {Yang}, \citenamefont {Liang}, \citenamefont {Xue},
  \citenamefont {Weng}, \citenamefont {Fang}, \citenamefont {Dai},\ and\
  \citenamefont {Chen}}]{PhysRevX.5.031023}%
  \BibitemOpen
  \bibfield  {author} {\bibinfo {author} {\bibfnamefont {X.}~\bibnamefont
  {Huang}}, \bibinfo {author} {\bibfnamefont {L.}~\bibnamefont {Zhao}},
  \bibinfo {author} {\bibfnamefont {Y.}~\bibnamefont {Long}}, \bibinfo {author}
  {\bibfnamefont {P.}~\bibnamefont {Wang}}, \bibinfo {author} {\bibfnamefont
  {D.}~\bibnamefont {Chen}}, \bibinfo {author} {\bibfnamefont {Z.}~\bibnamefont
  {Yang}}, \bibinfo {author} {\bibfnamefont {H.}~\bibnamefont {Liang}},
  \bibinfo {author} {\bibfnamefont {M.}~\bibnamefont {Xue}}, \bibinfo {author}
  {\bibfnamefont {H.}~\bibnamefont {Weng}}, \bibinfo {author} {\bibfnamefont
  {Z.}~\bibnamefont {Fang}}, \bibinfo {author} {\bibfnamefont {X.}~\bibnamefont
  {Dai}},\ and\ \bibinfo {author} {\bibfnamefont {G.}~\bibnamefont {Chen}},\
  }\bibfield  {title} {\bibinfo {title} {Observation of the
  chiral-anomaly-induced negative magnetoresistance in 3d {W}eyl semimetal
  {T}a{A}s},\ }\href {https://doi.org/10.1103/PhysRevX.5.031023} {\bibfield
  {journal} {\bibinfo  {journal} {Phys. Rev. X}\ }\textbf {\bibinfo {volume}
  {5}},\ \bibinfo {pages} {031023} (\bibinfo {year}
  {2015}{\natexlab{b}})}\BibitemShut {NoStop}%
\bibitem [{\citenamefont {Xu}\ \emph {et~al.}(2015{\natexlab{c}})\citenamefont
  {Xu}, \citenamefont {Alidoust}, \citenamefont {Belopolski}, \citenamefont
  {Yuan}, \citenamefont {Bian}, \citenamefont {Chang}, \citenamefont {Zheng},
  \citenamefont {Strocov}, \citenamefont {Sanchez}, \citenamefont {Chang},
  \citenamefont {Zhang}, \citenamefont {Mou}, \citenamefont {Wu}, \citenamefont
  {Huang}, \citenamefont {Lee}, \citenamefont {Huang}, \citenamefont {Wang},
  \citenamefont {Bansil}, \citenamefont {Jeng}, \citenamefont {Neupert},
  \citenamefont {Kaminski}, \citenamefont {Lin}, \citenamefont {Jia},\ and\
  \citenamefont {Zahid~Hasan}}]{Xu2015}%
  \BibitemOpen
  \bibfield  {author} {\bibinfo {author} {\bibfnamefont {S.-Y.}\ \bibnamefont
  {Xu}}, \bibinfo {author} {\bibfnamefont {N.}~\bibnamefont {Alidoust}},
  \bibinfo {author} {\bibfnamefont {I.}~\bibnamefont {Belopolski}}, \bibinfo
  {author} {\bibfnamefont {Z.}~\bibnamefont {Yuan}}, \bibinfo {author}
  {\bibfnamefont {G.}~\bibnamefont {Bian}}, \bibinfo {author} {\bibfnamefont
  {T.-R.}\ \bibnamefont {Chang}}, \bibinfo {author} {\bibfnamefont
  {H.}~\bibnamefont {Zheng}}, \bibinfo {author} {\bibfnamefont {V.~N.}\
  \bibnamefont {Strocov}}, \bibinfo {author} {\bibfnamefont {D.~S.}\
  \bibnamefont {Sanchez}}, \bibinfo {author} {\bibfnamefont {G.}~\bibnamefont
  {Chang}}, \bibinfo {author} {\bibfnamefont {C.}~\bibnamefont {Zhang}},
  \bibinfo {author} {\bibfnamefont {D.}~\bibnamefont {Mou}}, \bibinfo {author}
  {\bibfnamefont {Y.}~\bibnamefont {Wu}}, \bibinfo {author} {\bibfnamefont
  {L.}~\bibnamefont {Huang}}, \bibinfo {author} {\bibfnamefont {C.-C.}\
  \bibnamefont {Lee}}, \bibinfo {author} {\bibfnamefont {S.-M.}\ \bibnamefont
  {Huang}}, \bibinfo {author} {\bibfnamefont {B.}~\bibnamefont {Wang}},
  \bibinfo {author} {\bibfnamefont {A.}~\bibnamefont {Bansil}}, \bibinfo
  {author} {\bibfnamefont {H.-T.}\ \bibnamefont {Jeng}}, \bibinfo {author}
  {\bibfnamefont {T.}~\bibnamefont {Neupert}}, \bibinfo {author} {\bibfnamefont
  {A.}~\bibnamefont {Kaminski}}, \bibinfo {author} {\bibfnamefont
  {H.}~\bibnamefont {Lin}}, \bibinfo {author} {\bibfnamefont {S.}~\bibnamefont
  {Jia}},\ and\ \bibinfo {author} {\bibfnamefont {M.}~\bibnamefont
  {Zahid~Hasan}},\ }\bibfield  {title} {\bibinfo {title} {Discovery of a {W}eyl
  {F}ermion state with {F}ermi arcs in niobium arsenide},\ }\href
  {https://doi.org/10.1038/nphys3437} {\bibfield  {journal} {\bibinfo
  {journal} {Nature Physics}\ }\textbf {\bibinfo {volume} {11}},\ \bibinfo
  {pages} {748} (\bibinfo {year} {2015}{\natexlab{c}})}\BibitemShut {NoStop}%
\bibitem [{\citenamefont {Shekhar}\ \emph {et~al.}(2015)\citenamefont
  {Shekhar}, \citenamefont {Nayak}, \citenamefont {Sun}, \citenamefont
  {Schmidt}, \citenamefont {Nicklas}, \citenamefont {Leermakers}, \citenamefont
  {Zeitler}, \citenamefont {Skourski}, \citenamefont {Wosnitza}, \citenamefont
  {Liu}, \citenamefont {Chen}, \citenamefont {Schnelle}, \citenamefont
  {Borrmann}, \citenamefont {Grin}, \citenamefont {Felser},\ and\ \citenamefont
  {Yan}}]{Shekhar2015}%
  \BibitemOpen
  \bibfield  {author} {\bibinfo {author} {\bibfnamefont {C.}~\bibnamefont
  {Shekhar}}, \bibinfo {author} {\bibfnamefont {A.~K.}\ \bibnamefont {Nayak}},
  \bibinfo {author} {\bibfnamefont {Y.}~\bibnamefont {Sun}}, \bibinfo {author}
  {\bibfnamefont {M.}~\bibnamefont {Schmidt}}, \bibinfo {author} {\bibfnamefont
  {M.}~\bibnamefont {Nicklas}}, \bibinfo {author} {\bibfnamefont
  {I.}~\bibnamefont {Leermakers}}, \bibinfo {author} {\bibfnamefont
  {U.}~\bibnamefont {Zeitler}}, \bibinfo {author} {\bibfnamefont
  {Y.}~\bibnamefont {Skourski}}, \bibinfo {author} {\bibfnamefont
  {J.}~\bibnamefont {Wosnitza}}, \bibinfo {author} {\bibfnamefont
  {Z.}~\bibnamefont {Liu}}, \bibinfo {author} {\bibfnamefont {Y.}~\bibnamefont
  {Chen}}, \bibinfo {author} {\bibfnamefont {W.}~\bibnamefont {Schnelle}},
  \bibinfo {author} {\bibfnamefont {H.}~\bibnamefont {Borrmann}}, \bibinfo
  {author} {\bibfnamefont {Y.}~\bibnamefont {Grin}}, \bibinfo {author}
  {\bibfnamefont {C.}~\bibnamefont {Felser}},\ and\ \bibinfo {author}
  {\bibfnamefont {B.}~\bibnamefont {Yan}},\ }\bibfield  {title} {\bibinfo
  {title} {Extremely large magnetoresistance and ultrahigh mobility in the
  topological {W}eyl semimetal candidate {N}b{P}},\ }\href
  {https://doi.org/10.1038/nphys3372} {\bibfield  {journal} {\bibinfo
  {journal} {Nature Physics}\ }\textbf {\bibinfo {volume} {11}},\ \bibinfo
  {pages} {645} (\bibinfo {year} {2015})}\BibitemShut {NoStop}%
\bibitem [{\citenamefont {Lu}\ \emph {et~al.}(2013)\citenamefont {Lu},
  \citenamefont {Fu}, \citenamefont {Joannopoulos},\ and\ \citenamefont
  {Soljačić}}]{Lu2013}%
  \BibitemOpen
  \bibfield  {author} {\bibinfo {author} {\bibfnamefont {L.}~\bibnamefont
  {Lu}}, \bibinfo {author} {\bibfnamefont {L.}~\bibnamefont {Fu}}, \bibinfo
  {author} {\bibfnamefont {J.~D.}\ \bibnamefont {Joannopoulos}},\ and\ \bibinfo
  {author} {\bibfnamefont {M.}~\bibnamefont {Soljačić}},\ }\bibfield  {title}
  {\bibinfo {title} {{W}eyl points and line nodes in gyroid photonic
  crystals},\ }\href {https://doi.org/10.1038/nphoton.2013.42} {\bibfield
  {journal} {\bibinfo  {journal} {Nature Photonics}\ }\textbf {\bibinfo
  {volume} {7}},\ \bibinfo {pages} {294} (\bibinfo {year} {2013})}\BibitemShut
  {NoStop}%
\bibitem [{\citenamefont {Lu}\ \emph {et~al.}(2015)\citenamefont {Lu},
  \citenamefont {Wang}, \citenamefont {Ye}, \citenamefont {Ran}, \citenamefont
  {Fu}, \citenamefont {Joannopoulos},\ and\ \citenamefont {Solja{\v
  c}i{\'c}}}]{Lu622}%
  \BibitemOpen
  \bibfield  {author} {\bibinfo {author} {\bibfnamefont {L.}~\bibnamefont
  {Lu}}, \bibinfo {author} {\bibfnamefont {Z.}~\bibnamefont {Wang}}, \bibinfo
  {author} {\bibfnamefont {D.}~\bibnamefont {Ye}}, \bibinfo {author}
  {\bibfnamefont {L.}~\bibnamefont {Ran}}, \bibinfo {author} {\bibfnamefont
  {L.}~\bibnamefont {Fu}}, \bibinfo {author} {\bibfnamefont {J.~D.}\
  \bibnamefont {Joannopoulos}},\ and\ \bibinfo {author} {\bibfnamefont
  {M.}~\bibnamefont {Solja{\v c}i{\'c}}},\ }\bibfield  {title} {\bibinfo
  {title} {Experimental observation of {W}eyl points},\ }\href
  {https://doi.org/10.1126/science.aaa9273} {\bibfield  {journal} {\bibinfo
  {journal} {Science}\ }\textbf {\bibinfo {volume} {349}},\ \bibinfo {pages}
  {622} (\bibinfo {year} {2015})}\BibitemShut {NoStop}%
\bibitem [{\citenamefont {Soluyanov}\ \emph {et~al.}(2015)\citenamefont
  {Soluyanov}, \citenamefont {Gresch}, \citenamefont {Wang}, \citenamefont
  {Wu}, \citenamefont {Troyer}, \citenamefont {Dai},\ and\ \citenamefont
  {Bernevig}}]{Soluyanov2015}%
  \BibitemOpen
  \bibfield  {author} {\bibinfo {author} {\bibfnamefont {A.~A.}\ \bibnamefont
  {Soluyanov}}, \bibinfo {author} {\bibfnamefont {D.}~\bibnamefont {Gresch}},
  \bibinfo {author} {\bibfnamefont {Z.}~\bibnamefont {Wang}}, \bibinfo {author}
  {\bibfnamefont {Q.}~\bibnamefont {Wu}}, \bibinfo {author} {\bibfnamefont
  {M.}~\bibnamefont {Troyer}}, \bibinfo {author} {\bibfnamefont
  {X.}~\bibnamefont {Dai}},\ and\ \bibinfo {author} {\bibfnamefont {B.~A.}\
  \bibnamefont {Bernevig}},\ }\bibfield  {title} {\bibinfo {title} {Type-{II}
  {W}eyl semimetals},\ }\href {https://doi.org/10.1038/nature15768} {\bibfield
  {journal} {\bibinfo  {journal} {Nature}\ }\textbf {\bibinfo {volume} {527}},\
  \bibinfo {pages} {495} (\bibinfo {year} {2015})}\BibitemShut {NoStop}%
\bibitem [{\citenamefont {Xu}\ \emph {et~al.}(2015{\natexlab{d}})\citenamefont
  {Xu}, \citenamefont {Zhang},\ and\ \citenamefont
  {Zhang}}]{PhysRevLett.115.265304}%
  \BibitemOpen
  \bibfield  {author} {\bibinfo {author} {\bibfnamefont {Y.}~\bibnamefont
  {Xu}}, \bibinfo {author} {\bibfnamefont {F.}~\bibnamefont {Zhang}},\ and\
  \bibinfo {author} {\bibfnamefont {C.}~\bibnamefont {Zhang}},\ }\bibfield
  {title} {\bibinfo {title} {Structured {W}eyl points in spin-orbit coupled
  fermionic superfluids},\ }\href
  {https://doi.org/10.1103/PhysRevLett.115.265304} {\bibfield  {journal}
  {\bibinfo  {journal} {Phys. Rev. Lett.}\ }\textbf {\bibinfo {volume} {115}},\
  \bibinfo {pages} {265304} (\bibinfo {year} {2015}{\natexlab{d}})}\BibitemShut
  {NoStop}%
\bibitem [{\citenamefont {Nguyen}\ \emph {et~al.}(2020)\citenamefont {Nguyen},
  \citenamefont {Han}, \citenamefont {Andrejevic}, \citenamefont {Pablo-Pedro},
  \citenamefont {Apte}, \citenamefont {Tsurimaki}, \citenamefont {Ding},
  \citenamefont {Zhang}, \citenamefont {Alatas}, \citenamefont {Alp},
  \citenamefont {Chi}, \citenamefont {Fernandez-Baca}, \citenamefont {Matsuda},
  \citenamefont {Tennant}, \citenamefont {Zhao}, \citenamefont {Xu},
  \citenamefont {Lynn}, \citenamefont {Huang},\ and\ \citenamefont
  {Li}}]{PhysRevLett.124.236401}%
  \BibitemOpen
  \bibfield  {author} {\bibinfo {author} {\bibfnamefont {T.}~\bibnamefont
  {Nguyen}}, \bibinfo {author} {\bibfnamefont {F.}~\bibnamefont {Han}},
  \bibinfo {author} {\bibfnamefont {N.}~\bibnamefont {Andrejevic}}, \bibinfo
  {author} {\bibfnamefont {R.}~\bibnamefont {Pablo-Pedro}}, \bibinfo {author}
  {\bibfnamefont {A.}~\bibnamefont {Apte}}, \bibinfo {author} {\bibfnamefont
  {Y.}~\bibnamefont {Tsurimaki}}, \bibinfo {author} {\bibfnamefont
  {Z.}~\bibnamefont {Ding}}, \bibinfo {author} {\bibfnamefont {K.}~\bibnamefont
  {Zhang}}, \bibinfo {author} {\bibfnamefont {A.}~\bibnamefont {Alatas}},
  \bibinfo {author} {\bibfnamefont {E.~E.}\ \bibnamefont {Alp}}, \bibinfo
  {author} {\bibfnamefont {S.}~\bibnamefont {Chi}}, \bibinfo {author}
  {\bibfnamefont {J.}~\bibnamefont {Fernandez-Baca}}, \bibinfo {author}
  {\bibfnamefont {M.}~\bibnamefont {Matsuda}}, \bibinfo {author} {\bibfnamefont
  {D.~A.}\ \bibnamefont {Tennant}}, \bibinfo {author} {\bibfnamefont
  {Y.}~\bibnamefont {Zhao}}, \bibinfo {author} {\bibfnamefont {Z.}~\bibnamefont
  {Xu}}, \bibinfo {author} {\bibfnamefont {J.~W.}\ \bibnamefont {Lynn}},
  \bibinfo {author} {\bibfnamefont {S.}~\bibnamefont {Huang}},\ and\ \bibinfo
  {author} {\bibfnamefont {M.}~\bibnamefont {Li}},\ }\bibfield  {title}
  {\bibinfo {title} {Topological singularity induced chiral {K}ohn anomaly in a
  {W}eyl semimetal},\ }\href {https://doi.org/10.1103/PhysRevLett.124.236401}
  {\bibfield  {journal} {\bibinfo  {journal} {Phys. Rev. Lett.}\ }\textbf
  {\bibinfo {volume} {124}},\ \bibinfo {pages} {236401} (\bibinfo {year}
  {2020})}\BibitemShut {NoStop}%
\bibitem [{\citenamefont {Burkov}\ \emph {et~al.}(2011)\citenamefont {Burkov},
  \citenamefont {Hook},\ and\ \citenamefont {Balents}}]{PhysRevB.84.235126}%
  \BibitemOpen
  \bibfield  {author} {\bibinfo {author} {\bibfnamefont {A.~A.}\ \bibnamefont
  {Burkov}}, \bibinfo {author} {\bibfnamefont {M.~D.}\ \bibnamefont {Hook}},\
  and\ \bibinfo {author} {\bibfnamefont {L.}~\bibnamefont {Balents}},\
  }\bibfield  {title} {\bibinfo {title} {Topological nodal semimetals},\ }\href
  {https://doi.org/10.1103/PhysRevB.84.235126} {\bibfield  {journal} {\bibinfo
  {journal} {Phys. Rev. B}\ }\textbf {\bibinfo {volume} {84}},\ \bibinfo
  {pages} {235126} (\bibinfo {year} {2011})}\BibitemShut {NoStop}%
\bibitem [{\citenamefont {Yan}\ \emph {et~al.}(2017)\citenamefont {Yan},
  \citenamefont {Bi}, \citenamefont {Shen}, \citenamefont {Lu}, \citenamefont
  {Zhang},\ and\ \citenamefont {Wang}}]{PhysRevB.96.041103}%
  \BibitemOpen
  \bibfield  {author} {\bibinfo {author} {\bibfnamefont {Z.}~\bibnamefont
  {Yan}}, \bibinfo {author} {\bibfnamefont {R.}~\bibnamefont {Bi}}, \bibinfo
  {author} {\bibfnamefont {H.}~\bibnamefont {Shen}}, \bibinfo {author}
  {\bibfnamefont {L.}~\bibnamefont {Lu}}, \bibinfo {author} {\bibfnamefont
  {S.-C.}\ \bibnamefont {Zhang}},\ and\ \bibinfo {author} {\bibfnamefont
  {Z.}~\bibnamefont {Wang}},\ }\bibfield  {title} {\bibinfo {title} {Nodal-link
  semimetals},\ }\href {https://doi.org/10.1103/PhysRevB.96.041103} {\bibfield
  {journal} {\bibinfo  {journal} {Phys. Rev. B}\ }\textbf {\bibinfo {volume}
  {96}},\ \bibinfo {pages} {041103(R)} (\bibinfo {year} {2017})}\BibitemShut
  {NoStop}%
\bibitem [{\citenamefont {Bi}\ \emph {et~al.}(2017)\citenamefont {Bi},
  \citenamefont {Yan}, \citenamefont {Lu},\ and\ \citenamefont
  {Wang}}]{PhysRevB.96.201305}%
  \BibitemOpen
  \bibfield  {author} {\bibinfo {author} {\bibfnamefont {R.}~\bibnamefont
  {Bi}}, \bibinfo {author} {\bibfnamefont {Z.}~\bibnamefont {Yan}}, \bibinfo
  {author} {\bibfnamefont {L.}~\bibnamefont {Lu}},\ and\ \bibinfo {author}
  {\bibfnamefont {Z.}~\bibnamefont {Wang}},\ }\bibfield  {title} {\bibinfo
  {title} {Nodal-knot semimetals},\ }\href
  {https://doi.org/10.1103/PhysRevB.96.201305} {\bibfield  {journal} {\bibinfo
  {journal} {Phys. Rev. B}\ }\textbf {\bibinfo {volume} {96}},\ \bibinfo
  {pages} {201305(R)} (\bibinfo {year} {2017})}\BibitemShut {NoStop}%
\bibitem [{\citenamefont {Samanta}\ \emph {et~al.}(2021)\citenamefont
  {Samanta}, \citenamefont {Arovas},\ and\ \citenamefont
  {Auerbach}}]{PhysRevLett.126.076603}%
  \BibitemOpen
  \bibfield  {author} {\bibinfo {author} {\bibfnamefont {A.}~\bibnamefont
  {Samanta}}, \bibinfo {author} {\bibfnamefont {D.~P.}\ \bibnamefont
  {Arovas}},\ and\ \bibinfo {author} {\bibfnamefont {A.}~\bibnamefont
  {Auerbach}},\ }\bibfield  {title} {\bibinfo {title} {Hall coefficient of
  semimetals},\ }\href {https://doi.org/10.1103/PhysRevLett.126.076603}
  {\bibfield  {journal} {\bibinfo  {journal} {Phys. Rev. Lett.}\ }\textbf
  {\bibinfo {volume} {126}},\ \bibinfo {pages} {076603} (\bibinfo {year}
  {2021})}\BibitemShut {NoStop}%
\bibitem [{\citenamefont {Sengupta}\ \emph {et~al.}(2020)\citenamefont
  {Sengupta}, \citenamefont {Lhachemi},\ and\ \citenamefont
  {Garate}}]{PhysRevLett.125.146402}%
  \BibitemOpen
  \bibfield  {author} {\bibinfo {author} {\bibfnamefont {S.}~\bibnamefont
  {Sengupta}}, \bibinfo {author} {\bibfnamefont {M.~N.~Y.}\ \bibnamefont
  {Lhachemi}},\ and\ \bibinfo {author} {\bibfnamefont {I.}~\bibnamefont
  {Garate}},\ }\bibfield  {title} {\bibinfo {title} {Phonon magnetochiral
  effect of band-geometric origin in {W}eyl semimetals},\ }\href
  {https://doi.org/10.1103/PhysRevLett.125.146402} {\bibfield  {journal}
  {\bibinfo  {journal} {Phys. Rev. Lett.}\ }\textbf {\bibinfo {volume} {125}},\
  \bibinfo {pages} {146402} (\bibinfo {year} {2020})}\BibitemShut {NoStop}%
\bibitem [{\citenamefont {Lin}\ and\ \citenamefont
  {Hughes}(2018)}]{PhysRevB.98.241103}%
  \BibitemOpen
  \bibfield  {author} {\bibinfo {author} {\bibfnamefont {M.}~\bibnamefont
  {Lin}}\ and\ \bibinfo {author} {\bibfnamefont {T.~L.}\ \bibnamefont
  {Hughes}},\ }\bibfield  {title} {\bibinfo {title} {Topological quadrupolar
  semimetals},\ }\href {https://doi.org/10.1103/PhysRevB.98.241103} {\bibfield
  {journal} {\bibinfo  {journal} {Phys. Rev. B}\ }\textbf {\bibinfo {volume}
  {98}},\ \bibinfo {pages} {241103(R)} (\bibinfo {year} {2018})}\BibitemShut
  {NoStop}%
\bibitem [{\citenamefont {Wieder}\ \emph {et~al.}(2020)\citenamefont {Wieder},
  \citenamefont {Wang}, \citenamefont {Cano}, \citenamefont {Dai},
  \citenamefont {Schoop}, \citenamefont {Bradlyn},\ and\ \citenamefont
  {Bernevig}}]{Wieder2020}%
  \BibitemOpen
  \bibfield  {author} {\bibinfo {author} {\bibfnamefont {B.~J.}\ \bibnamefont
  {Wieder}}, \bibinfo {author} {\bibfnamefont {Z.}~\bibnamefont {Wang}},
  \bibinfo {author} {\bibfnamefont {J.}~\bibnamefont {Cano}}, \bibinfo {author}
  {\bibfnamefont {X.}~\bibnamefont {Dai}}, \bibinfo {author} {\bibfnamefont
  {L.~M.}\ \bibnamefont {Schoop}}, \bibinfo {author} {\bibfnamefont
  {B.}~\bibnamefont {Bradlyn}},\ and\ \bibinfo {author} {\bibfnamefont {B.~A.}\
  \bibnamefont {Bernevig}},\ }\bibfield  {title} {\bibinfo {title} {Strong and
  fragile topological dirac semimetals with higher-order fermi arcs},\ }\href
  {https://doi.org/10.1038/s41467-020-14443-5} {\bibfield  {journal} {\bibinfo
  {journal} {Nature Communications}\ }\textbf {\bibinfo {volume} {11}},\
  \bibinfo {pages} {627} (\bibinfo {year} {2020})}\BibitemShut {NoStop}%
\bibitem [{\citenamefont {Wu}\ \emph {et~al.}(2020{\natexlab{b}})\citenamefont
  {Wu}, \citenamefont {Yu}, \citenamefont {Zhou}, \citenamefont {Zhao},\ and\
  \citenamefont {Yang}}]{PhysRevB.101.205134}%
  \BibitemOpen
  \bibfield  {author} {\bibinfo {author} {\bibfnamefont {W.}~\bibnamefont
  {Wu}}, \bibinfo {author} {\bibfnamefont {Z.-M.}\ \bibnamefont {Yu}}, \bibinfo
  {author} {\bibfnamefont {X.}~\bibnamefont {Zhou}}, \bibinfo {author}
  {\bibfnamefont {Y.~X.}\ \bibnamefont {Zhao}},\ and\ \bibinfo {author}
  {\bibfnamefont {S.~A.}\ \bibnamefont {Yang}},\ }\bibfield  {title} {\bibinfo
  {title} {Higher-order {D}irac {F}ermions in three dimensions},\ }\href
  {https://doi.org/10.1103/PhysRevB.101.205134} {\bibfield  {journal} {\bibinfo
   {journal} {Phys. Rev. B}\ }\textbf {\bibinfo {volume} {101}},\ \bibinfo
  {pages} {205134} (\bibinfo {year} {2020}{\natexlab{b}})}\BibitemShut
  {NoStop}%
\bibitem [{\citenamefont {Roy}(2020)}]{PhysRevB.101.220506}%
  \BibitemOpen
  \bibfield  {author} {\bibinfo {author} {\bibfnamefont {B.}~\bibnamefont
  {Roy}},\ }\bibfield  {title} {\bibinfo {title} {Higher-order topological
  superconductors in $\mathcal{P}$-, $\mathcal{T}$-odd quadrupolar {D}irac
  materials},\ }\href {https://doi.org/10.1103/PhysRevB.101.220506} {\bibfield
  {journal} {\bibinfo  {journal} {Phys. Rev. B}\ }\textbf {\bibinfo {volume}
  {101}},\ \bibinfo {pages} {220506(R)} (\bibinfo {year} {2020})}\BibitemShut
  {NoStop}%
\bibitem [{\citenamefont {Wang}\ \emph
  {et~al.}(2020{\natexlab{a}})\citenamefont {Wang}, \citenamefont {Dai},
  \citenamefont {Shao}, \citenamefont {Yang},\ and\ \citenamefont
  {Zhao}}]{PhysRevLett.125.126403}%
  \BibitemOpen
  \bibfield  {author} {\bibinfo {author} {\bibfnamefont {K.}~\bibnamefont
  {Wang}}, \bibinfo {author} {\bibfnamefont {J.-X.}\ \bibnamefont {Dai}},
  \bibinfo {author} {\bibfnamefont {L.~B.}\ \bibnamefont {Shao}}, \bibinfo
  {author} {\bibfnamefont {S.~A.}\ \bibnamefont {Yang}},\ and\ \bibinfo
  {author} {\bibfnamefont {Y.~X.}\ \bibnamefont {Zhao}},\ }\bibfield  {title}
  {\bibinfo {title} {Boundary criticality of $\mathcal{PT}$-invariant topology
  and second-order nodal-line semimetals},\ }\href
  {https://doi.org/10.1103/PhysRevLett.125.126403} {\bibfield  {journal}
  {\bibinfo  {journal} {Phys. Rev. Lett.}\ }\textbf {\bibinfo {volume} {125}},\
  \bibinfo {pages} {126403} (\bibinfo {year} {2020}{\natexlab{a}})}\BibitemShut
  {NoStop}%
\bibitem [{\citenamefont {Wang}\ \emph
  {et~al.}(2020{\natexlab{b}})\citenamefont {Wang}, \citenamefont {Lin},
  \citenamefont {Jiang}, \citenamefont {Guo},\ and\ \citenamefont
  {Jiang}}]{PhysRevLett.125.146401}%
  \BibitemOpen
  \bibfield  {author} {\bibinfo {author} {\bibfnamefont {H.-X.}\ \bibnamefont
  {Wang}}, \bibinfo {author} {\bibfnamefont {Z.-K.}\ \bibnamefont {Lin}},
  \bibinfo {author} {\bibfnamefont {B.}~\bibnamefont {Jiang}}, \bibinfo
  {author} {\bibfnamefont {G.-Y.}\ \bibnamefont {Guo}},\ and\ \bibinfo {author}
  {\bibfnamefont {J.-H.}\ \bibnamefont {Jiang}},\ }\bibfield  {title} {\bibinfo
  {title} {Higher-order {W}eyl semimetals},\ }\href
  {https://doi.org/10.1103/PhysRevLett.125.146401} {\bibfield  {journal}
  {\bibinfo  {journal} {Phys. Rev. Lett.}\ }\textbf {\bibinfo {volume} {125}},\
  \bibinfo {pages} {146401} (\bibinfo {year} {2020}{\natexlab{b}})}\BibitemShut
  {NoStop}%
\bibitem [{\citenamefont {Ghorashi}\ \emph {et~al.}(2020)\citenamefont
  {Ghorashi}, \citenamefont {Li},\ and\ \citenamefont
  {Hughes}}]{PhysRevLett.125.266804}%
  \BibitemOpen
  \bibfield  {author} {\bibinfo {author} {\bibfnamefont {S.~A.~A.}\
  \bibnamefont {Ghorashi}}, \bibinfo {author} {\bibfnamefont {T.}~\bibnamefont
  {Li}},\ and\ \bibinfo {author} {\bibfnamefont {T.~L.}\ \bibnamefont
  {Hughes}},\ }\bibfield  {title} {\bibinfo {title} {Higher-order {W}eyl
  semimetals},\ }\href {https://doi.org/10.1103/PhysRevLett.125.266804}
  {\bibfield  {journal} {\bibinfo  {journal} {Phys. Rev. Lett.}\ }\textbf
  {\bibinfo {volume} {125}},\ \bibinfo {pages} {266804} (\bibinfo {year}
  {2020})}\BibitemShut {NoStop}%
\bibitem [{\citenamefont {Luo}\ \emph {et~al.}(2021)\citenamefont {Luo},
  \citenamefont {Wang}, \citenamefont {Lin}, \citenamefont {Jiang},
  \citenamefont {Wu}, \citenamefont {Li},\ and\ \citenamefont
  {Jiang}}]{Luo2021}%
  \BibitemOpen
  \bibfield  {author} {\bibinfo {author} {\bibfnamefont {L.}~\bibnamefont
  {Luo}}, \bibinfo {author} {\bibfnamefont {H.-X.}\ \bibnamefont {Wang}},
  \bibinfo {author} {\bibfnamefont {Z.-K.}\ \bibnamefont {Lin}}, \bibinfo
  {author} {\bibfnamefont {B.}~\bibnamefont {Jiang}}, \bibinfo {author}
  {\bibfnamefont {Y.}~\bibnamefont {Wu}}, \bibinfo {author} {\bibfnamefont
  {F.}~\bibnamefont {Li}},\ and\ \bibinfo {author} {\bibfnamefont {J.-H.}\
  \bibnamefont {Jiang}},\ }\bibfield  {title} {\bibinfo {title} {Observation of
  a phononic higher-order {W}eyl semimetal},\ }\href
  {https://doi.org/10.1038/s41563-021-00985-6} {\bibfield  {journal} {\bibinfo
  {journal} {Nature Materials}\ }\textbf {\bibinfo {volume} {20}},\ \bibinfo
  {pages} {794} (\bibinfo {year} {2021})}\BibitemShut {NoStop}%
\bibitem [{\citenamefont {Wei}\ \emph {et~al.}(2021)\citenamefont {Wei},
  \citenamefont {Zhang}, \citenamefont {Deng}, \citenamefont {Lu},
  \citenamefont {Huang}, \citenamefont {Yan}, \citenamefont {Chen},
  \citenamefont {Liu},\ and\ \citenamefont {Jia}}]{Wei2021}%
  \BibitemOpen
  \bibfield  {author} {\bibinfo {author} {\bibfnamefont {Q.}~\bibnamefont
  {Wei}}, \bibinfo {author} {\bibfnamefont {X.}~\bibnamefont {Zhang}}, \bibinfo
  {author} {\bibfnamefont {W.}~\bibnamefont {Deng}}, \bibinfo {author}
  {\bibfnamefont {J.}~\bibnamefont {Lu}}, \bibinfo {author} {\bibfnamefont
  {X.}~\bibnamefont {Huang}}, \bibinfo {author} {\bibfnamefont
  {M.}~\bibnamefont {Yan}}, \bibinfo {author} {\bibfnamefont {G.}~\bibnamefont
  {Chen}}, \bibinfo {author} {\bibfnamefont {Z.}~\bibnamefont {Liu}},\ and\
  \bibinfo {author} {\bibfnamefont {S.}~\bibnamefont {Jia}},\ }\bibfield
  {title} {\bibinfo {title} {Higher-order topological semimetal in acoustic
  crystals},\ }\href {https://doi.org/10.1038/s41563-021-00933-4} {\bibfield
  {journal} {\bibinfo  {journal} {Nature Materials}\ }\textbf {\bibinfo
  {volume} {20}},\ \bibinfo {pages} {812} (\bibinfo {year} {2021})}\BibitemShut
  {NoStop}%
\bibitem [{\citenamefont {Eckardt}(2017)}]{RevModPhys.89.011004}%
  \BibitemOpen
  \bibfield  {author} {\bibinfo {author} {\bibfnamefont {A.}~\bibnamefont
  {Eckardt}},\ }\bibfield  {title} {\bibinfo {title} {Colloquium: Atomic
  quantum gases in periodically driven optical lattices},\ }\href
  {https://doi.org/10.1103/RevModPhys.89.011004} {\bibfield  {journal}
  {\bibinfo  {journal} {Rev. Mod. Phys.}\ }\textbf {\bibinfo {volume} {89}},\
  \bibinfo {pages} {011004} (\bibinfo {year} {2017})}\BibitemShut {NoStop}%
\bibitem [{\citenamefont {Meinert}\ \emph {et~al.}(2016)\citenamefont
  {Meinert}, \citenamefont {Mark}, \citenamefont {Lauber}, \citenamefont
  {Daley},\ and\ \citenamefont {N\"agerl}}]{PhysRevLett.116.205301}%
  \BibitemOpen
  \bibfield  {author} {\bibinfo {author} {\bibfnamefont {F.}~\bibnamefont
  {Meinert}}, \bibinfo {author} {\bibfnamefont {M.~J.}\ \bibnamefont {Mark}},
  \bibinfo {author} {\bibfnamefont {K.}~\bibnamefont {Lauber}}, \bibinfo
  {author} {\bibfnamefont {A.~J.}\ \bibnamefont {Daley}},\ and\ \bibinfo
  {author} {\bibfnamefont {H.-C.}\ \bibnamefont {N\"agerl}},\ }\bibfield
  {title} {\bibinfo {title} {{F}loquet engineering of correlated tunneling in
  the {B}ose-{H}ubbard model with ultracold atoms},\ }\href
  {https://doi.org/10.1103/PhysRevLett.116.205301} {\bibfield  {journal}
  {\bibinfo  {journal} {Phys. Rev. Lett.}\ }\textbf {\bibinfo {volume} {116}},\
  \bibinfo {pages} {205301} (\bibinfo {year} {2016})}\BibitemShut {NoStop}%
\bibitem [{\citenamefont {Rechtsman}\ \emph {et~al.}(2013)\citenamefont
  {Rechtsman}, \citenamefont {Zeuner}, \citenamefont {Plotnik}, \citenamefont
  {Lumer}, \citenamefont {Podolsky}, \citenamefont {Dreisow}, \citenamefont
  {Nolte}, \citenamefont {Segev},\ and\ \citenamefont
  {Szameit}}]{Rechtsman2013}%
  \BibitemOpen
  \bibfield  {author} {\bibinfo {author} {\bibfnamefont {M.~C.}\ \bibnamefont
  {Rechtsman}}, \bibinfo {author} {\bibfnamefont {J.~M.}\ \bibnamefont
  {Zeuner}}, \bibinfo {author} {\bibfnamefont {Y.}~\bibnamefont {Plotnik}},
  \bibinfo {author} {\bibfnamefont {Y.}~\bibnamefont {Lumer}}, \bibinfo
  {author} {\bibfnamefont {D.}~\bibnamefont {Podolsky}}, \bibinfo {author}
  {\bibfnamefont {F.}~\bibnamefont {Dreisow}}, \bibinfo {author} {\bibfnamefont
  {S.}~\bibnamefont {Nolte}}, \bibinfo {author} {\bibfnamefont
  {M.}~\bibnamefont {Segev}},\ and\ \bibinfo {author} {\bibfnamefont
  {A.}~\bibnamefont {Szameit}},\ }\bibfield  {title} {\bibinfo {title}
  {Photonic {F}loquet topological insulators},\ }\href
  {https://doi.org/10.1038/nature12066} {\bibfield  {journal} {\bibinfo
  {journal} {Nature (London)}\ }\textbf {\bibinfo {volume} {496}},\ \bibinfo
  {pages} {196} (\bibinfo {year} {2013})}\BibitemShut {NoStop}%
\bibitem [{\citenamefont {Cheng}\ \emph {et~al.}(2019)\citenamefont {Cheng},
  \citenamefont {Pan}, \citenamefont {Wang}, \citenamefont {Zhang},
  \citenamefont {Yu}, \citenamefont {Gover}, \citenamefont {Zhang},
  \citenamefont {Li}, \citenamefont {Zhou},\ and\ \citenamefont
  {Zhu}}]{PhysRevLett.122.173901}%
  \BibitemOpen
  \bibfield  {author} {\bibinfo {author} {\bibfnamefont {Q.}~\bibnamefont
  {Cheng}}, \bibinfo {author} {\bibfnamefont {Y.}~\bibnamefont {Pan}}, \bibinfo
  {author} {\bibfnamefont {H.}~\bibnamefont {Wang}}, \bibinfo {author}
  {\bibfnamefont {C.}~\bibnamefont {Zhang}}, \bibinfo {author} {\bibfnamefont
  {D.}~\bibnamefont {Yu}}, \bibinfo {author} {\bibfnamefont {A.}~\bibnamefont
  {Gover}}, \bibinfo {author} {\bibfnamefont {H.}~\bibnamefont {Zhang}},
  \bibinfo {author} {\bibfnamefont {T.}~\bibnamefont {Li}}, \bibinfo {author}
  {\bibfnamefont {L.}~\bibnamefont {Zhou}},\ and\ \bibinfo {author}
  {\bibfnamefont {S.}~\bibnamefont {Zhu}},\ }\bibfield  {title} {\bibinfo
  {title} {Observation of anomalous $\ensuremath{\pi}$ modes in photonic
  {F}loquet engineering},\ }\href
  {https://doi.org/10.1103/PhysRevLett.122.173901} {\bibfield  {journal}
  {\bibinfo  {journal} {Phys. Rev. Lett.}\ }\textbf {\bibinfo {volume} {122}},\
  \bibinfo {pages} {173901} (\bibinfo {year} {2019})}\BibitemShut {NoStop}%
\bibitem [{\citenamefont {Roushan}\ \emph {et~al.}(2017)\citenamefont
  {Roushan}, \citenamefont {Neill}, \citenamefont {Megrant}, \citenamefont
  {Chen}, \citenamefont {Babbush}, \citenamefont {Barends}, \citenamefont
  {Campbell}, \citenamefont {Chen}, \citenamefont {Chiaro}, \citenamefont
  {Dunsworth}, \citenamefont {Fowler}, \citenamefont {Jeffrey}, \citenamefont
  {Kelly}, \citenamefont {Lucero}, \citenamefont {Mutus}, \citenamefont {OHuo
  Heng~alley}, \citenamefont {Neeley}, \citenamefont {Quintana}, \citenamefont
  {Sank}, \citenamefont {Vainsencher}, \citenamefont {Wenner}, \citenamefont
  {White}, \citenamefont {Kapit}, \citenamefont {Neven},\ and\ \citenamefont
  {Martinis}}]{Roushan2017}%
  \BibitemOpen
  \bibfield  {author} {\bibinfo {author} {\bibfnamefont {P.}~\bibnamefont
  {Roushan}}, \bibinfo {author} {\bibfnamefont {C.}~\bibnamefont {Neill}},
  \bibinfo {author} {\bibfnamefont {A.}~\bibnamefont {Megrant}}, \bibinfo
  {author} {\bibfnamefont {Y.}~\bibnamefont {Chen}}, \bibinfo {author}
  {\bibfnamefont {R.}~\bibnamefont {Babbush}}, \bibinfo {author} {\bibfnamefont
  {R.}~\bibnamefont {Barends}}, \bibinfo {author} {\bibfnamefont
  {B.}~\bibnamefont {Campbell}}, \bibinfo {author} {\bibfnamefont
  {Z.}~\bibnamefont {Chen}}, \bibinfo {author} {\bibfnamefont {B.}~\bibnamefont
  {Chiaro}}, \bibinfo {author} {\bibfnamefont {A.}~\bibnamefont {Dunsworth}},
  \bibinfo {author} {\bibfnamefont {A.}~\bibnamefont {Fowler}}, \bibinfo
  {author} {\bibfnamefont {E.}~\bibnamefont {Jeffrey}}, \bibinfo {author}
  {\bibfnamefont {J.}~\bibnamefont {Kelly}}, \bibinfo {author} {\bibfnamefont
  {E.}~\bibnamefont {Lucero}}, \bibinfo {author} {\bibfnamefont
  {J.}~\bibnamefont {Mutus}}, \bibinfo {author} {\bibfnamefont {P.~J.~J.}\
  \bibnamefont {OHuo Heng~alley}}, \bibinfo {author} {\bibfnamefont
  {M.}~\bibnamefont {Neeley}}, \bibinfo {author} {\bibfnamefont
  {C.}~\bibnamefont {Quintana}}, \bibinfo {author} {\bibfnamefont
  {D.}~\bibnamefont {Sank}}, \bibinfo {author} {\bibfnamefont {A.}~\bibnamefont
  {Vainsencher}}, \bibinfo {author} {\bibfnamefont {J.}~\bibnamefont {Wenner}},
  \bibinfo {author} {\bibfnamefont {T.}~\bibnamefont {White}}, \bibinfo
  {author} {\bibfnamefont {E.}~\bibnamefont {Kapit}}, \bibinfo {author}
  {\bibfnamefont {H.}~\bibnamefont {Neven}},\ and\ \bibinfo {author}
  {\bibfnamefont {J.}~\bibnamefont {Martinis}},\ }\bibfield  {title} {\bibinfo
  {title} {Chiral ground-state currents of interacting photons in a synthetic
  magnetic field},\ }\href {https://doi.org/10.1038/nphys3930} {\bibfield
  {journal} {\bibinfo  {journal} {Nature Physics}\ }\textbf {\bibinfo {volume}
  {13}},\ \bibinfo {pages} {146} (\bibinfo {year} {2017})}\BibitemShut
  {NoStop}%
\bibitem [{\citenamefont {McIver}\ \emph {et~al.}(2020)\citenamefont {McIver},
  \citenamefont {Schulte}, \citenamefont {Stein}, \citenamefont {Matsuyama},
  \citenamefont {Jotzu}, \citenamefont {Meier},\ and\ \citenamefont
  {Cavalleri}}]{McIver2020}%
  \BibitemOpen
  \bibfield  {author} {\bibinfo {author} {\bibfnamefont {J.~W.}\ \bibnamefont
  {McIver}}, \bibinfo {author} {\bibfnamefont {B.}~\bibnamefont {Schulte}},
  \bibinfo {author} {\bibfnamefont {F.-U.}\ \bibnamefont {Stein}}, \bibinfo
  {author} {\bibfnamefont {T.}~\bibnamefont {Matsuyama}}, \bibinfo {author}
  {\bibfnamefont {G.}~\bibnamefont {Jotzu}}, \bibinfo {author} {\bibfnamefont
  {G.}~\bibnamefont {Meier}},\ and\ \bibinfo {author} {\bibfnamefont
  {A.}~\bibnamefont {Cavalleri}},\ }\bibfield  {title} {\bibinfo {title}
  {Light-induced anomalous {H}all effect in graphene},\ }\href
  {https://doi.org/10.1038/s41567-019-0698-y} {\bibfield  {journal} {\bibinfo
  {journal} {Nature Physics}\ }\textbf {\bibinfo {volume} {16}},\ \bibinfo
  {pages} {38} (\bibinfo {year} {2020})}\BibitemShut {NoStop}%
\bibitem [{\citenamefont {Xiong}\ \emph {et~al.}(2016)\citenamefont {Xiong},
  \citenamefont {Gong},\ and\ \citenamefont {An}}]{PhysRevB.93.184306}%
  \BibitemOpen
  \bibfield  {author} {\bibinfo {author} {\bibfnamefont {T.-S.}\ \bibnamefont
  {Xiong}}, \bibinfo {author} {\bibfnamefont {J.}~\bibnamefont {Gong}},\ and\
  \bibinfo {author} {\bibfnamefont {J.-H.}\ \bibnamefont {An}},\ }\bibfield
  {title} {\bibinfo {title} {Towards large-{C}hern-number topological phases by
  periodic quenching},\ }\href {https://doi.org/10.1103/PhysRevB.93.184306}
  {\bibfield  {journal} {\bibinfo  {journal} {Phys. Rev. B}\ }\textbf {\bibinfo
  {volume} {93}},\ \bibinfo {pages} {184306} (\bibinfo {year}
  {2016})}\BibitemShut {NoStop}%
\bibitem [{\citenamefont {Liu}\ \emph {et~al.}(2019{\natexlab{b}})\citenamefont
  {Liu}, \citenamefont {Xiong}, \citenamefont {Zhang},\ and\ \citenamefont
  {An}}]{PhysRevA.100.023622}%
  \BibitemOpen
  \bibfield  {author} {\bibinfo {author} {\bibfnamefont {H.}~\bibnamefont
  {Liu}}, \bibinfo {author} {\bibfnamefont {T.-S.}\ \bibnamefont {Xiong}},
  \bibinfo {author} {\bibfnamefont {W.}~\bibnamefont {Zhang}},\ and\ \bibinfo
  {author} {\bibfnamefont {J.-H.}\ \bibnamefont {An}},\ }\bibfield  {title}
  {\bibinfo {title} {{F}loquet engineering of exotic topological phases in
  systems of cold atoms},\ }\href {https://doi.org/10.1103/PhysRevA.100.023622}
  {\bibfield  {journal} {\bibinfo  {journal} {Phys. Rev. A}\ }\textbf {\bibinfo
  {volume} {100}},\ \bibinfo {pages} {023622} (\bibinfo {year}
  {2019}{\natexlab{b}})}\BibitemShut {NoStop}%
\bibitem [{\citenamefont {Li}\ \emph {et~al.}(2018)\citenamefont {Li},
  \citenamefont {Lee},\ and\ \citenamefont {Gong}}]{PhysRevLett.121.036401}%
  \BibitemOpen
  \bibfield  {author} {\bibinfo {author} {\bibfnamefont {L.}~\bibnamefont
  {Li}}, \bibinfo {author} {\bibfnamefont {C.~H.}\ \bibnamefont {Lee}},\ and\
  \bibinfo {author} {\bibfnamefont {J.}~\bibnamefont {Gong}},\ }\bibfield
  {title} {\bibinfo {title} {Realistic {F}loquet semimetal with exotic
  topological linkages between arbitrarily many nodal loops},\ }\href
  {https://doi.org/10.1103/PhysRevLett.121.036401} {\bibfield  {journal}
  {\bibinfo  {journal} {Phys. Rev. Lett.}\ }\textbf {\bibinfo {volume} {121}},\
  \bibinfo {pages} {036401} (\bibinfo {year} {2018})}\BibitemShut {NoStop}%
\bibitem [{\citenamefont {Wu}\ and\ \citenamefont
  {An}(2020)}]{PhysRevB.102.041119}%
  \BibitemOpen
  \bibfield  {author} {\bibinfo {author} {\bibfnamefont {H.}~\bibnamefont
  {Wu}}\ and\ \bibinfo {author} {\bibfnamefont {J.-H.}\ \bibnamefont {An}},\
  }\bibfield  {title} {\bibinfo {title} {{F}loquet topological phases of
  non-{H}ermitian systems},\ }\href
  {https://doi.org/10.1103/PhysRevB.102.041119} {\bibfield  {journal} {\bibinfo
   {journal} {Phys. Rev. B}\ }\textbf {\bibinfo {volume} {102}},\ \bibinfo
  {pages} {041119(R)} (\bibinfo {year} {2020})}\BibitemShut {NoStop}%
\bibitem [{\citenamefont {Kundu}\ \emph {et~al.}(2014)\citenamefont {Kundu},
  \citenamefont {Fertig},\ and\ \citenamefont
  {Seradjeh}}]{PhysRevLett.113.236803}%
  \BibitemOpen
  \bibfield  {author} {\bibinfo {author} {\bibfnamefont {A.}~\bibnamefont
  {Kundu}}, \bibinfo {author} {\bibfnamefont {H.~A.}\ \bibnamefont {Fertig}},\
  and\ \bibinfo {author} {\bibfnamefont {B.}~\bibnamefont {Seradjeh}},\
  }\bibfield  {title} {\bibinfo {title} {Effective theory of {F}loquet
  topological transitions},\ }\href
  {https://doi.org/10.1103/PhysRevLett.113.236803} {\bibfield  {journal}
  {\bibinfo  {journal} {Phys. Rev. Lett.}\ }\textbf {\bibinfo {volume} {113}},\
  \bibinfo {pages} {236803} (\bibinfo {year} {2014})}\BibitemShut {NoStop}%
\bibitem [{\citenamefont {Wu}\ \emph {et~al.}(2021)\citenamefont {Wu},
  \citenamefont {Wang},\ and\ \citenamefont {An}}]{PhysRevB.103.L041115}%
  \BibitemOpen
  \bibfield  {author} {\bibinfo {author} {\bibfnamefont {H.}~\bibnamefont
  {Wu}}, \bibinfo {author} {\bibfnamefont {B.-Q.}\ \bibnamefont {Wang}},\ and\
  \bibinfo {author} {\bibfnamefont {J.-H.}\ \bibnamefont {An}},\ }\bibfield
  {title} {\bibinfo {title} {{F}loquet second-order topological insulators in
  non-{H}ermitian systems},\ }\href
  {https://doi.org/10.1103/PhysRevB.103.L041115} {\bibfield  {journal}
  {\bibinfo  {journal} {Phys. Rev. B}\ }\textbf {\bibinfo {volume} {103}},\
  \bibinfo {pages} {L041115} (\bibinfo {year} {2021})}\BibitemShut {NoStop}%
\bibitem [{\citenamefont {Peng}\ and\ \citenamefont
  {Refael}(2019)}]{PhysRevLett.123.016806}%
  \BibitemOpen
  \bibfield  {author} {\bibinfo {author} {\bibfnamefont {Y.}~\bibnamefont
  {Peng}}\ and\ \bibinfo {author} {\bibfnamefont {G.}~\bibnamefont {Refael}},\
  }\bibfield  {title} {\bibinfo {title} {{F}loquet second-order topological
  insulators from nonsymmorphic space-time symmetries},\ }\href
  {https://doi.org/10.1103/PhysRevLett.123.016806} {\bibfield  {journal}
  {\bibinfo  {journal} {Phys. Rev. Lett.}\ }\textbf {\bibinfo {volume} {123}},\
  \bibinfo {pages} {016806} (\bibinfo {year} {2019})}\BibitemShut {NoStop}%
\bibitem [{\citenamefont {Bomantara}\ \emph {et~al.}(2019)\citenamefont
  {Bomantara}, \citenamefont {Zhou}, \citenamefont {Pan},\ and\ \citenamefont
  {Gong}}]{PhysRevB.99.045441}%
  \BibitemOpen
  \bibfield  {author} {\bibinfo {author} {\bibfnamefont {R.~W.}\ \bibnamefont
  {Bomantara}}, \bibinfo {author} {\bibfnamefont {L.}~\bibnamefont {Zhou}},
  \bibinfo {author} {\bibfnamefont {J.}~\bibnamefont {Pan}},\ and\ \bibinfo
  {author} {\bibfnamefont {J.}~\bibnamefont {Gong}},\ }\bibfield  {title}
  {\bibinfo {title} {Coupled-wire construction of static and {F}loquet
  second-order topological insulators},\ }\href
  {https://doi.org/10.1103/PhysRevB.99.045441} {\bibfield  {journal} {\bibinfo
  {journal} {Phys. Rev. B}\ }\textbf {\bibinfo {volume} {99}},\ \bibinfo
  {pages} {045441} (\bibinfo {year} {2019})}\BibitemShut {NoStop}%
\bibitem [{\citenamefont {Seshadri}\ \emph {et~al.}(2019)\citenamefont
  {Seshadri}, \citenamefont {Dutta},\ and\ \citenamefont
  {Sen}}]{PhysRevB.100.115403}%
  \BibitemOpen
  \bibfield  {author} {\bibinfo {author} {\bibfnamefont {R.}~\bibnamefont
  {Seshadri}}, \bibinfo {author} {\bibfnamefont {A.}~\bibnamefont {Dutta}},\
  and\ \bibinfo {author} {\bibfnamefont {D.}~\bibnamefont {Sen}},\ }\bibfield
  {title} {\bibinfo {title} {Generating a second-order topological insulator
  with multiple corner states by periodic driving},\ }\href
  {https://doi.org/10.1103/PhysRevB.100.115403} {\bibfield  {journal} {\bibinfo
   {journal} {Phys. Rev. B}\ }\textbf {\bibinfo {volume} {100}},\ \bibinfo
  {pages} {115403} (\bibinfo {year} {2019})}\BibitemShut {NoStop}%
\bibitem [{\citenamefont {Hu}\ \emph {et~al.}(2020)\citenamefont {Hu},
  \citenamefont {Huang}, \citenamefont {Zhao},\ and\ \citenamefont
  {Liu}}]{PhysRevLett.124.057001}%
  \BibitemOpen
  \bibfield  {author} {\bibinfo {author} {\bibfnamefont {H.}~\bibnamefont
  {Hu}}, \bibinfo {author} {\bibfnamefont {B.}~\bibnamefont {Huang}}, \bibinfo
  {author} {\bibfnamefont {E.}~\bibnamefont {Zhao}},\ and\ \bibinfo {author}
  {\bibfnamefont {W.~V.}\ \bibnamefont {Liu}},\ }\bibfield  {title} {\bibinfo
  {title} {Dynamical singularities of {F}loquet higher-order topological
  insulators},\ }\href {https://doi.org/10.1103/PhysRevLett.124.057001}
  {\bibfield  {journal} {\bibinfo  {journal} {Phys. Rev. Lett.}\ }\textbf
  {\bibinfo {volume} {124}},\ \bibinfo {pages} {057001} (\bibinfo {year}
  {2020})}\BibitemShut {NoStop}%
\bibitem [{\citenamefont {Huang}\ and\ \citenamefont
  {Liu}(2020)}]{PhysRevLett.124.216601}%
  \BibitemOpen
  \bibfield  {author} {\bibinfo {author} {\bibfnamefont {B.}~\bibnamefont
  {Huang}}\ and\ \bibinfo {author} {\bibfnamefont {W.~V.}\ \bibnamefont
  {Liu}},\ }\bibfield  {title} {\bibinfo {title} {Floquet higher-order
  topological insulators with anomalous dynamical polarization},\ }\href
  {https://doi.org/10.1103/PhysRevLett.124.216601} {\bibfield  {journal}
  {\bibinfo  {journal} {Phys. Rev. Lett.}\ }\textbf {\bibinfo {volume} {124}},\
  \bibinfo {pages} {216601} (\bibinfo {year} {2020})}\BibitemShut {NoStop}%
\bibitem [{\citenamefont {Ghosh}\ \emph {et~al.}(2021)\citenamefont {Ghosh},
  \citenamefont {Nag},\ and\ \citenamefont {Saha}}]{PhysRevB.103.045424}%
  \BibitemOpen
  \bibfield  {author} {\bibinfo {author} {\bibfnamefont {A.~K.}\ \bibnamefont
  {Ghosh}}, \bibinfo {author} {\bibfnamefont {T.}~\bibnamefont {Nag}},\ and\
  \bibinfo {author} {\bibfnamefont {A.}~\bibnamefont {Saha}},\ }\bibfield
  {title} {\bibinfo {title} {{F}loquet generation of a second-order topological
  superconductor},\ }\href {https://doi.org/10.1103/PhysRevB.103.045424}
  {\bibfield  {journal} {\bibinfo  {journal} {Phys. Rev. B}\ }\textbf {\bibinfo
  {volume} {103}},\ \bibinfo {pages} {045424} (\bibinfo {year}
  {2021})}\BibitemShut {NoStop}%
\bibitem [{\citenamefont {Nag}\ \emph {et~al.}(2021)\citenamefont {Nag},
  \citenamefont {Juri\ifmmode \check{c}\else \v{c}\fi{}i\ifmmode~\acute{c}\else
  \'{c}\fi{}},\ and\ \citenamefont {Roy}}]{PhysRevB.103.115308}%
  \BibitemOpen
  \bibfield  {author} {\bibinfo {author} {\bibfnamefont {T.}~\bibnamefont
  {Nag}}, \bibinfo {author} {\bibfnamefont {V.}~\bibnamefont {Juri\ifmmode
  \check{c}\else \v{c}\fi{}i\ifmmode~\acute{c}\else \'{c}\fi{}}},\ and\
  \bibinfo {author} {\bibfnamefont {B.}~\bibnamefont {Roy}},\ }\bibfield
  {title} {\bibinfo {title} {Hierarchy of higher-order {F}loquet topological
  phases in three dimensions},\ }\href
  {https://doi.org/10.1103/PhysRevB.103.115308} {\bibfield  {journal} {\bibinfo
   {journal} {Phys. Rev. B}\ }\textbf {\bibinfo {volume} {103}},\ \bibinfo
  {pages} {115308} (\bibinfo {year} {2021})}\BibitemShut {NoStop}%
\bibitem [{\citenamefont {Rodriguez-Vega}\ \emph {et~al.}(2019)\citenamefont
  {Rodriguez-Vega}, \citenamefont {Kumar},\ and\ \citenamefont
  {Seradjeh}}]{PhysRevB.100.085138}%
  \BibitemOpen
  \bibfield  {author} {\bibinfo {author} {\bibfnamefont {M.}~\bibnamefont
  {Rodriguez-Vega}}, \bibinfo {author} {\bibfnamefont {A.}~\bibnamefont
  {Kumar}},\ and\ \bibinfo {author} {\bibfnamefont {B.}~\bibnamefont
  {Seradjeh}},\ }\bibfield  {title} {\bibinfo {title} {Higher-order {F}loquet
  topological phases with corner and bulk bound states},\ }\href
  {https://doi.org/10.1103/PhysRevB.100.085138} {\bibfield  {journal} {\bibinfo
   {journal} {Phys. Rev. B}\ }\textbf {\bibinfo {volume} {100}},\ \bibinfo
  {pages} {085138} (\bibinfo {year} {2019})}\BibitemShut {NoStop}%
\bibitem [{\citenamefont {Rui}\ \emph {et~al.}(2021)\citenamefont {Rui},
  \citenamefont {Zhang}, \citenamefont {Hirschmann}, \citenamefont {Zheng},
  \citenamefont {Schnyder}, \citenamefont {Trauzettel},\ and\ \citenamefont
  {Wang}}]{PhysRevB.103.184510}%
  \BibitemOpen
  \bibfield  {author} {\bibinfo {author} {\bibfnamefont {W.~B.}\ \bibnamefont
  {Rui}}, \bibinfo {author} {\bibfnamefont {S.-B.}\ \bibnamefont {Zhang}},
  \bibinfo {author} {\bibfnamefont {M.~M.}\ \bibnamefont {Hirschmann}},
  \bibinfo {author} {\bibfnamefont {Z.}~\bibnamefont {Zheng}}, \bibinfo
  {author} {\bibfnamefont {A.~P.}\ \bibnamefont {Schnyder}}, \bibinfo {author}
  {\bibfnamefont {B.}~\bibnamefont {Trauzettel}},\ and\ \bibinfo {author}
  {\bibfnamefont {Z.~D.}\ \bibnamefont {Wang}},\ }\bibfield  {title} {\bibinfo
  {title} {Higher-order {W}eyl superconductors with anisotropic {W}eyl-point
  connectivity},\ }\href {https://doi.org/10.1103/PhysRevB.103.184510}
  {\bibfield  {journal} {\bibinfo  {journal} {Phys. Rev. B}\ }\textbf {\bibinfo
  {volume} {103}},\ \bibinfo {pages} {184510} (\bibinfo {year}
  {2021})}\BibitemShut {NoStop}%
\bibitem [{SMP()}]{SMP}%
  \BibitemOpen
  \href@noop {} {}See the Supplemental Material for the real-space Hamiltonian,
  the definitions of the mirror-graded winding number and the chirality, the
  symmetry analysis, the derivation of the condition to form the {D}irac/{W}eyl
  nodes, and the definition of Wilson loop.\BibitemShut {Stop}%
\bibitem [{\citenamefont {Benalcazar}\ \emph
  {et~al.}(2017{\natexlab{b}})\citenamefont {Benalcazar}, \citenamefont
  {Bernevig},\ and\ \citenamefont {Hughes}}]{PhysRevB.96.245115}%
  \BibitemOpen
  \bibfield  {author} {\bibinfo {author} {\bibfnamefont {W.~A.}\ \bibnamefont
  {Benalcazar}}, \bibinfo {author} {\bibfnamefont {B.~A.}\ \bibnamefont
  {Bernevig}},\ and\ \bibinfo {author} {\bibfnamefont {T.~L.}\ \bibnamefont
  {Hughes}},\ }\bibfield  {title} {\bibinfo {title} {Electric multipole
  moments, topological multipole moment pumping, and chiral hinge states in
  crystalline insulators},\ }\href {https://doi.org/10.1103/PhysRevB.96.245115}
  {\bibfield  {journal} {\bibinfo  {journal} {Phys. Rev. B}\ }\textbf {\bibinfo
  {volume} {96}},\ \bibinfo {pages} {245115} (\bibinfo {year}
  {2017}{\natexlab{b}})}\BibitemShut {NoStop}%
\bibitem [{\citenamefont {Sambe}(1973)}]{PhysRevA.7.2203}%
  \BibitemOpen
  \bibfield  {author} {\bibinfo {author} {\bibfnamefont {H.}~\bibnamefont
  {Sambe}},\ }\bibfield  {title} {\bibinfo {title} {Steady states and
  quasienergies of a quantum-mechanical system in an oscillating field},\
  }\href {https://doi.org/10.1103/PhysRevA.7.2203} {\bibfield  {journal}
  {\bibinfo  {journal} {Phys. Rev. A}\ }\textbf {\bibinfo {volume} {7}},\
  \bibinfo {pages} {2203} (\bibinfo {year} {1973})}\BibitemShut {NoStop}%
\bibitem [{\citenamefont {Chen}\ \emph
  {et~al.}(2015{\natexlab{b}})\citenamefont {Chen}, \citenamefont {An},
  \citenamefont {Luo}, \citenamefont {Sun},\ and\ \citenamefont
  {Oh}}]{PhysRevA.91.052122}%
  \BibitemOpen
  \bibfield  {author} {\bibinfo {author} {\bibfnamefont {C.}~\bibnamefont
  {Chen}}, \bibinfo {author} {\bibfnamefont {J.-H.}\ \bibnamefont {An}},
  \bibinfo {author} {\bibfnamefont {H.-G.}\ \bibnamefont {Luo}}, \bibinfo
  {author} {\bibfnamefont {C.~P.}\ \bibnamefont {Sun}},\ and\ \bibinfo {author}
  {\bibfnamefont {C.~H.}\ \bibnamefont {Oh}},\ }\bibfield  {title} {\bibinfo
  {title} {{F}loquet control of quantum dissipation in spin chains},\ }\href
  {https://doi.org/10.1103/PhysRevA.91.052122} {\bibfield  {journal} {\bibinfo
  {journal} {Phys. Rev. A}\ }\textbf {\bibinfo {volume} {91}},\ \bibinfo
  {pages} {052122} (\bibinfo {year} {2015}{\natexlab{b}})}\BibitemShut
  {NoStop}%
\bibitem [{\citenamefont {Lu}\ \emph {et~al.}(2020)\citenamefont {Lu},
  \citenamefont {Wang},\ and\ \citenamefont {Liu}}]{LU20202080}%
  \BibitemOpen
  \bibfield  {author} {\bibinfo {author} {\bibfnamefont {Y.-H.}\ \bibnamefont
  {Lu}}, \bibinfo {author} {\bibfnamefont {B.-Z.}\ \bibnamefont {Wang}},\ and\
  \bibinfo {author} {\bibfnamefont {X.-J.}\ \bibnamefont {Liu}},\ }\bibfield
  {title} {\bibinfo {title} {Ideal {W}eyl semimetal with 3d spin-orbit coupled
  ultracold quantum gas},\ }\href
  {https://doi.org/https://doi.org/10.1016/j.scib.2020.09.036} {\bibfield
  {journal} {\bibinfo  {journal} {Science Bulletin}\ }\textbf {\bibinfo
  {volume} {65}},\ \bibinfo {pages} {2080} (\bibinfo {year}
  {2020})}\BibitemShut {NoStop}%
\bibitem [{\citenamefont {Li}\ \emph {et~al.}(2020)\citenamefont {Li},
  \citenamefont {Wang}, \citenamefont {Li}, \citenamefont {Zheng},
  \citenamefont {Brinkman}, \citenamefont {Yu},\ and\ \citenamefont
  {Liao}}]{PhysRevLett.124.156601}%
  \BibitemOpen
  \bibfield  {author} {\bibinfo {author} {\bibfnamefont {C.-Z.}\ \bibnamefont
  {Li}}, \bibinfo {author} {\bibfnamefont {A.-Q.}\ \bibnamefont {Wang}},
  \bibinfo {author} {\bibfnamefont {C.}~\bibnamefont {Li}}, \bibinfo {author}
  {\bibfnamefont {W.-Z.}\ \bibnamefont {Zheng}}, \bibinfo {author}
  {\bibfnamefont {A.}~\bibnamefont {Brinkman}}, \bibinfo {author}
  {\bibfnamefont {D.-P.}\ \bibnamefont {Yu}},\ and\ \bibinfo {author}
  {\bibfnamefont {Z.-M.}\ \bibnamefont {Liao}},\ }\bibfield  {title} {\bibinfo
  {title} {Reducing electronic transport dimension to topological hinge states
  by increasing geometry size of dirac semimetal josephson junctions},\ }\href
  {https://doi.org/10.1103/PhysRevLett.124.156601} {\bibfield  {journal}
  {\bibinfo  {journal} {Phys. Rev. Lett.}\ }\textbf {\bibinfo {volume} {124}},\
  \bibinfo {pages} {156601} (\bibinfo {year} {2020})}\BibitemShut {NoStop}%
\bibitem [{\citenamefont {Mahmood}\ \emph {et~al.}(2016)\citenamefont
  {Mahmood}, \citenamefont {Chan}, \citenamefont {Alpichshev}, \citenamefont
  {Gardner}, \citenamefont {Lee}, \citenamefont {Lee},\ and\ \citenamefont
  {Gedik}}]{Mahmood2016}%
  \BibitemOpen
  \bibfield  {author} {\bibinfo {author} {\bibfnamefont {F.}~\bibnamefont
  {Mahmood}}, \bibinfo {author} {\bibfnamefont {C.-K.}\ \bibnamefont {Chan}},
  \bibinfo {author} {\bibfnamefont {Z.}~\bibnamefont {Alpichshev}}, \bibinfo
  {author} {\bibfnamefont {D.}~\bibnamefont {Gardner}}, \bibinfo {author}
  {\bibfnamefont {Y.}~\bibnamefont {Lee}}, \bibinfo {author} {\bibfnamefont
  {P.~A.}\ \bibnamefont {Lee}},\ and\ \bibinfo {author} {\bibfnamefont
  {N.}~\bibnamefont {Gedik}},\ }\bibfield  {title} {\bibinfo {title} {Selective
  scattering between floquet-bloch and volkov states in a topological
  insulator},\ }\href {https://doi.org/10.1038/nphys3609} {\bibfield  {journal}
  {\bibinfo  {journal} {Nat. Phys.}\ }\textbf {\bibinfo {volume} {12}},\
  \bibinfo {pages} {306} (\bibinfo {year} {2016})}\BibitemShut {NoStop}%
\bibitem [{\citenamefont {Wintersperger}\ \emph {et~al.}(2020)\citenamefont
  {Wintersperger}, \citenamefont {Braun}, \citenamefont {Ünal}, \citenamefont
  {Eckardt}, \citenamefont {Liberto}, \citenamefont {Goldman}, \citenamefont
  {Bloch},\ and\ \citenamefont {Aidelsburger}}]{Wintersperger2020}%
  \BibitemOpen
  \bibfield  {author} {\bibinfo {author} {\bibfnamefont {K.}~\bibnamefont
  {Wintersperger}}, \bibinfo {author} {\bibfnamefont {C.}~\bibnamefont
  {Braun}}, \bibinfo {author} {\bibfnamefont {F.~N.}\ \bibnamefont {Ünal}},
  \bibinfo {author} {\bibfnamefont {A.}~\bibnamefont {Eckardt}}, \bibinfo
  {author} {\bibfnamefont {M.~D.}\ \bibnamefont {Liberto}}, \bibinfo {author}
  {\bibfnamefont {N.}~\bibnamefont {Goldman}}, \bibinfo {author} {\bibfnamefont
  {I.}~\bibnamefont {Bloch}},\ and\ \bibinfo {author} {\bibfnamefont
  {M.}~\bibnamefont {Aidelsburger}},\ }\bibfield  {title} {\bibinfo {title}
  {Realization of an anomalous floquet topological system with ultracold
  atoms},\ }\href {https://doi.org/10.1038/s41567-020-0949-y} {\bibfield
  {journal} {\bibinfo  {journal} {Nat. Phys.}\ }\textbf {\bibinfo {volume}
  {16}},\ \bibinfo {pages} {1058} (\bibinfo {year} {2020})}\BibitemShut
  {NoStop}%
\bibitem [{\citenamefont {Mukherjee}\ \emph {et~al.}(2017)\citenamefont
  {Mukherjee}, \citenamefont {Spracklen}, \citenamefont {Valiente},
  \citenamefont {Andersson}, \citenamefont {Öhberg}, \citenamefont {Goldman},\
  and\ \citenamefont {Thomson}}]{Mukherjee2017}%
  \BibitemOpen
  \bibfield  {author} {\bibinfo {author} {\bibfnamefont {S.}~\bibnamefont
  {Mukherjee}}, \bibinfo {author} {\bibfnamefont {A.}~\bibnamefont
  {Spracklen}}, \bibinfo {author} {\bibfnamefont {M.}~\bibnamefont {Valiente}},
  \bibinfo {author} {\bibfnamefont {E.}~\bibnamefont {Andersson}}, \bibinfo
  {author} {\bibfnamefont {P.}~\bibnamefont {Öhberg}}, \bibinfo {author}
  {\bibfnamefont {N.}~\bibnamefont {Goldman}},\ and\ \bibinfo {author}
  {\bibfnamefont {R.~R.}\ \bibnamefont {Thomson}},\ }\bibfield  {title}
  {\bibinfo {title} {Experimental observation of anomalous topological edge
  modes in a slowly driven photonic lattice},\ }\href
  {https://doi.org/10.1038/ncomms13918} {\bibfield  {journal} {\bibinfo
  {journal} {Nature Communications}\ }\textbf {\bibinfo {volume} {8}},\
  \bibinfo {pages} {13918} (\bibinfo {year} {2017})}\BibitemShut {NoStop}%
\bibitem [{\citenamefont {Maczewsky}\ \emph {et~al.}(2017)\citenamefont
  {Maczewsky}, \citenamefont {Zeuner}, \citenamefont {Nolte},\ and\
  \citenamefont {Szameit}}]{Maczewsky2017}%
  \BibitemOpen
  \bibfield  {author} {\bibinfo {author} {\bibfnamefont {L.~J.}\ \bibnamefont
  {Maczewsky}}, \bibinfo {author} {\bibfnamefont {J.~M.}\ \bibnamefont
  {Zeuner}}, \bibinfo {author} {\bibfnamefont {S.}~\bibnamefont {Nolte}},\ and\
  \bibinfo {author} {\bibfnamefont {A.}~\bibnamefont {Szameit}},\ }\bibfield
  {title} {\bibinfo {title} {Observation of photonic anomalous floquet
  topological insulators},\ }\href {https://doi.org/10.1038/ncomms13756}
  {\bibfield  {journal} {\bibinfo  {journal} {Nat. Commun.}\ }\textbf {\bibinfo
  {volume} {8}},\ \bibinfo {pages} {13756} (\bibinfo {year}
  {2017})}\BibitemShut {NoStop}%
\end{thebibliography}%


\begin{thebibliography}{5}%
\makeatletter
\providecommand \@ifxundefined [1]{%
 \@ifx{#1\undefined}
}%
\providecommand \@ifnum [1]{%
 \ifnum #1\expandafter \@firstoftwo
 \else \expandafter \@secondoftwo
 \fi
}%
\providecommand \@ifx [1]{%
 \ifx #1\expandafter \@firstoftwo
 \else \expandafter \@secondoftwo
 \fi
}%
\providecommand \natexlab [1]{#1}%
\providecommand \enquote  [1]{``#1''}%
\providecommand \bibnamefont  [1]{#1}%
\providecommand \bibfnamefont [1]{#1}%
\providecommand \citenamefont [1]{#1}%
\providecommand \href@noop [0]{\@secondoftwo}%
\providecommand \href [0]{\begingroup \@sanitize@url \@href}%
\providecommand \@href[1]{\@@startlink{#1}\@@href}%
\providecommand \@@href[1]{\endgroup#1\@@endlink}%
\providecommand \@sanitize@url [0]{\catcode `\\12\catcode `\$12\catcode
  `\&12\catcode `\#12\catcode `\^12\catcode `\_12\catcode `\%12\relax}%
\providecommand \@@startlink[1]{}%
\providecommand \@@endlink[0]{}%
\providecommand \url  [0]{\begingroup\@sanitize@url \@url }%
\providecommand \@url [1]{\endgroup\@href {#1}{\urlprefix }}%
\providecommand \urlprefix  [0]{URL }%
\providecommand \Eprint [0]{\href }%
\providecommand \doibase [0]{https://doi.org/}%
\providecommand \selectlanguage [0]{\@gobble}%
\providecommand \bibinfo  [0]{\@secondoftwo}%
\providecommand \bibfield  [0]{\@secondoftwo}%
\providecommand \translation [1]{[#1]}%
\providecommand \BibitemOpen [0]{}%
\providecommand \bibitemStop [0]{}%
\providecommand \bibitemNoStop [0]{.\EOS\space}%
\providecommand \EOS [0]{\spacefactor3000\relax}%
\providecommand \BibitemShut  [1]{\csname bibitem#1\endcsname}%
\let\auto@bib@innerbib\@empty
%</preamble>
\bibitem [{\citenamefont {Benalcazar}\ \emph
  {et~al.}(2017{\natexlab{a}})\citenamefont {Benalcazar}, \citenamefont
  {Bernevig},\ and\ \citenamefont {Hughes}}]{Benalcazar61}%
  \BibitemOpen
  \bibfield  {author} {\bibinfo {author} {\bibfnamefont {W.~A.}\ \bibnamefont
  {Benalcazar}}, \bibinfo {author} {\bibfnamefont {B.~A.}\ \bibnamefont
  {Bernevig}},\ and\ \bibinfo {author} {\bibfnamefont {T.~L.}\ \bibnamefont
  {Hughes}},\ }\bibfield  {title} {\bibinfo {title} {Quantized electric
  multipole insulators},\ }\href {https://doi.org/10.1126/science.aah6442}
  {\bibfield  {journal} {\bibinfo  {journal} {Science}\ }\textbf {\bibinfo
  {volume} {357}},\ \bibinfo {pages} {61} (\bibinfo {year}
  {2017}{\natexlab{a}})}\BibitemShut {NoStop}%
\bibitem [{\citenamefont {Benalcazar}\ \emph
  {et~al.}(2017{\natexlab{b}})\citenamefont {Benalcazar}, \citenamefont
  {Bernevig},\ and\ \citenamefont {Hughes}}]{PhysRevB.96.245115}%
  \BibitemOpen
  \bibfield  {author} {\bibinfo {author} {\bibfnamefont {W.~A.}\ \bibnamefont
  {Benalcazar}}, \bibinfo {author} {\bibfnamefont {B.~A.}\ \bibnamefont
  {Bernevig}},\ and\ \bibinfo {author} {\bibfnamefont {T.~L.}\ \bibnamefont
  {Hughes}},\ }\bibfield  {title} {\bibinfo {title} {Electric multipole
  moments, topological multipole moment pumping, and chiral hinge states in
  crystalline insulators},\ }\href {https://doi.org/10.1103/PhysRevB.96.245115}
  {\bibfield  {journal} {\bibinfo  {journal} {Phys. Rev. B}\ }\textbf {\bibinfo
  {volume} {96}},\ \bibinfo {pages} {245115} (\bibinfo {year}
  {2017}{\natexlab{b}})}\BibitemShut {NoStop}%
\bibitem [{\citenamefont {Imhof}\ \emph {et~al.}(2018)\citenamefont {Imhof},
  \citenamefont {Berger}, \citenamefont {Bayer}, \citenamefont {Brehm},
  \citenamefont {Molenkamp}, \citenamefont {Kiessling}, \citenamefont
  {Schindler}, \citenamefont {Lee}, \citenamefont {Greiter}, \citenamefont
  {Neupert},\ and\ \citenamefont {Thomale}}]{Imhof2018}%
  \BibitemOpen
  \bibfield  {author} {\bibinfo {author} {\bibfnamefont {S.}~\bibnamefont
  {Imhof}}, \bibinfo {author} {\bibfnamefont {C.}~\bibnamefont {Berger}},
  \bibinfo {author} {\bibfnamefont {F.}~\bibnamefont {Bayer}}, \bibinfo
  {author} {\bibfnamefont {J.}~\bibnamefont {Brehm}}, \bibinfo {author}
  {\bibfnamefont {L.~W.}\ \bibnamefont {Molenkamp}}, \bibinfo {author}
  {\bibfnamefont {T.}~\bibnamefont {Kiessling}}, \bibinfo {author}
  {\bibfnamefont {F.}~\bibnamefont {Schindler}}, \bibinfo {author}
  {\bibfnamefont {C.~H.}\ \bibnamefont {Lee}}, \bibinfo {author} {\bibfnamefont
  {M.}~\bibnamefont {Greiter}}, \bibinfo {author} {\bibfnamefont
  {T.}~\bibnamefont {Neupert}},\ and\ \bibinfo {author} {\bibfnamefont
  {R.}~\bibnamefont {Thomale}},\ }\bibfield  {title} {\bibinfo {title}
  {Topolectrical-circuit realization of topological corner modes},\ }\href
  {https://doi.org/10.1038/s41567-018-0246-1} {\bibfield  {journal} {\bibinfo
  {journal} {Nature Physics}\ }\textbf {\bibinfo {volume} {14}},\ \bibinfo
  {pages} {925} (\bibinfo {year} {2018})}\BibitemShut {NoStop}%
\bibitem [{\citenamefont {Burkov}\ \emph {et~al.}(2011)\citenamefont {Burkov},
  \citenamefont {Hook},\ and\ \citenamefont {Balents}}]{PhysRevB.84.235126}%
  \BibitemOpen
  \bibfield  {author} {\bibinfo {author} {\bibfnamefont {A.~A.}\ \bibnamefont
  {Burkov}}, \bibinfo {author} {\bibfnamefont {M.~D.}\ \bibnamefont {Hook}},\
  and\ \bibinfo {author} {\bibfnamefont {L.}~\bibnamefont {Balents}},\
  }\bibfield  {title} {\bibinfo {title} {Topological nodal semimetals},\ }\href
  {https://doi.org/10.1103/PhysRevB.84.235126} {\bibfield  {journal} {\bibinfo
  {journal} {Phys. Rev. B}\ }\textbf {\bibinfo {volume} {84}},\ \bibinfo
  {pages} {235126} (\bibinfo {year} {2011})}\BibitemShut {NoStop}%
\bibitem [{\citenamefont {Wang}\ \emph {et~al.}(2020)\citenamefont {Wang},
  \citenamefont {Lin}, \citenamefont {Jiang}, \citenamefont {Guo},\ and\
  \citenamefont {Jiang}}]{PhysRevLett.125.146401}%
  \BibitemOpen
  \bibfield  {author} {\bibinfo {author} {\bibfnamefont {H.-X.}\ \bibnamefont
  {Wang}}, \bibinfo {author} {\bibfnamefont {Z.-K.}\ \bibnamefont {Lin}},
  \bibinfo {author} {\bibfnamefont {B.}~\bibnamefont {Jiang}}, \bibinfo
  {author} {\bibfnamefont {G.-Y.}\ \bibnamefont {Guo}},\ and\ \bibinfo {author}
  {\bibfnamefont {J.-H.}\ \bibnamefont {Jiang}},\ }\bibfield  {title} {\bibinfo
  {title} {Higher-order {W}eyl semimetals},\ }\href
  {https://doi.org/10.1103/PhysRevLett.125.146401} {\bibfield  {journal}
  {\bibinfo  {journal} {Phys. Rev. Lett.}\ }\textbf {\bibinfo {volume} {125}},\
  \bibinfo {pages} {146401} (\bibinfo {year} {2020})}\BibitemShut {NoStop}%
\end{thebibliography}%
\end{document}

\title{Supplemental material for ``Engineering exotic second-order topological semimetals by periodic driving''}
\author{Bao-Qin Wang}\thanks{These two authors contributed equally.}
\affiliation{Lanzhou Center for Theoretical Physics, Key Laboratory of Theoretical Physics of Gansu Province, Lanzhou University, Lanzhou 730000, China}
\author{Hong Wu}\thanks{These two authors contributed equally.}
\affiliation{Lanzhou Center for Theoretical Physics, Key Laboratory of Theoretical Physics of Gansu Province, Lanzhou University, Lanzhou 730000, China}
\author{Jun-Hong An}
\email{anjhong@lzu.edu.cn}
\affiliation{Lanzhou Center for Theoretical Physics, Key Laboratory of Theoretical Physics of Gansu Province, Lanzhou University, Lanzhou 730000, China}
\maketitle

\tableofcontents
\section{Real-space Hamiltonian and mirror-graded winding number}
We investigate a system of spinless fermions moving on a 3D lattice. Its real-space Hamiltonian reads
\begin{eqnarray}
\hat{H}=&&\sum_{\mathbf{r}}\{\gamma[\hat{c}_{\mathbf{r},1}^{\dagger}(\hat{c}_{\mathbf{r},3}+\hat{c}_{\mathbf{r},4})+\hat{c}_{\mathbf{r},2}^{\dagger}(\hat{c}_{\mathbf{r},4}
-\hat{c}_{\mathbf{r},3})]\nonumber\\
&&+\lambda[\hat{c}_{\mathbf{r},1}^{\dagger}(\hat{c}_{\mathbf{r}+x,3}+\hat{c}_{\mathbf{r}+y,4})+\hat{c}_{\mathbf{r},2}^{\dagger}(\hat{c}_{\mathbf{r}-x,4}
-\hat{c}_{\mathbf{r}-y,3})]\nonumber\\
&&+\frac{a}{2}[\hat{c}_{\mathbf{r},1}^{\dagger}(\hat{c}_{\mathbf{r}+x+z,3}+\hat{c}_{\mathbf{r}+y+z,4})+\hat{c}_{\mathbf{r},2}^{\dagger}(\hat{c}_{\mathbf{r}-x+z,4}\nonumber\\
&&-\hat{c}_{\mathbf{r}-y+z,3})+\hat{c}_{\mathbf{r},3}^{\dagger}(\hat{c}_{\mathbf{r}-x+z,1}-\hat{c}_{\mathbf{r}+y+z,2})\nonumber\\
&&+\hat{c}_{\mathbf{r},4}^{\dagger}(\hat{c}_{\mathbf{r}+x+z,2}+\hat{c}_{\mathbf{r}-y+z,1})]+\text{H.c.}\},\label{Hamilr}
\end{eqnarray}
where $\hat{c}_{\mathbf{r},i} (i=1,2,3,4)$ is the annihilation operator of the fermion at sublattice $i$ of unit-cell site $\mathbf{r}=(x,y,z)$, $\lambda$, $\gamma$, and $a$ are the intercell, intracell, and interlayer hopping rates, respectively. Our system is a 3D generalization of the Benalcazar-Bernevig-Hughes (BBH) model \cite{Benalcazar61,PhysRevB.96.245115}, which is a 2D second-order topological insulators (SOTIs), by further considering the coupling between different layers. The 3D second-order topological semimetal (SOTSM) can be sliced into the stacking of 2D SOTIs and normal insulators.

The momentum-space Hamiltonian under the periodic boundary condition along all the three directions reads $\hat{H}=\sum_{\bf k}\hat{\bf C}_{\bf k}^\dag\mathcal{H}({\bf k})\hat{\bf C}_{\bf k}$ with $\hat{\bf C}^\dag_{\bf k}=( \begin{array}{cccc} \hat{c}_{{\bf k},1}^\dag & \hat{c}_{{\bf k},2}^\dag  & \hat{c}_{{\bf k},3}^\dag  & \hat{c}_{{\bf k},4}^\dag  \\\end{array})$ and
\begin{eqnarray}
\mathcal{H}(\mathbf{k})&=&[\gamma+\chi(k_{z})\cos k_{x}]\Gamma_5-\chi(k_{z})\sin k_{x}\Gamma_3\nonumber\\
&&-[\gamma+\chi(k_{z})\cos k_{y}]\Gamma_2-\chi(k_{z})\sin k_{y}\Gamma_1,\label{bloch}
\end{eqnarray}
where $\chi(k_z)=\lambda+a\cos k_z$, $\Gamma_1=\tau_{y}\sigma_{x}$, $\Gamma_2=\tau_{y}\sigma_{y}$, $\Gamma_3=\tau_{y}\sigma_{z}$, $\Gamma_4=\tau_z\sigma_0$, and $\Gamma_5=\tau_{x}\sigma_{0}$, with $\tau_i$ and $\sigma_i$ being Pauli matrices, $\tau_0$ and $\sigma_0$ being identity matrices.

The second-order topology of the system with mirror-rotation and chiral symmetries can be described by the mirror-graded winding number \cite{Imhof2018}. The mirror-graded winding number is defined in the Hamiltonian $\mathcal{H}(\theta,\theta,k_z)$ along the high-symmetry line $k_x=k_y\equiv\theta$. After diagonalizing $\mathcal{H}(\theta,\theta,k_z)$ into diag[$\mathcal{H}^{+}(\theta,k_z)$,$\mathcal{H}^{-}(\theta,k_z)$]with $\mathcal{H}^\pm(\theta,k_z)={\bf h}^\pm\cdot{\pmb\sigma}$ and ${\bf h}^\pm=\sqrt{2}[\gamma+\chi(k_{z})\cos \theta,\pm\chi(k_{z})\sin \theta,0]$, we define the mirror-graded winding number as \begin{equation}\mathcal{W}(k_z)=(\mathcal{W}_{+}-\mathcal{W}_{-})/2.
\end{equation}Here $\mathcal{W}_\pm$ are the winding number associated with $\mathcal{H}^{\pm}(\theta,k_z)$ as
\begin{eqnarray}
\mathcal{W}_{\pm}&=&\frac{i}{2\pi}\int_{0}^{2\pi}\langle u_{\pm}(\theta,k_z)|\partial_{\theta}|u_{\pm}(\theta,k_z)\rangle d\theta\nonumber\\
&=&\frac{1}{2\pi}\int_{0}^{2\pi}(\underline{\mathbf{h}}^\pm \times \partial_\theta\underline{\mathbf{h}}^\pm) d\theta,
\end{eqnarray}
where $\underline{\mathbf{h}}^\pm=\mathbf{h}^\pm/|\mathbf{h}^\pm|$ and $|u_{\pm}(\theta,k_z)\rangle$ are the eigenstates of $\mathcal{H}^{\pm}(\theta,k_z)$.

\section{Chirality of Dirac points}
Each Dirac node has a well-defined chirality. The chirality for the first-order node has been defined in Ref. \cite{PhysRevB.84.235126}. We here give a definition of the chirality for a second-order one. Choosing a closed path $c$ encircling the Dirac node $(k_0,k_0,k_{z,0})$, we define its chirality as
\begin{equation}
\mathcal{Q}=\frac{-i}{4\pi}\oint_{c}[\langle u_{+}(\mathbf{k})\lvert {\pmb\nabla}_{\mathbf{k}}\lvert u_{+}(\mathbf{k})\rangle - \langle u_{-}(\mathbf{k})\lvert {\pmb\nabla}_{\mathbf{k}}\lvert u_{-}(\mathbf{k})\rangle]\cdot d\mathbf{k},\label{smq}
\end{equation}
where $\lvert u_{\pm}(\mathbf{k})\rangle$ are the eigenstates of $\mathcal{H}^{\pm}(\mathbf{k})$ with the mirror-rotation symmetry.

\begin{figure}[tbp]
\centering
\includegraphics[width=.98\columnwidth]{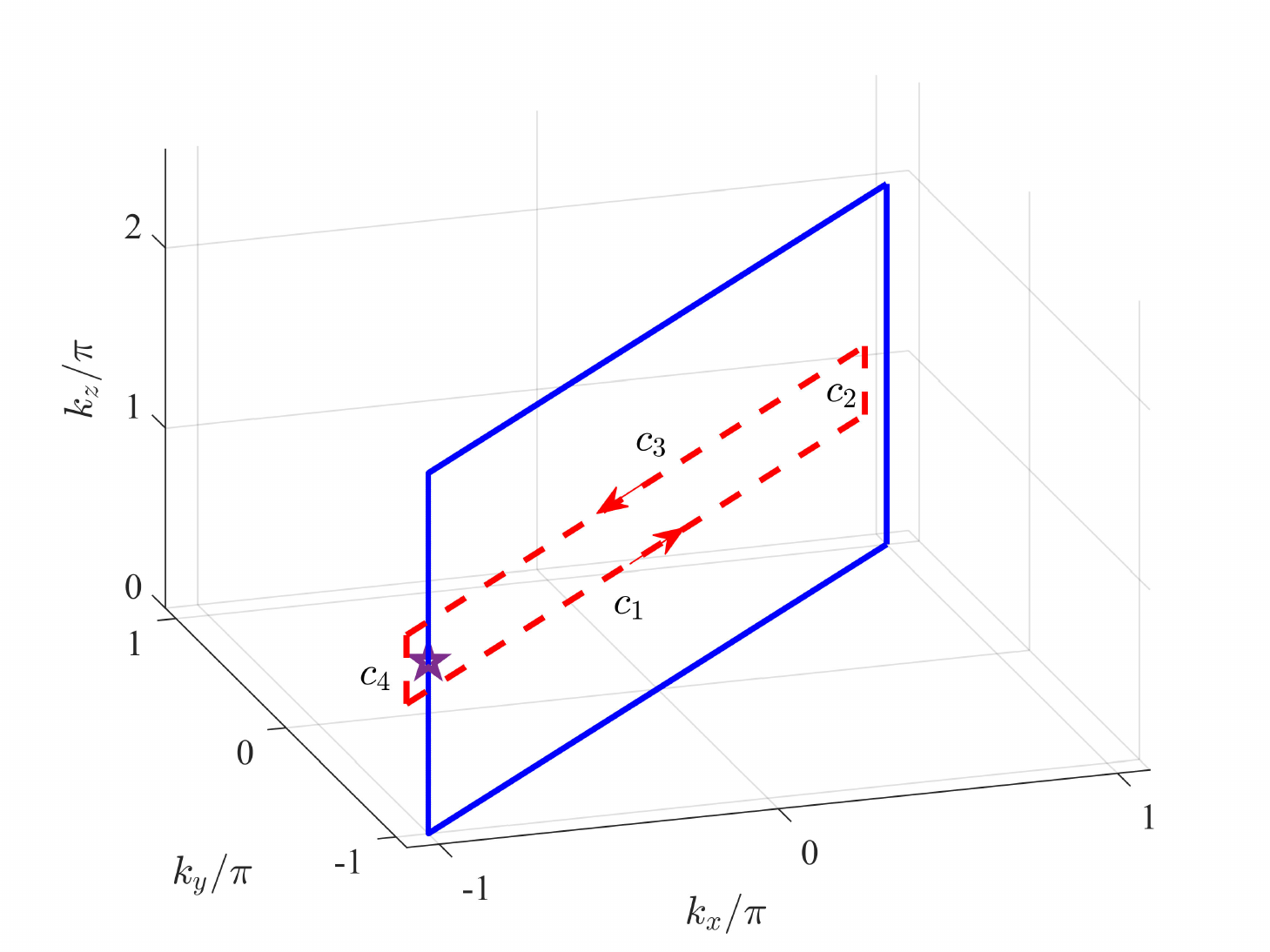}
 \caption{Closed path (red dashed line) to calculate the chirality of the Dirac points marked by purple star. The blue solid line is the boundary of the Brillouin zone.}
 \label{splcdp}
\end{figure}
It can be proven that the chirality of the second-order Dirac node equals exactly to the difference  between the mirror-graded winding numbers of the two topological phases separated by this Dirac node. In order to prove this, we choose for convenience a rectangle path depicted in Fig. \ref{splcdp} as $c_1$: from $(-\pi-\delta,-\pi-\delta,k_{z,0}-\delta)$ to $(\pi-\delta,\pi-\delta,k_{z,0}-\delta)$, $c_2$: from $(\pi-\delta,\pi-\delta,k_{z,0}-\delta)$ to $(\pi-\delta,\pi-\delta,k_{z,0}+\delta)$, $c_3$: from $(\pi-\delta,\pi-\delta,k_{z,0}+\delta)$ to $(-\pi-\delta,-\pi-\delta,k_{z,0}+\delta)$, and $c_4$: from $(-\pi-\delta,-\pi-\delta,k_{z,0}+\delta)$ to $(-\pi-\delta,-\pi-\delta,k_{z,0}-\delta)$, where $\delta$ is an infinitesimal. Then the chirality reads $\mathcal{Q}=\sum_{j=1}^4\mathcal{Q}_j$ with $\mathcal{Q}_j$ being the chirality contributed by the path $c_j$. It can be readily see that $\mathcal{Q}_2$ and $\mathcal{Q}_4$ have the same integral function but along opposite integral paths $c_2$ and $c_4$. Thus we have $\mathcal{Q}_2+\mathcal{Q}_4=0$. According to the definition of mirror-graded winding number, we have $\mathcal{Q}_1=\mathcal{W}(k_{z,0}+\delta)$ and $\mathcal{Q}_3=-\mathcal{W}(k_{z,0}-\delta)$. Therefore, we obtain $\mathcal{Q}=\mathcal{W}(k_{z,0}+\delta)-\mathcal{W}(k_{z,0}-\delta)$.

To numerically confirm this conclusion, we use Eq. \eqref{smq} to calculate the chirality of the Dirac point at ${\bf k}_0=(0, 0,\frac{\pi}{2})$ when $\gamma=-0.2f$ in Fig. 1(b) of the main text. Making the low-energy expansion to $\mathcal{H}^{\pm}(\theta,k_z)$ near the Dirac node ${\bf k}_0$, we obtain
\begin{equation}
\mathcal{H}^{\pm}(\theta,k_z)=\sqrt{2}[-a(k_z-{\pi\over2})\sigma_x\pm\lambda\theta\sigma_y].
\end{equation}Their eigenstates $|u_\pm(\theta,k_z)\rangle$ corresponding to the respective smaller eigenvalues can be readily calculated.
By choosing the closed path $c$ enclosing the Dirac node ${\bf k}_0$, i.e., $\theta=A\cos\beta$ and $k_z-{\pi\over 2}=A\sin\beta$, we can calculate from Eq. \eqref{smq}
\begin{eqnarray}
\mathcal{Q}_{{\bf k}_0}=\frac{1}{2\pi}\int_{0}^{2\pi}\frac{a\lambda}{a^2\sin^2\beta+\lambda^2\cos^2\beta}d\beta=1.
\end{eqnarray}
It is the same as the value calculating from the difference of the mirror-graded winding numbers of its neighboring topological phases, where $\mathcal{W}(k_{z,0}+\delta)=0$ and $\mathcal{W}(k_{z,0}-\delta)=-1$.

\section{Floquet Hamiltonian}

According to $\Gamma _{i}\Gamma _{j}+\Gamma _{j}\Gamma _{i}=2\delta_{ij}I_{4\times 4}$ and $\Gamma _{i}^{2}=I_{4\times 4}$, we have $(\mathbf{n}\cdot \mathbf{\Gamma })^{2}=n^{2}=E^{2}$ and $e^{-i\alpha \mathbf{n}\cdot\mathbf{\Gamma }}=\cos (\alpha E)-i\underline{\mathbf{n}}\cdot \mathbf{\Gamma }\sin (\alpha E)$ with $\mathbf{n}=E\underline{\mathbf{n}}$.
Therefore, the one-period evolution operator can be expanded as\begin{widetext}
\begin{eqnarray}
U_{T} &=&e^{-i\mathbf{n}_{2}\cdot \mathbf{\Gamma }T_{2}}e^{-i\mathbf{n}%
_{1}\cdot \mathbf{\Gamma }T_{1}}=\cos (E_{1}T_{1})\cos (E_{2}T_{2})-\sin (E_{1}T_{1})\sin (E_{2}T_{2})\Big[%
\underline{\mathbf{n}}_{1}\cdot \underline{\mathbf{n}}_{2}+\sum_{j\neq k}^{5}%
\underline{n}_{1j}\underline{n}_{2k}\Gamma _{j}\Gamma _{k}\Big] \nonumber\\
&&-i\underline{\mathbf{n}}_{1}\cdot \mathbf{\Gamma }\sin (E_{1}T_{1})\cos
(E_{2}T_{2})-i\underline{\mathbf{n}}_{2}\cdot \mathbf{\Gamma }\cos
(E_{1}T_{1})\sin (E_{2}T_{2})\equiv A I_{4\times 4}-iB,
\end{eqnarray}
where $A$ and $B$ are%
\begin{eqnarray}
A &=&\cos (E_{1}T_{1})\cos (E_{2}T_{2})-\sin (E_{1}T_{1})\sin (E_{2}T_{2})%
\underline{\mathbf{n}}_{1}\cdot \underline{\mathbf{n}}_{2},\label{smaa} \\
B &=&-i\sin (E_{1}T_{1})\sin (E_{2}T_{2})\sum_{j\neq k}^{5}\underline{n}_{1j}%
\underline{n}_{2k}\Gamma _{j}\Gamma _{k}+\underline{\mathbf{n}}_{1}\cdot
\mathbf{\Gamma }\sin (E_{1}T_{1})\cos (E_{2}T_{2})+\underline{\mathbf{n}}%
_{2}\cdot \mathbf{\Gamma }\cos (E_{1}T_{1})\sin (E_{2}T_{2}).
\end{eqnarray}\end{widetext}
The unitariness of $U_{T}$ requires
that $A^{2}I_{4\times 4}+B^{2}=I_{4\times 4}$. It indicates that $B^2=\mathcal{B}^2I_{4\times 4}$ and $A^2+\mathcal{B}^2=1$. Thus we have
\begin{eqnarray}
U_T&=&\cos (\arccos A)I_{4\times 4}-i\frac{B}{\mathcal{B}}\sin (\arccos A)\nonumber\\
&=&\exp \big[-i\frac{B}{\mathcal{B}}\arccos A\big].
\end{eqnarray}%
Then according to $\mathcal{H}_\text{eff}={i\over T}\ln U_T$, the effective Hamiltonian reads%
\begin{equation}
\mathcal{H}_{\text{eff}}=\frac{\arccos A}{T}\frac{B}{%
\mathcal{B}}.
\end{equation}%
Using the fact that the eigenvalues of $B$ are $\pm\mathcal{B}$, we readily obtain the eigenvalues of $\mathcal{H}_{\text{eff}}$ are%
\begin{equation}
\varepsilon =\pm \frac{\arccos A}{T}.
\end{equation}%
When $A=1$, the band touching point occurs at $\varepsilon=0$. When $A=-1$, the band touching point occurs at $\varepsilon=\pi/T$. According to Eq. \eqref{smaa}, the bands touch at $\varepsilon=0$ for the points of $\mathbf{k}$ and driving parameters which satisfy either
\begin{eqnarray}
&&T_{j}E_{j}=z_{j}\pi , \\
\text{or}~ &&%
\begin{cases}
\underline{\mathbf{n}}_{1}\cdot \underline{\mathbf{n}}_{2}=\pm 1 \\
T_{1}{E}_{1}\pm T_{2}{E}_{2}=z\pi ,%
\end{cases}%
\end{eqnarray}%
where $z_{1}$ and $z_{2}$ are integers with same parity, and $z$ is even
number. The bands touch at $\varepsilon =\pi /T$ for the points of $\mathbf{k%
}$ and the driving parameters which satisfy either
\begin{eqnarray}
&&T_{j}E_{j}=z_{j}\pi , \\
\text{or}~ &&%
\begin{cases}
\underline{\mathbf{n}}_{1}\cdot \underline{\mathbf{n}}_{2}=\pm 1 \\
T_{1}{E}_{1}\pm T_{2}{E}_{2}=z\pi ,%
\end{cases}%
\end{eqnarray}%
where $z_{1}$ and $z_{2}$ are integers with different parities, and $z$ is
odd number.

\section{Symmetry analysis}
It can be verified that the static Hamiltonian (1) in the main text possesses the time-reversal symmetry $\mathcal{T}\mathcal{H}_j({\bf k})\mathcal{T}^{-1}=\mathcal{H}^*_j({\bf k})=\mathcal{H}_j(-{\bf k})$, the chiral symmetry $\mathcal{S}\mathcal{H}_j({\bf k})=-\mathcal{H}_j({\bf k})$, the inversion symmetry $\mathcal{P}\mathcal{H}_j({\bf k})\mathcal{P}^{-1}=\mathcal{H}_j(-{\bf k})$, and the mirror-rotation symmetry $\mathcal{M}_{xy}\mathcal{H}_j({\bf k})\mathcal{M}^{-1}_{xy}=\mathcal{H}_j(k_y,k_x,k_z)$.
It can be proven that the latter two symmetries are preserved, while the former two symmetries are broken by our periodic driving. According to $\mathcal{H}_\text{eff}({\bf k})\equiv {i\over T}\ln U_T({\bf k})$ and $U_T({\bf k})=e^{-i\mathcal{H}_2({\bf k})T_2}e^{-i\mathcal{H}_1({\bf k})T_1}$, we have
\begin{eqnarray}
\mathcal{P} U_T({\bf k})\mathcal{P}^{-1}&=&e^{-i\mathcal{H}_2(-{\bf k})T_2}e^{-i\mathcal{H}_1(-{\bf k})T_1}\nonumber\\
&=& U_T(-{\bf k}),~~~~~~\\
\mathcal{M}_{xy} U_T({\bf k})\mathcal{M}_{xy}^{-1}&=&e^{-i\mathcal{H}_2(k_y,k_x,k_z)T_2}e^{-i\mathcal{H}_1(k_y,k_x,k_z)T_1}\nonumber\\
&=& U_T(k_y,k_x,k_z),~
\end{eqnarray} which results in that the inversion and mirror-rotation symmetries are inherited by $\mathcal{H}_\text{eff}({\bf k})$. On the other hand, because
\begin{eqnarray}
\mathcal{T} U_T({\bf k})\mathcal{T}^{-1}&=&e^{i\mathcal{H}_2(-{\bf k})T_2}e^{i\mathcal{H}_1(-{\bf k})T_1}\neq U^\dag_T(-{\bf k}),~~\\
\mathcal{S} U_T({\bf k})\mathcal{S}^{-1}&=&e^{i\mathcal{H}_2({\bf k})T_2}e^{i\mathcal{H}_1({\bf k})T_1}\neq U^{-1}_T({\bf k}),
\end{eqnarray} the time-reversal and chiral symmetries are broken by $\mathcal{H}_\text{eff}({\bf k})$.

However, the time-reversal and chiral symmetries can be simultaneously recovered by the unitary transformation $G_1({\bf k})=e^{-i\mathcal{H}_1({\bf k})T_1/2}$ and $G_2({\bf k})=e^{i\mathcal{H}_2({\bf k})T_2/2}$. Using $\tilde{\mathcal H}_{\text{eff},j}({\bf k})={i\over T}\ln\tilde{U}_{T,j}({\bf k})$ with $\tilde{U}_{T,j}({\bf k})=G_j({\bf k})U_T({\bf k})G_j^{-1}({\bf k})$, we have
\begin{eqnarray}
\mathcal{T}\tilde{U}_{T,1}({\bf k})\mathcal{T}^{-1}&=&e^{i\mathcal{H}_1(-{\bf k})T_1/2}e^{i\mathcal{H}_2(-{\bf k})T_2}e^{i\mathcal{H}_1(-{\bf k})T_1/2}\nonumber\\
&=&\tilde{U}^\dag_{T,1}(-{\bf k}),\label{smtt1}\\
\mathcal{T}\tilde{U}_{T,2}({\bf k})\mathcal{T}^{-1}&=&e^{i\mathcal{H}_2(-{\bf k})T_2/2}e^{i\mathcal{H}_1(-{\bf k})T_1}e^{i\mathcal{H}_2(-{\bf k})T_2/2}\nonumber\\
&=&\tilde{U}^\dag_{T,2}(-{\bf k}),\label{smtt2}
\end{eqnarray}which means that the time-reversal symmetry $\mathcal{T}\tilde{\mathcal H}_{\text{eff},j}({\bf k})\mathcal{T}^{-1}=\tilde{\mathcal H}_{\text{eff},j}(-{\bf k})$ is recovered in $\tilde{H}_{\text{eff},j}({\bf k})$. In the similar manner, we can prove
\begin{eqnarray}
\mathcal{S}\tilde{U}_{T,1}({\bf k})\mathcal{S}^{-1}&=&e^{i\mathcal{H}_1({\bf k})T_1/2}e^{i\mathcal{H}_2({\bf k})T_2}e^{i\mathcal{H}_1({\bf k})T_1/2}\nonumber\\
&=&\tilde{U}^{-1}_{T,1}({\bf k}),\label{smcr1}\\
\mathcal{S}\tilde{U}_{T,2}({\bf k})\mathcal{S}^{-1}&=&e^{i\mathcal{H}_2({\bf k})T_2/2}e^{i\mathcal{H}_1({\bf k})T_1}e^{i\mathcal{H}_2({\bf k})T_2/2}\nonumber\\
&=&\tilde{U}^{-1}_{T,2}({\bf k}),\label{smcr2}
\end{eqnarray}which means that the chiral symmetry $\mathcal{S}\tilde{\mathcal H}_{\text{eff},j}({\bf k})\mathcal{S}^{-1}=-\tilde{\mathcal H}_{\text{eff},j}({\bf k})$ is also recovered in $\tilde{\mathcal H}_{\text{eff},j}({\bf k})$. With these two symmetries recovered, we can use the similar way as the static system to establish the bulk-corner correspondence by defining the proper topological invariants for our periodically driven system.

It is interesting to note that the original effective Hamiltonian $\mathcal{H}_\text{eff}({\bf k})$ equivalently possesses a hidden time-reversal symmetry and a hidden chiral symmetry after performing the inverse unitary transformations. After rewriting Eqs. \eqref{smtt1} and \eqref{smtt2} as
\begin{equation}
\mathcal{T}G_j({\bf k})U_{T}({\bf k})G^{-1}_j({\bf k})\mathcal{T}^{-1}=G_j(-{\bf k})U^\dag_{T}(-{\bf k})G_j^{-1}(-{\bf k}),
\end{equation}we readily obtain
\begin{equation}
G^{-1}_j(-{\bf k})\mathcal{T}G_j({\bf k})U_{T}({\bf k})G^{-1}_j({\bf k})\mathcal{T}^{-1}G_j(-{\bf k})=U^\dag_{T}(-{\bf k}).
\end{equation}It implies that the original effective $\mathcal{H}_\text{eff}({\bf k})$ possesses the time-reversal symmetry under the newly defined time-reversal operator $G^{-1}_j(-{\bf k})\mathcal{T}G_j({\bf k})$. Similarly, Eqs. \eqref{smcr1} and \eqref{smcr2} result in
\begin{equation}
G^{-1}_j({\bf k})\mathcal{S}G_j({\bf k})U_{T}({\bf k})G^{-1}_j({\bf k})\mathcal{S}^{-1}G_j({\bf k})=U^{-1}_{T}({\bf k}).
\end{equation}It implies that $\mathcal{H}_\text{eff}({\bf k})$ possesses the chiral symmetry under the newly defined chiral operator $G^{-1}_j({\bf k})\mathcal{S}G_j({\bf k})$.

\section{Weyl nodes}

\begin{figure}[tbp]
\centering
\includegraphics[width=.98 \columnwidth]{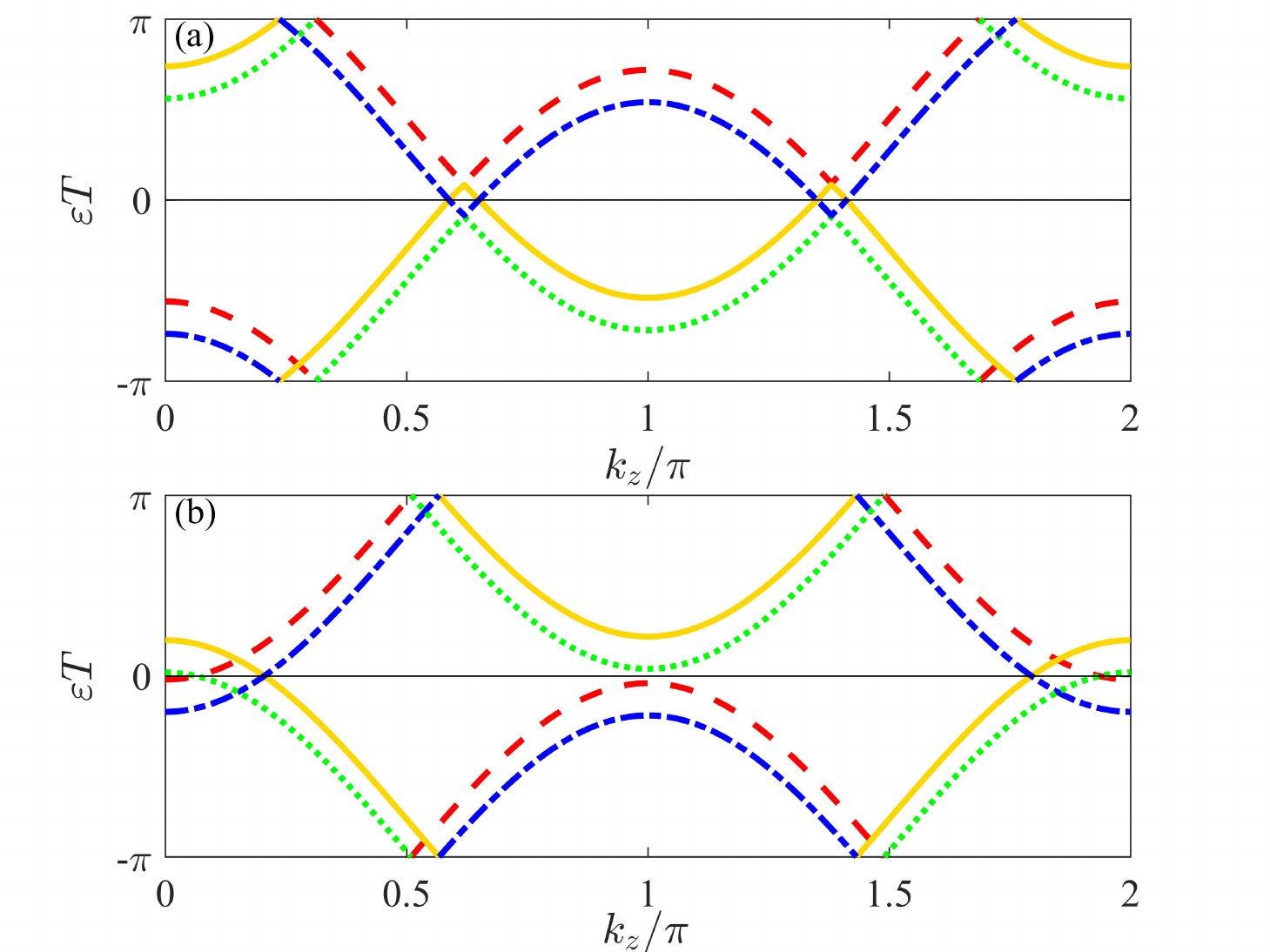}
 \caption{Quasienergies $\pm \varepsilon_\pm(\theta,k_z)$ with $\theta=0$ (a) and $\pi$ (b) of $\mathcal{H}_\text{eff}(\theta,k_z)$. We use the parameters same as Fig. 4 in the main text.}
 \label{splwp}
\end{figure}
The Hamiltonian along the high-symmetry lines $k_x=k_y=\theta=0$ or $\pi$ satisfies $[\mathcal{H}_1(\theta,k_z)+ip\Gamma_1\Gamma_3,\mathcal{H}_2(\theta,k_z)+ip\Gamma_1\Gamma_3]=0$. Thus we have
\begin{eqnarray}
\mathcal{H}_\text{eff}(\theta,k_z)&=&\sum_{j=1,2}[\gamma_j+\chi(k_z)\cos \theta](\Gamma_5-\Gamma_2)T_j/T\nonumber\\
&&+ip\Gamma_1\Gamma_3(T_1+T_2)/T.
\end{eqnarray}
Its eigenvalues are $\pm \varepsilon_\pm(\theta,k_z)$, where
\begin{eqnarray}
 \varepsilon_\pm(\theta,k_z)&=& p(T_1+T_2)/T\pm\sqrt{2}|\gamma_1T_1+\gamma_2T_2\nonumber\\
 &&+e^{i\theta}\chi(k_z)(T_1+T_2)|/T.
\end{eqnarray}
The Weyl points are present if
\begin{equation}
\varepsilon_\pm(\theta,k_z)=n_{\theta,\pm}\pi/T.\label{smtcond}
\end{equation}

We plot in Fig. \ref{splwp} the four quasienergies $\pm \varepsilon_\pm(\theta,k_z)$ along the high-symmetry line $\theta=0$ and $\pi$. They explain well the Weyl points formed in Fig. 4(a) of the main text. The Weyl points at $k_z=0.06\pi$ and $0.51\pi$ are reproduced by Eq. \eqref{smtcond} with $n_{\pi,+}=2$ and $3$, respectively. The ones at $k_z=0.21\pi$ and $0.57\pi$ are reproduced by Eq. \eqref{smtcond} with $n_{\pi,-}=-2$ and $-3$, respectively. The ones at $k_z=0.24\pi$ and $0.59\pi$ are reproduced by Eq. \eqref{smtcond} with $n_{0,-}=-1$ and $0$, respectively. The ones $k_z=0.31\pi$ and $0.65\pi$ are reproduced by Eq. \eqref{smtcond} with $n_{0,+}=1$ and $n_{0,-}=0$, respectively.
\section{Wilson loop}
The Chern insulator is characterized by the Chern number. The Chern number relates to the Wannier center $\frac{-i}{2\pi}\log[W(k_y,k_z)]$ as \cite{PhysRevLett.125.146401}
\begin{equation}
\mathcal{C}(k_z)=\frac{-i}{2\pi}\int_0^{2\pi}\partial_{k_y}\log[W(k_y,k_z)]dk_y,
\end{equation}
where $W(k_y,k_z)$ is called Wilson loop. The Wilson loop is defined by the multiplication of the discretized Berry connections along $k_x$, i.e.
\begin{equation}
W(k_y,k_z)=\prod_{j=0}^{N-1}\langle u(k_x+(j+1)\Delta,k_y,k_z) \lvert u(k_x+j\Delta ,k_y,k_z) \rangle
\end{equation}
where $\lvert u(k_x,k_y,k_z)\rangle$ is the eigen state of $\mathcal{H}_\text{eff}({\bf k})$ and $\Delta=2\pi/N$. The Wannier center itself can also act as a quantification of the topological phase. If the Wannier center $\frac{-i}{2\pi}\log[W(k_y,k_z)]$ changes from $-0.5$ to $0.5$ when $k_y$ runs over the full Brillouin zone, then the system is a Chern insulator with one pair of chiral boundary state formed.

\begin{figure}[tbp]
\centering
\includegraphics[width=.98 \columnwidth]{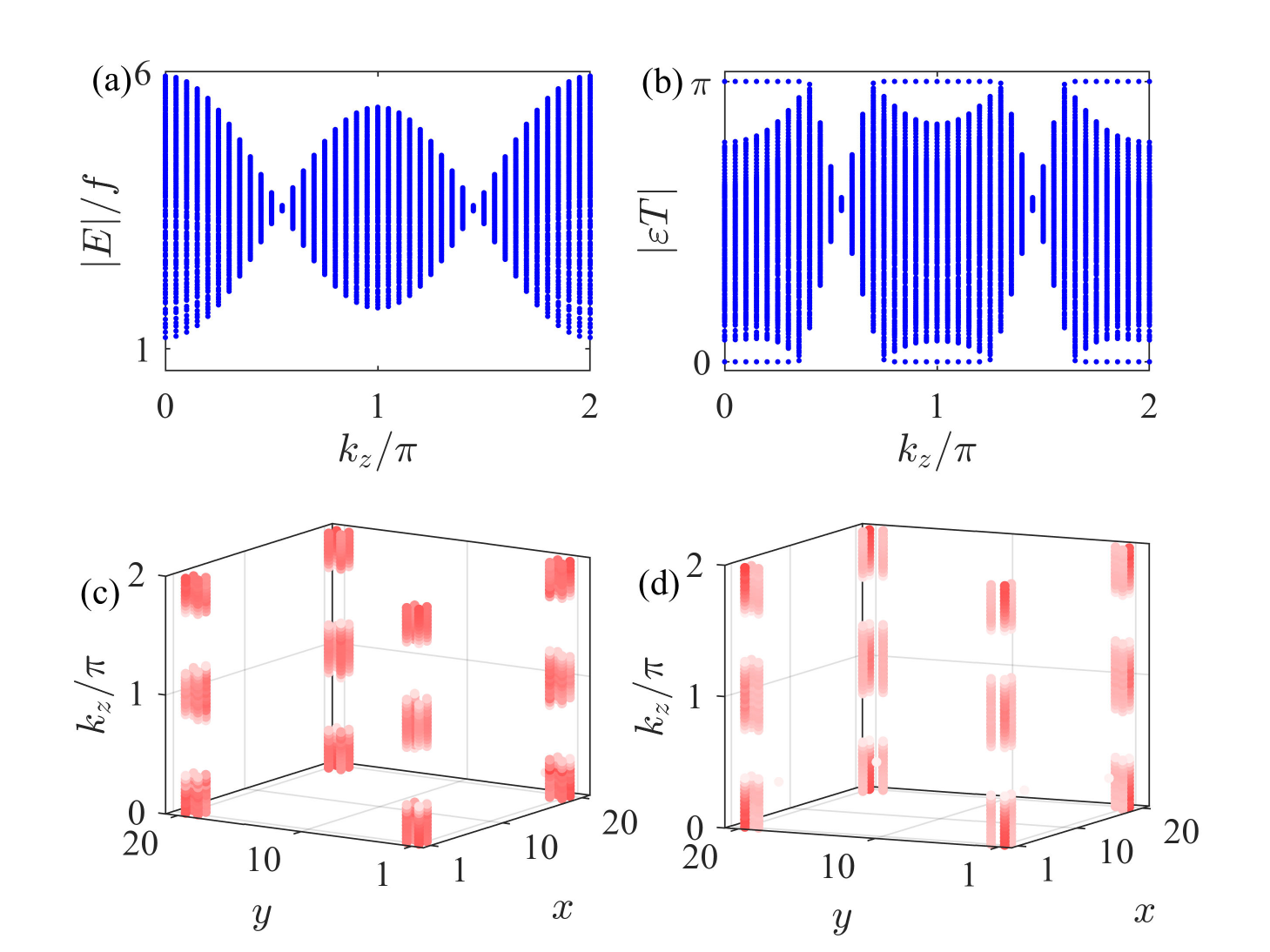}
 \caption{(a) Energy spectrum of the static systems $H_1$ and $H_2$ under the open-boundary condition along $x$ and $y$ directions. (b) Quasienergy spectrum of the periodically changed system between $H_1$ and $H_2$ under the open-boundary condition along $x$ and $y$ directions. Hinge Fermi arcs of the zero modes in (c) and the $\pi/T$ mode in (d). The parameters are $T_1=0.5f^{-1}$, $T_2=f^{-1}$, $a=1.5f$, $\lambda=0.2f$, and $q_2=q_1=-2.5$.      }
 \label{sptrivtosec}
\end{figure}

\section{Conversion of different orders of semimetals}

\begin{figure}[tbp]
\centering
\includegraphics[width=.98 \columnwidth]{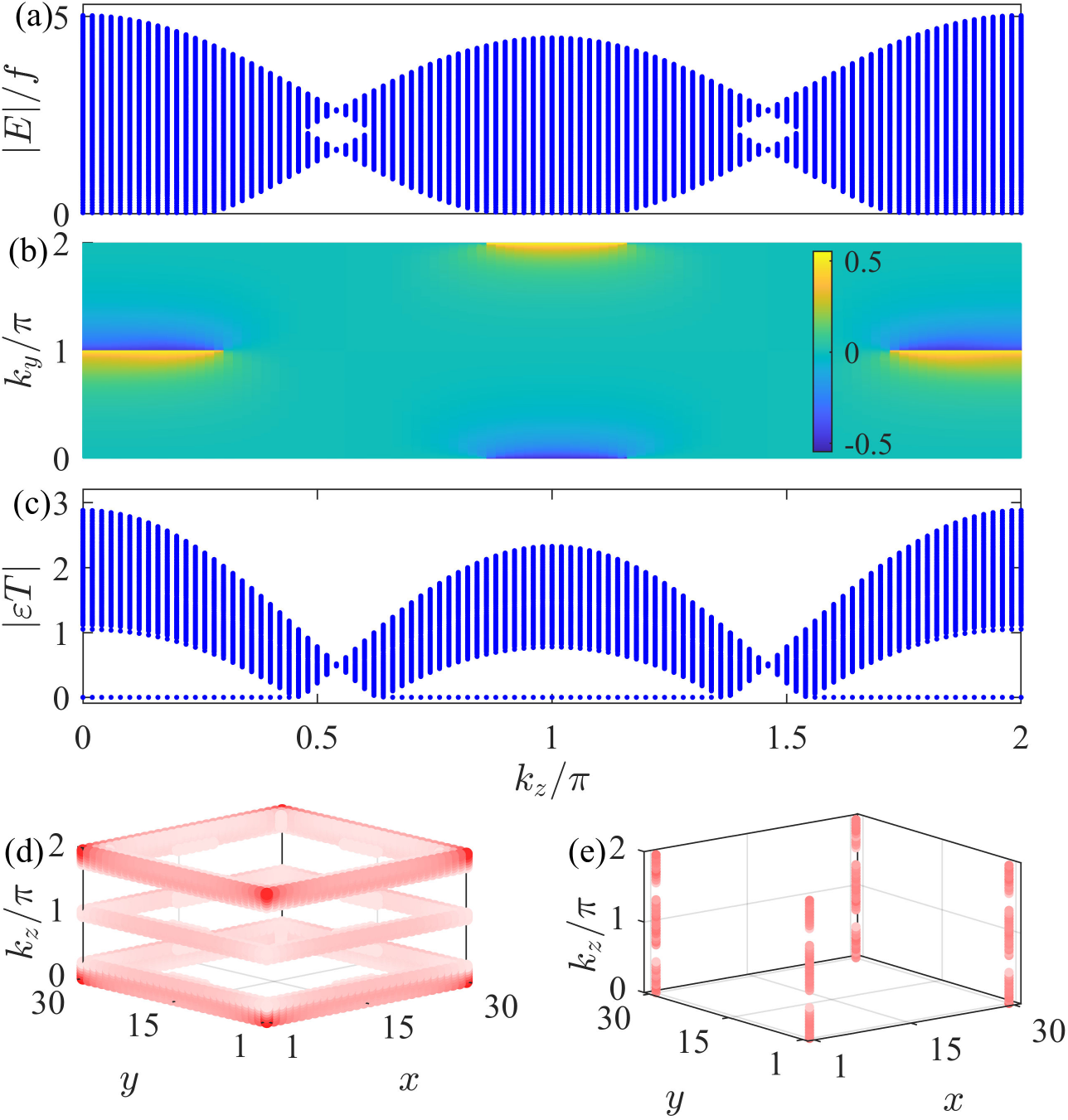}
 \caption{(a) Energy spectrum under the open-boundary condition along $x$ and $y$ directions and (b) Wannier centers for the zero mode of the static systems $H_1$ and $H_2$. (c) Quasienergy spectrum of the periodically changed system between $H_1$ and $H_2$ under the open-boundary condition along $x$ and $y$ directions. Surface Fermi arcs of the zero modes corresponding to the static systems $H_1$ and $H_2$ in (d). Hinge Fermi arcs of the zero modes corresponding to the periodically driven system in (e). The parameters are $T_1=T_2=0.5f^{-1}$, $a=1.5f$, $\lambda=0.2f$, $p=0.5f$, and $q_2=-q_1=1.5$.}
 \label{spfsttosec}
\end{figure}

The periodic driving also possesses the ability to realize the inter-conversion of different orders of topological semimetals. More interesting, not only the second-order topological semimetal from the static first-order semimetal, but also the one from the static normal insulator can be realized by the periodic driving.

Figure \ref{sptrivtosec} shows the result that a second-order Dirac semimetal is produced from the static normal insulator by the periodic driving. The energy spectrum of the static Hamiltonian $H_1$ in Fig. \ref{sptrivtosec}(a), which is the same as the one of $H_2$, indicates that there is no band touching point and thus both of $H_1$ and $H_2$ are topologically trivial. When $H_1$ and $H_2$ are periodically inter-changed under the protocol (2) in the main text, a series of Dirac nodal points and zero- and $\pi/T$-mode bound states are formed in the quasienergy spectrum [see Fig. \ref{sptrivtosec}(b)]. The corresponding probability distributions of the zero- and $\pi/T$-mode bound states in Figs. \ref{sptrivtosec}(c) and \ref{sptrivtosec}(d) signify the presence of the hinge Fermi arcs. It confirms the second-order nature of the formed Dirac nodal points in Fig. \ref{sptrivtosec}(b). Therefore, we succeed in converting the static normal insulator to a second-order Dirac semimetal by our periodic driving protocol.

Figure \ref{sptrivtosec} shows the result that a second-order Weyl semimetal is produced from the static first-order Weyl semimetal by the periodic driving. The energy spectrum of the static system $H_1$ in Fig. \ref{sptrivtosec}(a), which is the same as the one of $H_2$, reveals three continuous band touching lines. The 2D Chern insulators characterized by the gapless chiral boundary states are formed within these band touching lines. They can be topologically witnessed by the Wannier center in Fig. \ref{spfsttosec}(b), where the jump of the Wannier center from $-0.5$ to $0.5$ when $k_y$ runs from $0$ to $2\pi$ verifies the formation of a first-order Chern band. The probability distribution of the gapless chiral boundary states contributes to the surface Fermi arcs [see Fig. \ref{sptrivtosec}(d)]. It demonstrates that both of the static systems $H_1$ and $H_2$ are the first-order Weyl semimetal, which is sliced into a family of 2D Chern insulators and normal insulators separated by the Weyl points. When $H_1$ and $H_2$ are periodically inter-changed under the protocol (2) in the main text, it is interesting to find from the quasienergy spectrum in Fig. \ref{spfsttosec}(c) that a series of 2D zero-mode corner states are formed. The probability distributions of these corner states form the hinge Fermi arcs [see Fig. \ref{sptrivtosec}(e)]. Thus, the periodically driven system is a second-order Weyl semimetal. By this example, we succeed in converting the static first-order Weyl semimetal to a second-order one by the periodic driving.

\bibliography{references}